\documentclass[a4paper,11pt]{article}
\pdfoutput=1 
\usepackage{jcappub}
\usepackage[T1]{fontenc}
\author[a,1]{Asta~Heinesen,\note{Corresponding author.}}
\author[b]{Chris~Blake,}
\author[a]{David~L.~Wiltshire}

\affiliation[a]{School of Physical \& Chemical Sciences, University of Canterbury,\\
Private Bag 4800, Christchurch 8140, New Zealand}
\affiliation[b]{Centre for Astrophysics \& Supercomputing, Swinburne University of Technology,\\ P.O. Box 218, Hawthorn, VIC 3122, Australia}

\emailAdd{asta.heinesen@pg.canterbury.ac.nz}
\emailAdd{cblake@swin.edu.au}
\emailAdd{david.wiltshire@canterbury.ac.nz}

\usepackage{braket}
\usepackage{dirtytalk}
\usepackage{changes}
\usepackage{subcaption}
\usepackage[section]{placeins}
\usepackage[titletoc]{appendix}
\usepackage{xcolor}
\usepackage{amssymb}
\usepackage{dsfont}
\usepackage{bm}
\usepackage{mathtools}
\usepackage{enumerate}
\let\mathbb\mathds

\DeclarePairedDelimiter\abs{\lvert}{\rvert}%

\providecommand{\href}[2]{#2}\def\link#1{\href{http://arxiv.org/abs/#1}{{\tt #1}}}
\def\etal{{et al}.}
\def\sgn{\mathop{\rm sgn}\nolimits}
\def\lsim{\mathop{\hbox{${\lower3.8pt\hbox{$<$}}\atop{\raise0.2pt\hbox{$\sim$}}
$}}}
\def\LCDM{$\Lambda$CDM}

\def\fv{f_{\rm v}}

\def\hm{\;\hbox{Mpc}/h} 
\def\transpose{{\raise2pt\hbox{$\scriptstyle\intercal$}}}

\def\ovo#1{\overset{\circ}{#1}}
\definecolor{MyB}{rgb}{0.1,0.1,1.0}

\definecolor{MyDarkRed}{rgb}{0.71,0.24,0.57}

\definecolor{MyG}{rgb}{0.,0.5,0.}

\title{Quantifying the accuracy of the Alcock-Paczy\'nski scaling of baryon acoustic oscillation measurements}

\arxivnumber{1908.11508}
\abstract{
We investigate -- in a generic setting -- the regime of applicability of the Alcock-Paczy\'nski (AP) scaling conventionally applied to test different cosmological models, given a fiducial measurement of the baryon acoustic oscillation (BAO) characteristic scale in the galaxy 2-point correlation function. 
We quantify the error in conventional AP scaling methods, for which our ignorance about the true cosmology is parameterised in terms of two constant AP scaling parameters. 
We propose a new, and as it turns out, improved version of the constant AP scaling, also consisting of two scaling parameters. 

The two constant AP scaling methods are almost indistinguishable when the fiducial model used in data reduction and the \say{true} underlying cosmology are not differing substantially in terms of metric gradients, but are otherwise expected to differ. 
Our new methods can be applied to existing analyses through a reinterpretation of the results of the conventional AP scaling. 
This reinterpretation might be important in model universes where curvature gradients above the scale of galaxies are significant.

We test our theoretical findings on \LCDM\ mock catalogues.
The conventional constant AP scaling methods are surprisingly successful for pairs of large-scale metrics, but eventually break down when toy models allowing for large metric gradients are tested. 
The new constant AP scaling methods proposed in this paper are efficient for all test models examined. 
We find systematic errors of $\sim$1\% in the recovery of the BAO scale when the true model is distant from the fiducial, which cannot be attributed to any constant AP approximation. 
The level of systematic uncertainty is robust to the exact fitting method employed. 
This indicates that caution must be taken with the error budget when extrapolating the BAO acoustic scale measurements obtained in the standard literature.
}
\keywords{gravity, baryon acoustic oscillations, galaxy clustering}
\catcode`\@=11
\gdef\@fpheader{Published:\ \href{https://doi.org/10.1088/1475-7516/2020/01/038}{JCAP 01 (2020) 038}\hfill \href{https://doi.org/10.1088/1475-7516/2020/01/038}{doi: 10.1088/1475-7516/2020/01/038}}
\catcode`\@=12
\begin{document}
\maketitle

\def\PRL#1{{\em Phys.\ Rev.\ Lett.}\ {\bf#1}}
\def\JCAP#1{{\em J.\ Cosmol.\ Astropart.\ Phys.}\ {\bf#1}}
\def\ApJ#1{{\em Astrophys.\ J.}\ {\bf#1}}
\def\PR#1{{\em Phys.\ Rev.}\ {\bf#1}}
\def\MNRAS#1{{\em Mon.\ Not.\ R.\ Astr.\ Soc.}\ {\bf#1}}
\def\CQG#1{{\em Class.\ Quantum Grav.}\ {\bf#1}}
\def\GRG#1{{\em Gen.\ Relativ.\ Grav.}\ {\bf#1}}
\def\IJMP#1{{\em Int.\ J.\ Mod.\ Phys.}\ {\bf#1}}
\def\AaA#1{{\em Astron.\ Astrophys}.\ {\bf#1}}
\def\AJ#1{{\em Astron.\ J}.\ {\bf#1}}
\def\ApJs#1{{\em Astrophys.\ J.\ Suppl}.\ {\bf#1}}
\def\PLB#1{{\em Phys.\ Lett.}\ {\bf B #1}}
\def\APSS#1{{\em Astrophys. \ Space \ Sci.}\ {\bf#1}}
\def\NatA#1{{\em Nature Astron.}\ {\bf#1}}
\def\ARNPS#1{{\em Ann.\ Rev.\ Nucl.\ Part.\ Sci.}\ {\bf#1}}

\def\beq{\begin{equation}} \def\eeq{\end{equation}}
\def\bea{\begin{eqnarray}} \def\eea{\end{eqnarray}}
\def\e{\mathop{\rm e}\nolimits}
\def \domain{\mathcal{D}}
\def\doubleunderline#1{\underline{\underline{#1}}}

\DeclareRobustCommand{\orderof}{\ensuremath{\mathcal{O}}}

\tableofcontents
\newpage

\section{Introduction}
The distribution of gravitationally-bound matter in space serves as an important probe of large-scale dynamics of the Universe.
In the Lambda Cold Dark Matter (\LCDM) cosmology -- or indeed any Friedmann-Lema\^{\i}tre-Robertson-Walker (FLRW) model beginning in a hot big bang --  the primordial plasma exhibits sound waves which are imprinted as a characteristic scale in the  matter distribution after the decoupling of the baryons from the photons \cite{peeplesBAO,SunZeldovich}. 
This characteristic baryon acoustic oscillation (BAO) scale is predicted to be seen as an excess in the autocorrelation of the matter distribution as seen today. 
The BAO scale constitutes an important link between the physics before and around the drag epoch and the present epoch, by its property as a statistical \say{standard ruler} \cite{SeoEisenstein,BlakeGlazebrook}.

The signature of the BAO characteristic scale as an excess in the 2-point correlation function of the matter distribution was first detected using galaxy surveys in the mid 2000's \cite{EisensteinDetection,Cole}, and has later been measured more accurately in large surveys such as the WiggleZ
Dark Energy Survey \cite{BlakeCov} and the Baryon Oscillation
Spectroscopic Survey \cite{Deff,BOSSquasars}. 
The BAO scale has also been
measured using the Lyman-$\alpha$ absorption line of hydrogen as a
tracer of the matter distribution \cite{Busca,Delubac}.

Most work on the 2-point correlation function (theoretical and observational) has been done assuming homogeneous spatially-flat FLRW models. 
While cosmological data, when interpreted within the \LCDM\ cosmology, suggests that the universe is spatially flat on large scales, there is nothing preventing significant large-scale spatial curvature if the universe is more accurately described by a model outside the class of the conventionally studied FLRW models which may still be consistent with the data. This is the case, e.g., in the timescape model \cite{clocks,sol,obstimescape,dnw} which is used as a test case in this analysis.

In large-scale structure analyses there are strong motivations for
assuming a fiducial cosmological model in data reduction, such as the
use of of N-body mock catalogues to investigate non-linear effects.  In
  the context of BAOs, applying a fiducial cosmological model allows
  the computation of an accurate template for the BAO peak and all galaxy
  pairs to be binned by their estimated co-moving spatial separation.
Reconstruction methods \cite{reconstruction} based on
\LCDM\ perturbation theory can further enhance the signal.  An obvious
draw-back of imposing fiducial model cosmologies in data reduction is
that the assumptions of a model cosmology are then implicitly present
in the conclusions drawn. This may in some cases
bias the results, lead to an underestimation of the error budget, and will in a worst-case scenario lead to circular verification of the
assumed fiducial cosmological model. 

Alcock and Paczy\'nski \cite{APoriginal} introduced a geometric test to compare radial and transverse distance measures for a spherical region that is expanding with the Hubble flow in a FLRW model. This provided the means to distinguish FLRW models with a cosmological constant from those with $\Lambda=0$. Recent analyses of the BAO scale build on the ideas of Alcock and Paczy\'nski \cite{APoriginal} and its early applications \cite{APBallinger,APMatsubara}, and are now described as AP scaling methods. In modern analysis these methods are applied to parametrise a FLRW trial cosmology in terms of a different fiducial FLRW cosmology to \say{first order} \cite{APmethods,APscale}.  

The AP scaling used in BAO analysis, (see, e.g., \cite{Standardresults,wedgefit}), makes use of this reparametrisation in order to test cosmological models different to the fiducial model. 
The extent to which the AP scaling methods, which rely on the scaling of a fiducial template-metric by two constant \say{AP scaling parameters}, can be thought of as independent of the fiducial model cosmology has not been thoroughly tested in the literature. 
This question is important for the range of validity of the distance measurements inferred from such procedures, and for constraining alternative cosmological models to that of the fiducial template-metric used to extract them. 

While the systematic errors related to the AP-distortion of conventional BAO analysis have been quantified by some studies such as \cite{BOSSsystematics,Carter}, such analyses usually only examine the cases of a few $\Lambda$CDM models which are close in terms of model parameters. 
In this paper we will test the extent to which this underestimates the error for constraining models which are outside the narrow space of cosmological models assessed for systematics, using a framework already developed to study the 2-point correlation function and the BAO feature in spherically-symmetric template metrics \cite{HBLW}.

{
If the actual Universe is 
in all respects very close to the $\Lambda$CDM model with parameters as determined from Planck \cite{planckParams}, then there is little to be gained by doing (semi-)model independent analysis. 
However, the abundance of freedom in full general relativity (not to mention modified theories of gravity), allows for an
infinity of ways of geometrically constructing cosmological spacetimes. 
Given the poor constraining power of cosmological data relative to this abundance, the only way to obtain tight constraints on cosmological dynamics is to reduce the space of metric solutions tremendously a priori, e.g., by imposing the usual global FLRW ansatz.  
It is therefore possible that models which are not contained within the usual class of FLRW geometries -- and which might not be close to a $\Lambda$CDM model in all respects -- can fit data as well. 
Irrespective of the closeness of the \say{true} metric description of the Universe to a particular $\Lambda$CDM model, care must be taken in data analysis so as not to impose implicit priors.  }

In section
\ref{theory} we outline some theoretical results and definitions on
which the analysis of this work is built.  In section \ref{errorAP} we
provide general results for the effect of the AP scaling on the
2-point correlation function, as viewed in the fiducial cosmological
model as compared to the \say{true} underlying cosmological model, and
we propose a new and improved AP scaling approximation.  In section
\ref{APBAO} we apply our results to a concrete model of the 2-point
correlation function, and investigate how the BAO feature depends on
the redshift-dependent AP scaling. In section \ref{testsmocks} we test
our predictions by applying them to the 2-point correlation function
based on \LCDM\ mock catalogues and formulated in a selection of
fiducial model cosmologies, some of which are \say{physical}
cosmological models built from general relativistic modelling and some
of which are \say{artificial} models.  We assess systematic errors
associated with the AP scaling approximations and additional
systematic errors.  We discuss our results in section
\ref{discussion}.

\section{The framework}
\label{theory}

\subsection{Models under investigation}
\label{models}
We follow \cite{HBLW} and consider the observer-adapted spherically-symmetric template metrics\footnote{The metric considered might be an exact solution to the Einstein equations (e.g., a Lema\^{\i}tre-Tolman-Bondi space-time metric), a solution to other specified field equations from modified gravity theories, or an effective metric which is not necessarily a space-time metric substituted into the Einstein equations or any set of local field equations. The spherically-symmetric metrics allow for defining the Alcock-Paczy\'nski (AP) scaling in section \ref{APscaling}.}
\begin{align} \label{eq:metriclagrangian2}
&ds^2 = - \alpha(t,r)^2 c^2 dt^2 + g_{rr}(t,r)\, dr^2 + g_{\theta \theta}(t,r) \left(d\theta^2 + \cos^2(\theta) d \phi^2 \right) , 
\end{align}
where $\theta$ and $\phi$ are angular coordinates on the observer's sky, $r$ is a radial coordinate, and $t$ is a time-coordinate labelling surfaces orthogonal to the \say{matter frame} with which the galaxies of the survey are (statistically) comoving. 
We shall further assume that the model redshift $z$ of radially propagating null rays is monotonic in the radial coordinate $r$, in which case the adapted metric on a given 3-surface selected by $t=T$ can be written 
\begin{align} \label{eq:metriclagrangiansphericalsymmetry}
&ds_T^2 = g_{zz}(t=T,r) dz^2 + g_{\theta \theta}(t=T,r) \left(d\theta^2 + \cos^2(\theta) d \phi^2 \right) ,
\end{align}
where
\begin{align} \label{eq:gzz}
g_{zz}(t=T,r) \equiv g_{rr}(t=T,r) \left( \frac{dr}{dz} \right)^2 .
\end{align}
As outlined in appendix A of \cite{HBLW}, for small separations of points $P_1$ and $P_2$ on the $t=T$ hypersurface as compared to variations of the adapted spatial metric
(\ref{eq:metriclagrangiansphericalsymmetry}), the geodesic distance
$D_{T}(P_1, P_2)$ between the points $P_1$ and $P_2$ represented by coordinates $(z_1,\theta_1,\phi_1)$ and
$(z_2,\theta_2,\phi_2)$ 
is
\begin{align} \label{eq:lagrangiandistapprox}
D_{T}^2(P_1, P_2)& \approx g_{zz}(t=T,\bar{z}) (\delta z)^2 +
g_{\theta \theta}(t=T,\bar{z}) (\delta \Theta)^2 ,
\end{align}
where $\bar{z} = (z_1 + z_2)/2$ is the intermediate redshift, $\delta z = z_2 - z_1$ is the separation in redshift, and $\delta \Theta$ is the separation in angle 
\begin{align} \label{eq:angularsep}
\delta \Theta &= \arccos \left[ \sin(\theta _{1})\sin(\theta
_{2})+\cos(\theta _{1})\cos(\theta _{2})\cos(\phi _{2}-\phi _{1})
\right]\\ &\approx \sqrt{(\theta_2 - \theta_1)^2 +
\cos^2(\bar{\theta} )(\phi_2 - \phi_1)^2 } , \qquad \bar{\theta} =
(\theta_1 + \theta_2)/2 \nonumber
\end{align}
As an example, for the FLRW and timescape models with reasonable model parameters, we find that higher-order corrections to eq.~(\ref{eq:lagrangiandistapprox}) are of order $\lsim 10^{-3}$ for galaxy separations of order $100\hm$.

From the approximation (\ref{eq:lagrangiandistapprox}) it is natural to define the \say{radial fraction} of the separation as
\begin{align} \label{eq:mu}
\mu_T(P_1,P_2) = \frac{ \sqrt{ g_{zz}(t=T,\bar{z}) (\delta z)^2} }{D_{T}(P_1, P_2)} . 
\end{align}
It is conventional to take the surface of evaluation $t = T$ to be that of the 
\say{present epoch}. When we refer to evaluation at the present epoch we
shall omit the $T$ subscript on eq.~(\ref{eq:lagrangiandistapprox})
and (\ref{eq:mu}). 
We shall also sometimes omit the reference to the points $P_1$,$P_2$ for ease of notation, and refer to $D_{T}(P_1,P_2)$ and $\mu_T(P_1,P_2)$ as $D$ and $\mu$ respectively.

{In the present paper we perform concrete investigations using mock catalogues for a few chosen FLRW models, a set of toy models constructed from distorting the distance redshift relation of a spatially flat FLRW cosmology, and the timescape cosmological model.} 
The timescape model \cite{clocks,sol,obstimescape,dnw} is a cosmology with backreaction of inhomogeneities. In backreaction models \cite{buchert00,BRreview} the Einstein equations with dust matter are assumed to apply to small scale structures, but their global average is not a common FLRW background. Generically, as density contrasts grow, so do spatial curvature gradients. At late epochs void regions of negative spatial curvature dominate the global average volume expansion and expand faster than the denser ``wall regions'' where galaxy clusters are located. The timescape model is a phenomenological interpretation of generic Buchert averages \cite{buchert00}, whereby observers in galaxies with a mass--biased view of the Universe interpret cosmological distances by extrapolating the local geometry to which their rulers and clocks are tied. For a review, see \cite{TSreview}. 

\subsection{Alcock-Paczy\'nski scaling}
\label{APscaling}
The conventional Alcock-Paczy\'nski (AP) scaling as outlined in \cite{APmethods,APscale} exploits the fact that a geodesic distance between two points in a spherically-symmetric large-scale metric can be approximated by (\ref{eq:lagrangiandistapprox}), as long as second-order metric variations within the distance spanned between the points are negligible. 

The geodesic distance between \say{closely separated} points in a model cosmology of the type described in section \ref{models} can be parametrised in terms of an unknown \say{true} model cosmology of the same type, by associating points of the same observational
coordinates $(z,\theta,\phi)$, as
\begin{align} \label{eq:Dtransformation}
D^2 & \approx g_{zz}(t=T_0,z) (\delta z)^2 + g_{\theta \theta}(t=T_0,z) (\delta \Theta)^2 \\
& = \frac{1}{\alpha^2_{\parallel} (z) }g^{\rm tr}_{zz}(t^{\rm tr}=T^{\rm tr}_0,z) (\delta z)^2 + \frac{1}{\alpha^2_{\perp}(z) } g^{\rm tr}_{\theta \theta}(t^{\rm tr}=T^{\rm tr}_0,z) (\delta \Theta)^2 \nonumber
\end{align}
where \say{tr} stands for the \say{true} cosmology, $t=T_0$ and $t^{\rm tr}=T_0^{\rm tr}$ are the \say{present epoch} hypersurfaces of the trial model cosmology and the \say{true} model cosmology respectively, and the redshift of evaluation $z$ is the mean redshift of the points. The AP scaling functions
\begin{align} \label{eq:APdef}
\alpha_{\parallel}(z) & \equiv \sqrt{ \frac{g^{\rm tr}_{zz}(t^{\rm tr}=T_0^{\rm tr},z) }{g_{zz}(t=T_0,z)} } , \qquad \alpha_{\perp}(z) \equiv \sqrt{ \frac{ g^{\rm tr}_{\theta \theta}(t^{\rm tr}=T_0^{\rm tr},z) }{ g_{\theta \theta}(t=T_0,z) }},
\end{align}
describe the relative radial and transverse distortion between the \say{true} cosmology and the trial cosmology. 
We can re-express the information of $\alpha_{\parallel}(z)$ and $\alpha_{\perp}(z)$ in terms of the isotropic scaling function $\alpha(z)$ and the anisotropic scaling function $\epsilon(z)$
\begin{align} \label{eq:alphaepsilon}
\alpha(z) \equiv (\alpha^2_{\perp}(z) \alpha_{\parallel}(z))^{1/3} , \qquad (1+\epsilon(z))^3 \equiv \frac{\alpha_{\parallel}(z)} {\alpha_{\perp}(z)} .
\end{align}
The definitions (\ref{eq:alphaepsilon}) are analogous to those presented in \cite{APmethods}, except that we keep the redshift dependence instead of assuming $\alpha(z)$ and $\epsilon(z)$ to be constant.
The function $\alpha(z)$ describes how the volume measure of a
small coordinate volume $\delta z \, \cos(\theta) \, \delta \theta \,
\delta \phi$ centred at $z$ differs between the \say{true} and the model cosmology, while $\epsilon(z)$ quantifies the relative scaling of the angular and transverse metric components between the \say{true} and the model cosmologies. 

Using the definitions (\ref{eq:mu}) and (\ref{eq:alphaepsilon}), we can rewrite the approximation (\ref{eq:Dtransformation}) for points with mean redshift $z$ as \cite{HBLW}
\begin{align} \label{eq:DAP}
(D^{\rm tr})^2 & \approx  \alpha^2(z) D^2 \frac{  1 + \psi(z) \mu^2  }{(1 + \epsilon(z))^2} \, , \qquad  \psi(z) \equiv  \left(1+\epsilon(z) \right)^6 - 1 . 
\end{align}
Similarly, using the definitions (\ref{eq:mu}) and (\ref{eq:alphaepsilon}), and the result (\ref{eq:DAP}), we have the relation
\begin{align} \label{eq:muAP}
(\mu^{\rm tr})^2& \approx  \mu^2 \frac{   (1 + \epsilon(z))^6 }{1 + \psi(z) \mu^2} .
\end{align}
When the AP scaling is applied in standard analysis it is assumed that $\alpha(z)$ and $\epsilon(z)$ can be considered constant and equal to their evaluation at the effective redshift of the survey, i.e., that the replacement $\alpha(z) \mapsto \alpha(\bar{z}) \, , \,\epsilon(z) \mapsto \epsilon(\bar{z})$ is accurate.
This replacement is expected to be a reasonable approximation if the survey volume has a
relatively narrow redshift distribution, and if both the \say{true} and the model metric are slowly changing in redshift. 

In the present analysis we will investigate the correction terms that arise when we take into account the variation of $\alpha(z)$ and $\epsilon(z)$ over the survey volume, and quantify the accuracy of the usual constant AP scaling approximation $\alpha(z) \mapsto \alpha(\bar{z}) \, , \,\epsilon(z) \mapsto \epsilon(\bar{z})$ when applied in the 2-point correlation function to extract the parameters of the BAO feature.

\subsection{Empirical model for the correlation function}
\label{empiricalBAOmain}
In this section we introduce the empirical fitting function we use for examining the BAO feature of the two-point correlation function, its dependence on the fiducial cosmology, and the accuracy of the constant AP scaling approximation. 
We could have used the fiducial $\Lambda$CDM template fitting function outlined, e.g., in \cite{wedgefit}, where the BAO feature is derived from a model power spectrum. However, we expect the conclusions about the accuracy of the constant AP scaling approximation to be similar between the two fitting functions. The advantage of considering the simple empirical fitting function is that it does not assume a particular cosmological model. Specifically, for non-FLRW models where no well-defined perturbation theory exists, but where we nevertheless expect a statistical standard ruler to be present in form of a BAO scale, we must rely on empirical extraction methods of the BAO characteristic scale. 
Furthermore, the simple form of the empirical fitting function presented here allows us to obtain useful analytical results.

We follow \cite{HBLW} and consider the model for the 2-point correlation function (\ref{eq:ximuDz}) 
\begin{align} \label{eq:xifit}
& \xi^{\rm tr}(D^{\rm tr}, \mu^{\rm tr},z) = (D^{\rm tr})^2 A(z)\e^{\frac{-\left(D^{\rm tr} - r \right)^2}{2 \sigma^2(z)}} + C^{\rm tr}_0(\mu^{\rm tr},z) + \frac{C^{\rm tr}_1(\mu^{\rm tr},z) }{D^{\rm tr}} + \frac{C^{\rm tr}_2(\mu^{\rm tr},z )}{(D^{\rm tr})^2} \, , 
\end{align}
as formulated in the underlying \say{true} cosmology, where $r$ denotes the BAO scale or a characteristic scale shifted with respect to the BAO scale. (See the discussion below on calibration of the BAO scale.) 
The polynomial terms account for the \say{background} shape of the correlation function without the BAO feature and are
equivalent in form to those of \cite{wedgefit}. The scaled Gaussian models the BAO feature, and replaces the \LCDM\
power spectrum model of \cite{wedgefit}.
Empirical models of similar form to (\ref{eq:xifit}) have been considered in, e.g., \cite{MotionAccousticScale,sancheztransverse,sanchezradial}.

Note that in contrast to previous analyses we are modelling the redshift-dependent 2-point correlation function $\xi^{\rm tr}(D^{\rm tr}, \mu^{\rm tr},z)$ (\ref{eq:ximuDz}). This is done to examine the impact of the redshift dependence of the generalised AP-scaling functions (\ref{eq:alphaepsilon}).
For simplicity we assume a model where $A$ and $\sigma$ are constant for the numerical investigations in this paper, while no assumption is made on the redshift dependence of the polynomial coefficients.

We assume the peak of the BAO feature $r$ to be a \say{standard ruler} independent of redshift. 
This approximation is good in $\Lambda$CDM cosmology as confirmed with $\Lambda$CDM mock catalogues in \cite{ShiftPeakBAO} using standard \LCDM\ template procedures and in \cite{HBLW} using the empirical model presented here. 
From the results in \cite{ShiftPeakBAO} and \cite{HBLW} we can expect shifts of the BAO scale of $\lsim 0.5\%$ in a \LCDM\ universe at redshifts $\gtrsim 0.3$. 
The approximation is less obviously good in non-\LCDM\ cosmology, where environmental dependence of the BAO peak is expected \cite{environmentRoukema}. 
However, as long as: (i) data is not binned according to environmental factors such as density, and (ii) each redshift slice represents the volume average for the corresponding approximate cosmic epoch, then we might make the ansatz that $r$ is an approximate statistical standard ruler for volume measures.

The scaling of the Gaussian part of the model 2-point correlation function in (\ref{eq:xifit}) is an approximation that accounts for calibration issues in BAO physics: that the local maximum of the 2-point correlation function does not in general correspond to the BAO scale. (In \LCDM\ cosmology these two scales differ by roughly $\sim 2-3\%$.) The scaling by the factor $(D^{\rm tr})^2$ of the Gaussian feature allows us to interpret the mean of the Gaussian $r$ as the BAO scale with a precision of $< 1 \%$ within the $\Lambda$CDM concordance cosmology, as verified with $\Lambda$CDM mock catalogues in \cite{HBLW}. 
Note that the degree to which $r$ can be interpreted as a BAO scale for other models must be assessed for each particular case, or simply be posed as an ansatz of the analysis.\footnote{For models where perturbation theory has yet to be developed, we cannot predict how the sound horizon scale of the drag epoch will appear in the galaxy distribution, and an ansatz is needed in order to constrain the sound horizon scale at the drag epoch with galaxy catalogues.}

\section{Theoretical investigation of the redshift-dependent Alcock-Paczy\'nski scaling}

\label{errorAP}

In this section we investigate how the redshift-dependent Alcock-Paczy\'nski scaling enters in the 2-point correlation function. 
We quantify the accuracy of the conventional constant Alcock-Paczy\'nski (AP) scaling approximation $\alpha(z) \mapsto \alpha(\bar{z})$, $\epsilon(z) \mapsto \epsilon(\bar{z})$ over the survey volume.  Based on our investigations, we propose a new and improved version of the constant Alcock-Paczy\'nski (AP) scaling.

In standard BAO analysis, as described in, e.g., \cite{APmethods,APscale} for $\Lambda$CDM cosmology, and in the generalisation of such analyses to generic geometries \cite{HBLW}, the fitting procedure is based on making an ansatz for the form of the 2-point correlation function as formulated in the unknown \say{true} cosmological model. Furthermore, the assumed function is parameterised in a given fiducial cosmology using AP scaling methods as outlined in section \ref{APscaling}.

\subsection{Re-parametrisation of the 2-point correlation function}
\label{reparametrisation}
Suppose that the 2-point correlation function (\ref{eq:ximuDz}) in $D$, $\mu$, and $z$ has the form 
\begin{align} \label{eq:xitrue}
& \xi^{\rm tr}(D^{\rm tr},\mu^{\rm tr},z) = \frac{f^{\rm tr}(D^{\rm tr},\mu^{\rm tr},z)}{f^{\rm tr}_{\text{Poisson}}(D^{\rm tr},\mu^{\rm tr},z) } - 1 , 
\end{align}
in the \say{true} spherically-symmetric cosmology, 
where $f^{\rm tr}(D^{\rm tr},\mu^{\rm tr},z)$ and $f^{\rm tr}_{\text{Poisson}}(D^{\rm tr},\mu^{\rm tr},z)$ are the probability densities of finding a pair of galaxies separated by $D^{\rm tr}$ and $\mu^{\rm tr}$ with one of the galaxies centred at $z$, in the catalogue and random catalogue respectively. 
We can express $\xi(D,\mu,z)$ of any other given spherically-symmetric model in terms of $\xi^{\rm tr}(D^{\rm tr},\mu^{\rm tr},z)$ in the following way
\begin{align} \label{eq:ximodel}
\xi(D,\mu,z) &= \frac{f(D,\mu,z)}{f_{\text{Poisson}}(D,\mu,z) } - 1  =  \frac{J \, f^{\rm tr}\left(D^{\rm tr}\left(D,\mu,\alpha(z), \epsilon(z) \right),\mu^{\rm tr}(\mu,\epsilon(z)),z \right)}{J \,  f^{\rm tr}_{\text{Poisson}}\left(D^{\rm tr}(D,\mu,\alpha(z), \epsilon(z)),\mu^{\rm tr}(\mu,\epsilon(z)),z\right) }  - 1   \nonumber  \\ 
&= \xi^{\rm tr}\left(D^{\rm tr}(D,\mu,\alpha(z), \epsilon(z)),\mu^{\rm tr}(\mu,\epsilon(z)),z\right) , 
\end{align}
where $J$ is the determinant of the Jacobian of the transformation $(D^{\rm tr},\mu^{\rm tr}) \, \mapsto \, (D, \mu)$ which can be derived from (\ref{eq:DAP}) and (\ref{eq:muAP}). The first line follows from the transformation of a density under a change of variables by the determinant of the Jacobian of the transformation, and the second line follows from the cancellation of $J$ in the numerator and denominator of the expression. 
Note that $D^{\rm tr}$ and $\mu^{\rm tr}$ introduce redshift dependence in $\xi(D,\mu,z)$ through $\alpha(z)$ and $\epsilon(z)$. 
{
We shall sometimes be interested in evaluating the right hand side of (\ref{eq:ximodel}) for parameters $\alpha$ and $\epsilon$ which do not necessarily correspond to the redshift dependent AP scaling functions $\alpha(z)$ and $\epsilon(z)$.
In such cases we simply write $\xi^{\rm tr}\left(D^{\rm tr}(D,\mu,{\alpha},{\epsilon}),\mu^{\rm tr}(\mu,{\epsilon}),{z}\right)$ for evaluation for any given point $z,\alpha,\epsilon$. 
}

As outlined in appendix \ref{2PCFestimatorsmain}, $\xi(D,\mu)$ (\ref{eq:ximudef1}) can be obtained as a weighted integral in redshift over $\xi(D,\mu,z)$ if the condition of almost multiplicative separability (\ref{eq:fDmuzapprox}) is satisfied. 
If this is the case the result in (\ref{eq:xidef1int}) holds, and to first order in the non-multiplicatively separable functions $\delta(D,\mu,z)$ and $\delta_{\text{Poisson}}(D,\mu,z)$ defined in (\ref{eq:fDmuzapprox}), we have 
\begin{align} \label{eq:xiapprox}
\xi(D,\mu)  &\approx \int \, dz \,P(z)\, \xi(D,\mu,z)   , 
\end{align}
where $P(z)$ is the normalised galaxy distribution in redshift (\ref{eq:numbercountz}). 
We might further expand $\xi(D,\mu,z) = \xi^{\rm tr}\left(D^{\rm tr}(D,\mu,\alpha,\epsilon),\mu^{\rm tr}(\mu,\epsilon),z\right)$ to first order in $z$, $\alpha$, and $\epsilon$, (leaving $\alpha$ and $\epsilon$ as exact functions in $z$, rather than their approximations in terms of expansions in $z$), around some appropriate point $z=\ovo{z}$ to obtain 
\begin{align} \label{eq:xiapproxexpand}
\xi(D,\mu) & \approx  \int \, dz \,P(z)\, \xi(D,\mu,z) =  \nonumber \\
&\approx  \int \, dz \,P(z)\,  \left( \left. \xi^{\rm tr} \right|_{\ovo{z}, \ovo{\alpha}, \ovo{\epsilon}} + \left. \frac{\partial \xi^{\rm tr} }{\partial z} \right|_{\ovo{z}, \ovo{\alpha}, \ovo{\epsilon}} (z - \ovo{z})  + \left. \frac{\partial \xi^{\rm tr} }{\partial \alpha} \right|_{\ovo{z}, \ovo{\alpha}, \ovo{\epsilon}} (\alpha - \ovo{\alpha}) +  \left. \frac{\partial \xi^{\rm tr} }{\partial \epsilon} \right|_{\ovo{z}, \ovo{\alpha}, \ovo{\epsilon}}  (\epsilon - \ovo{\epsilon})  \right)   \nonumber \\
&=  \left. \xi^{\rm tr} \right|_{\ovo{z}, \ovo{\alpha}, \ovo{\epsilon}} + \left. \frac{\partial \xi^{\rm tr} }{\partial z} \right|_{\ovo{z}, \ovo{\alpha}, \ovo{\epsilon}} (\bar{z} - \ovo{z})  + \left. \frac{\partial \xi^{\rm tr} }{\partial \alpha} \right|_{\ovo{z}, \ovo{\alpha}, \ovo{\epsilon}} (\bar{\alpha} - \ovo{\alpha}) +  \left. \frac{\partial \xi^{\rm tr} }{\partial \epsilon} \right|_{\ovo{z}, \ovo{\alpha}, \ovo{\epsilon}}  (\bar{\epsilon} - \ovo{\epsilon})   \nonumber \\
&\approx  \left. \xi^{\rm tr} \right|_{\bar{z}, \bar{\alpha}, \bar{\epsilon}}  \, , 
\end{align}
where $\{ \ovo{z}, \ovo{\alpha}, \ovo{\epsilon} \} = \{ \ovo{z}, \alpha(\ovo{z}), \epsilon(\ovo{z}) \}$, and where we use the short hand notation $\left. \xi^{\rm tr} \right|_{{z}, {\alpha}, {\epsilon}} \equiv \xi^{\rm tr}\left(D^{\rm tr}(D,\mu,{\alpha},{\epsilon}),\mu^{\rm tr}(\mu,{\epsilon}),{z}\right)$ where the dependence on $D,\mu$ is implicit.
In the third line of (\ref{eq:xiapproxexpand}) we have used the short-hand notation for the averages in redshift
\begin{align} \label{eq:barsdef}
\bar{z} \equiv \int \, dz \,P(z)\, z \, , \qquad  \bar{\alpha} \equiv \int \, dz \,P(z)\, \alpha(z) \, , \qquad  \bar{\epsilon} \equiv \int \, dz \,P(z)\, \epsilon(z) \, , 
\end{align}
which we use throughout this analysis.
The accuracy of the expansion (\ref{eq:xiapproxexpand}) depends on the magnitude of the deviations of $z$, $\alpha$, $\epsilon$ 
over the survey and on the function $\xi(D,\mu,z) = \xi^{\rm tr}\left(D^{\rm tr}(D,\mu,\alpha,\epsilon),\mu^{\rm tr}(\mu,\epsilon),z\right)$. 

The result in (\ref{eq:xiapproxexpand}) {suggests that the re-parametrising of a given physical 2-point correlation function in terms of a distorted fiducial cosmology is \emph{more accurately described by the survey averages $\bar{\alpha}$ and $\bar{\epsilon}$ of the AP-scaling functions}, rather than by the same AP-scaling functions evaluated at the mean redshift of the survey $\alpha(\bar{z})$ and $\epsilon(\bar{z})$.
This conjecture is substantiated in appendix \ref{conjecture}. 
We shall examine this hypothesis for a set of concrete empirical models for the 2-point correlation function of mock catalogues in section \ref{testsmocks}. }

We will denote the replacement $\alpha(z) \mapsto \bar{\alpha}$, $\epsilon(z) \mapsto \bar{\epsilon}$ the \emph{modified} constant AP scaling approximation, in order to distinguish it from the standard constant AP scaling approximation $\alpha(z) \mapsto \alpha(\bar{z})$, $\epsilon(z) \mapsto \epsilon(\bar{z})$. 
The modified constant AP scaling approximation is intuitive, and formalises that statistical estimators built from a survey probe \emph{averaged} distance scales over the survey volume.

\subsection{Bounding the difference between the constant AP scaling approximations}
\label{bounds}
We now quantify the difference between the modified constant AP scaling approximation $\alpha(z) \mapsto \bar{\alpha}$, $\epsilon(z) \mapsto \bar{\epsilon}$ and the standard constant AP scaling approximation $\alpha(z) \mapsto \alpha(\bar{z})$, $\epsilon(z) \mapsto \epsilon(\bar{z})$.
In the ideal case, where the modified constant AP scaling approximation $\alpha(z) \mapsto \bar{\alpha}$, $\epsilon(z) \mapsto \bar{\epsilon}$ can be made with no error in the approximation (\ref{eq:xiapproxexpand}), we can view the difference between the two AP approximations as quantifying the error in the conventional AP approximation $\alpha(z) \mapsto \alpha(\bar{z})$, $\epsilon(z) \mapsto \epsilon(\bar{z})$.\footnote{Indeed, the modified constant AP scaling approximation $\alpha(z) \mapsto \bar{\alpha}$, $\epsilon(z) \mapsto \bar{\epsilon}$ turns out to be very accurate for the broad sample of tested models in section \ref{testsmocks}.}

Assuming that $\alpha(z)$ and $\epsilon(z)$ are both twice differentiable over the redshift range of the survey we can use the following approximations
\begin{align} \label{eq:taylor}
\alpha(z) = \alpha(\bar{z}) + \frac{ \partial \alpha}{\partial z} (\bar{z}) \, (z-\bar{z}) + \mathcal{R}^{\alpha}_{1}(z) \, , \qquad \epsilon(z) = \epsilon(\bar{z}) + \frac{ \partial \epsilon}{\partial z} (\bar{z}) \, (z-\bar{z}) + \mathcal{R}^{\epsilon}_{1}(z) \, ,
\end{align}
where $\mathcal{R}^{\alpha}_{1}(z)$ and $\mathcal{R}^{\epsilon}_{1}(z)$ are the remainder terms of the first order expansions in $\alpha$ and $\epsilon$ respectively. 
Let us first consider the $\alpha$ parameter. Its integral reads 
\begin{equation} \label{eq:alphaexpand}
  \bar{\alpha} = \int \, dz \,P(z)\, \alpha(z) =  \alpha(\bar{z}) \left( 1 + \Delta_{\alpha} \right) \, ,
\end{equation}
where we define the \say{error term} as
\begin{equation} \label{eq:alphaerror}
  \Delta_{\alpha} \equiv  \frac{1}{\alpha(\bar{z})}  \int \, dz \,P(z)\, \mathcal{R}^{\alpha}_{1}(z)  \, ,
\end{equation}
where it has been used that the first order term in (\ref{eq:taylor}) vanishes by construction, since $\overline{(z-\bar{z})} \equiv  \int \, dz \,P(z)\, (z-\bar{z}) = 0 \,$. 
We can bound the error term (\ref{eq:alphaerror}) by bounding the remainder $\mathcal{R}^{\alpha}_{1}(z)$ of the first order expansion (\ref{eq:taylor}). {The detailed derivations of a bound on the error term (\ref{eq:alphaerror}) and the corresponding error term for $\epsilon$ are given in appendix \ref{boundsAP}.}
We obtain the following bound, derived in appendix \ref{deltaalpha}:
\begin{align} \label{eq:alphaexpandbound}
\abs*{\Delta_{\alpha} } &\, \leq \,  \frac{1}{2} \frac{ M^{\text{max}}_{L \, 0}}{ M^{\text{min}}_{L \, 0} } \left(  \beta_{L \,2} M_{L \,2} + 2 \beta^2_{L \,1} M_{L \,1}   \right) \, \overline{(z-\bar{z})^2}  , 
\end{align}
where $M^{\text{min}}_{L \, 0}$, $M^{\text{max}}_{L \, 0}$, $M_{L \,1}$, $M_{L \,2}$, $\beta_{L \,1}$, and $\beta_{L \,2}$ are all positive dimensionless constants bounding the metric combinations $L\equiv  \left( g^2_{\theta \theta} g_{z z}\right)^{\frac{1}{6}}$, $L^{\rm tr} \equiv  \left( (g^{\rm tr}_{\theta \theta})^2 g^{\rm tr}_{z z}  \right)^{\frac{1}{6}}$ and their derivatives in the following way  
\begin{align} \label{eq:boundsLLtr}
\hspace*{-0.1cm} M^{\text{min}}_{L \, 0} \leq \frac{L^{\rm tr}}{L} \leq M^{\text{max}}_{L \,0} , \qquad  \abs*{ \frac{  \left(  \frac{\partial L^{\rm tr}/ \partial z  }{L^{\rm tr}} \right)   }{  \left( \frac{ \partial L / \partial z  }{L} \right) } \, - \, 1 }  \leq M_{L \,1} \, , \qquad \abs*{ \frac{  \left(  \frac{\partial^2 L^{\rm tr}/ \partial z^2  }{L^{\rm tr}} \right)   }{  \left( \frac{ \partial^2 L / \partial z^2  }{L} \right) } \, - \, 1 }  \leq M_{L \,2} \, , 
\end{align}
\begin{align} \label{eq:boundsL}
\abs*{ \frac{\frac{ \partial L}{\partial z}  }{L} } \leq  \beta_{L \,1}  \, , \qquad  \abs*{ \frac{\frac{ \partial^2 L}{\partial z^2}  }{L} } \leq  \beta_{L \,2} .
\end{align}
We note that it is possible to have order of magnitude $\sim 1$ deviations between the models such that $M^{\text{max}}_{L \, 0} \sim 1/M^{\text{min}}_{L \, 0} \sim 2$ and $M_{L \,1}\sim M_{L \,2}\sim1$ while still having $\Delta_{\alpha}$ $\lsim$ a few percent, depending on the survey and of the first and second-order derivatives of $L$. 
We shall investigate bounds for various choices of trial cosmologies in subsection \ref{boundsexamples} below. 
Let us next consider the corresponding integral for $\epsilon$,
\begin{equation} \label{eq:epsilonexpand}
\bar{\epsilon} = \int \, dz \,P(z)\, \epsilon(z) =  \epsilon(\bar{z}) +  \Delta_{\epsilon}  \, ,
\end{equation}
where we define the error term,
\begin{equation} \label{eq:epsilonerror}
\Delta_{\epsilon} \equiv  \int \, dz \,P(z)\, \mathcal{R}^{\epsilon}_{1}(z)  \, 
\end{equation}
which is obtained in a similar way as the error term $\Delta_{\alpha}$ in (\ref{eq:alphaerror}). 
Note that, as opposed to $\alpha$ which is strictly larger than zero since it describes the ratio of two positive distance scales, $\epsilon$ can be zero, and thus, $\Delta_{\epsilon}$ is defined as an absolute error rather than a relative error. 
We obtain the following bound on $\Delta_{\epsilon}$, derived in appendix \ref{deltaepsilon}:
\begin{align} \label{eq:epsilonerrorbound}
\hspace*{-0.3cm} \abs*{\Delta_{\epsilon}}  \, \leq \,  \frac{1}{6} M^{\text{max}}_{R \, 0} \left( \beta_{R \,2} M_{R \,2}  + 2 \beta^2_{R \,1} \left(  M_{R \,1} + \frac{1}{3} M_{R \,1}^2  \right)   \right)  \, \overline{(z-\bar{z})^2}        \, , 
\end{align}
where $M^{\text{max}}_{R \, 0}$, $M_{R \,1}$, $M_{R \,2}$, $\beta_{R \,1}$, and $\beta_{R \,2}$ are all positive dimensionless constants bounding the metric combinations $R \equiv ( g_{z z} / g_{\theta \theta} )^{1/2}$ and $R^{\rm tr} \equiv ( g^{\rm tr}_{z z} / g^{\rm tr}_{\theta \theta} )^{1/2}$ in the following way
\begin{align} \label{eq:boundsRRtr} 
\hspace*{-0.3cm} \left( \frac{R^{\rm tr}}{R} \right)^{\frac{1}{3}}\leq M^{\text{max}}_{R \, 0}  , \qquad  \abs*{ \frac{  \left(  \frac{\partial R^{\rm tr}/ \partial z  }{R^{\rm tr}} \right)   }{  \left( \frac{ \partial R / \partial z  }{R} \right) } \, - \, 1 }  \leq M_{R \,1} \, , \qquad \abs*{ \frac{  \left(  \frac{\partial^2 R^{\rm tr}/ \partial z^2  }{R^{\rm tr}} \right)   }{  \left( \frac{ \partial^2 R / \partial z^2  }{R} \right) } \, - \, 1 }  \leq M_{R \,2} \, , 
\end{align}
\begin{align} \label{eq:boundsR}
\abs*{ \frac{\frac{ \partial R}{\partial z}  }{R} } \leq  \beta_{R \,1}  \, , \qquad  \abs*{ \frac{\frac{ \partial^2 R}{\partial z^2}  }{R} } \leq  \beta_{R \,2} \, . 
\end{align}
The bound in (\ref{eq:epsilonerrorbound}) shows that it is possible to have order of magnitude $\sim 1$ deviations between the models such that $M^{\text{max}}_{R \, 0} \sim 2$ and $M_{R \,1}\sim M_{R \,2}\sim1$ while still having $\Delta_{\alpha}$ $\lsim$ a few percent, depending on the survey and of the first and second-order derivatives of $R$. 


Assuming that the modified constant AP approximation is accurate -- which is indeed the case for the broad sample of tested models in section \ref{testsmocks} -- the bounds in (\ref{eq:alphaexpandbound}) and (\ref{eq:epsilonerrorbound}) are useful for quantifying which models are expected to be well-approximated by the usual constant AP scaling approximation: $\alpha \mapsto \alpha(\bar{z})$, $\epsilon \mapsto \epsilon(\bar{z})$. 
For models that have metric combinations $L$ and $R$ with derivatives up to second order within order $\sim 1$ from the corresponding derivatives of the \say{true} metric combinations $L^{\rm tr}$ and $R^{\rm tr}$, we expect the usual constant AP scaling approximation to be reasonable for typical galaxy surveys. 

For example, the CMASS NGC catalogue \cite{BOSS} has $\overline{(z-\bar{z})^2} = 4.0 \times 10^{-3}$ when including galaxies in the interval $0.43 < z < 0.7$, and the LOWZ NGC catalogue has $\overline{(z-\bar{z})^2} = 5.7 \times 10^{-3}$ when including galaxies in the interval $0.15 < z < 0.43$. Thus the terms multiplying (\ref{eq:alphaexpandbound}) and (\ref{eq:epsilonerrorbound}) must be larger than $1$ in order to facilitate a correction of more than 1\% to the standard constant AP scaling approximation for these surveys. 
Such large terms can only be obtained if one considers models with large bounding coefficients in (\ref{eq:boundsLLtr}), (\ref{eq:boundsL}), (\ref{eq:boundsRRtr}), and (\ref{eq:boundsR}).
This could for instance happen for a \say{true} model differing by more than order $\sim 1$ from the fiducial model -- e.g., if the fiducial model is \say{smooth} in its distance measures while the \say{true} model is rapidly oscillating.

\subsection{Quantitative results for selected models}
\label{boundsexamples}
We now consider a few model cosmologies for which we will compute the error terms $\Delta_\alpha$ and $\Delta_\epsilon$ and their corresponding bounds as given by the results in section \ref{bounds}. 
{The models investigated are the spatially-flat $\Lambda$CDM model with $\Omega_M = 0.99$, the Milne universe model, the negatively curved FLRW model with $\Omega_M = 0.25$ and $\Omega_{\Lambda} = 0.65$ (differing by $\sim 0.1$ in cosmological parameters as compared to the best fit $\Lambda$CDM model as determined by Planck \cite{planckParams}),} and the timescape cosmological model with\footnote{Here $\Omega_M$ refers to the present epoch value of the ``dressed matter density parameter'' in the timescape model. It is {\em not} related to the Friedmann equation in the usual way and is not a fundamental parameter of the model. Rather it is defined for convenience to take numerical values of similar order to those of the matter density parameter in the $\Lambda$CDM model. At late epochs it is related to fundamental parameter of the model, the void fraction $\fv$, according to $\Omega_M=\frac12(1-\fv)(2+\fv)$.} $\Omega_M = 0.3$.
In addition we consider a class of unphysical models which are bounded with respect to a fiducial $\Lambda$CDM model but which allows for large metric gradients. 

We consider the typical redshift range used for the LOWZ catalogue $0.15 < z < 0.43$ and the CMASS catalogue $0.43 < z < 0.7$ respectively. 

We imagine that the given model cosmology is the \say{true} underlying cosmology, and take the spatially-flat $\Lambda$CDM model with $\Omega_M = 0.3, \, \Omega_{\Lambda} = 0.7$ to be the fiducial cosmological model. 
The derivations below could easily be reversed in terms of \say{true} and fiducial cosmology, by making the replacements $L \leftrightarrow L^{\rm tr}$ and $R \leftrightarrow R^{\rm tr}$ in all expressions of section \ref{bounds}. 

For a given underlying \say{true} cosmological model and for a given redshift distribution of a survey, we can compute $\Delta_{\alpha}$ (\ref{eq:alphaerror}) and $\Delta_{\epsilon}$ (\ref{eq:epsilonerror}).
We might also compute the associated bounds on $\abs*{\Delta_{\alpha}}$ (\ref{eq:alphaexpandbound}) and $\abs*{\Delta_{\epsilon}}$ (\ref{eq:epsilonerrorbound}), assuming knowledge only on the realised bounds (\ref{eq:boundsLLtr}), (\ref{eq:boundsL}), (\ref{eq:boundsRRtr}), and (\ref{eq:boundsR}), but no additional knowledge of the functions $\alpha(z)$ and $\epsilon(z)$. 

\begin{figure}[!htb]
\centering
\includegraphics[scale=0.7]{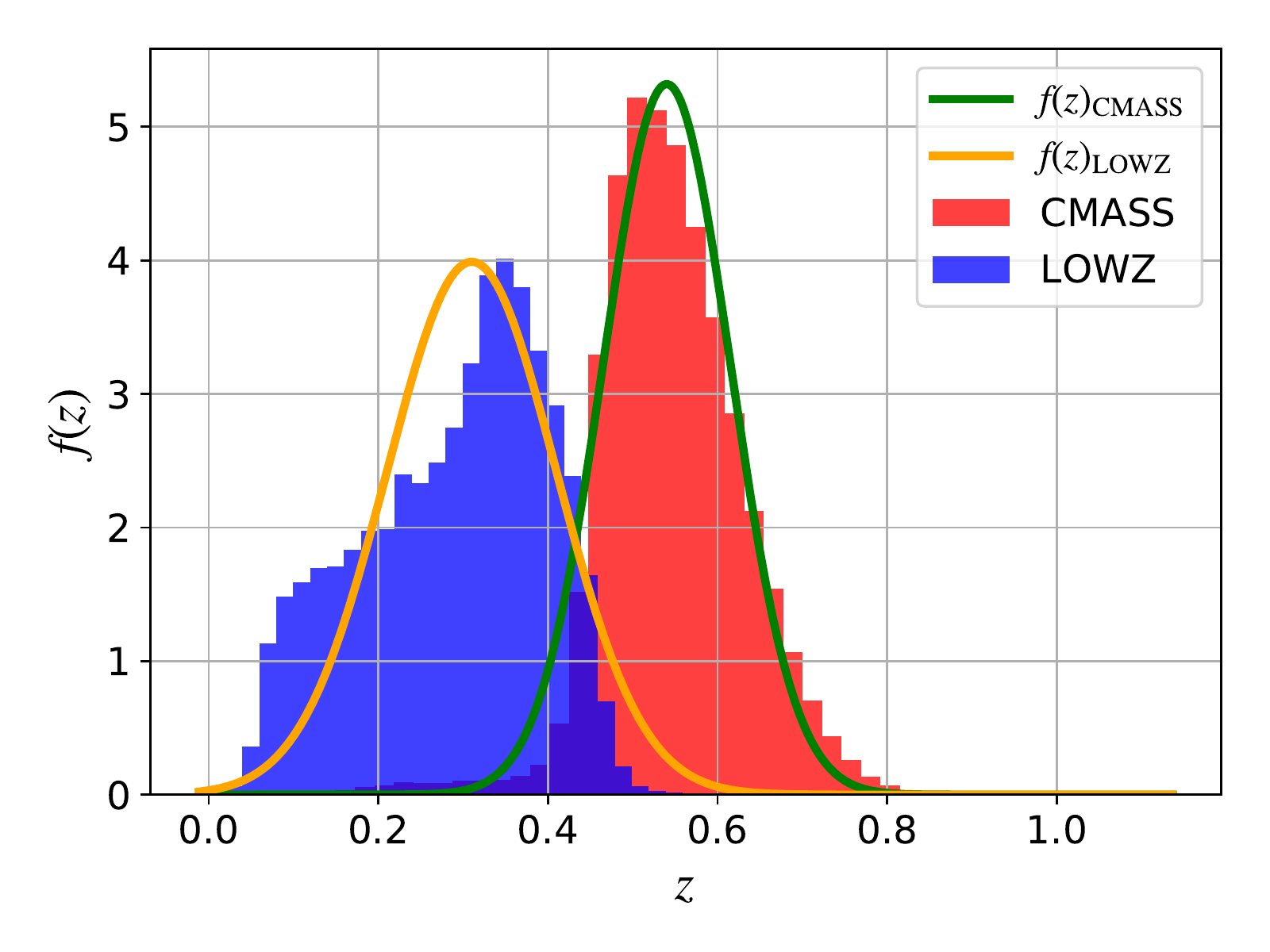}
\caption{Normalised redshift distributions of CMASS (red) and LOWZ (blue) along with superimposed Gaussian probability density distributions $f(z)_{\rm CMASS}$ and $f(z)_{\rm LOWZ}$ which roughly sample the redshift distributions. $f(z)_{\rm CMASS}$ has mean $\mu_{\rm CMASS} = 0.54$ and standard deviation $\sigma_{\rm CMASS} = 0.075$, and $f(z)_{\rm LOWZ}$ has $\mu_{\rm LOWZ} = 0.31$ and standard deviation $\sigma_{\rm LOWZ} = 0.10$.}
\label{fig:zdistModel}
\end{figure}

For convenience, we model the redshift distributions of the galaxy catalogues as truncated Gaussian distributions\footnote{{This approximation will become convenient in the analysis of mock catalogues, where integrals of the type (\ref{eq:xiapprox}) are performed at each point of iteration over the parameter space of the empirical fitting function used to extract the BAO characteristic scale. We might have used a spline function for more precision, but for the purpose of this paper the approximation by a Gaussian distribution is sufficiently accurate.}} 
\begin{align} \label{eq:Pgaussian}
P(z) \equiv 
    \begin{cases}
      \frac{1}{\mathcal{N}}  \frac{1}{\sqrt{2 \pi} \sigma} e^{\frac{-(z-\mu)^2}{2 \sigma^2}}   &  z \in [z_1, z_2] \, ,  \\
      0  & \text{otherwise} \, ,
    \end{cases}
    \qquad \quad \mathcal{N} \equiv \int_{z_1}^{z_2} dz \, \frac{1}{\sqrt{2 \pi} \sigma} e^{\frac{-(z-\mu)^2}{2 \sigma^2}} \, , 
\end{align}
noting that using the exact redshift distributions produce nearly identical results.
The normalised redshift distributions of CMASS and LOWZ are shown with superimposed Gaussian models with suitable parameters $\mu$ and $\sigma$ in figure \ref{fig:zdistModel}. 

We compute $\bar{\alpha}$ and $\bar{\epsilon}$ and compare these to $\alpha$ and $\epsilon$ evaluated at the mean redshifts of the truncated artificial distributions $P(z)_{\rm CMASS}$ and $P(z)_{\rm LOWZ}$ in order to compute $\Delta_{\alpha}$ (\ref{eq:alphaerror}) and $\Delta_{\epsilon}$ (\ref{eq:epsilonerror}).

The exact results for the error terms $\Delta_{\alpha}$, $\Delta_{\epsilon}$ and their upper bounds -- assuming knowledge only of the bounds on the distance combinations (\ref{eq:boundsLLtr}), (\ref{eq:boundsL}), (\ref{eq:boundsRRtr}), and (\ref{eq:boundsR}) over the surveys and using the inequalities (\ref{eq:alphaexpandbound}) and (\ref{eq:epsilonerrorbound}) -- are shown in table \ref{table:alphaerror} for four cosmological test-models.  All of these models have error terms $\Delta_{\alpha}$ of order $0.2$\% or smaller when compared to the fiducial spatially-flat FLRW model with $\Omega_M = 0.3, \, \Omega_{\Lambda} = 0.7$. The corresponding upper bounds on $\abs*{\Delta_{\alpha}}$ are of order $5$\% or smaller. 
The value of
$\Delta_{\epsilon}$ for the models tested is of order $0.0005$ or
smaller.  The bounds on $\abs*{\Delta_{\epsilon}}$ are of order
$0.02$ or smaller.
{For the negatively curved FLRW model with $\Omega_M = 0.25, \, \Omega_{\Lambda} = 0.65$ -- which differ by $\sim 0.1$ in cosmological parameters as compared to the Planck \cite{planckParams} best fit $\Lambda$CDM model -- the upper bounds on $\abs*{\Delta_{\alpha}}$ and $\abs*{\Delta_{\epsilon}}$ are particularly small and of size $\lesssim 0.1\%$ and $\lesssim 0.0005$ respectively.}

We note that even though the models investigated in table \ref{table:alphaerror} are significantly different to the fiducial spatially-flat $\Lambda$CDM model with $\Omega_M = 0.3, \, \Omega_{\Lambda} = 0.7$, $\Delta_{\alpha}$ and $\Delta_{\epsilon}$ -- of order $\lsim 0.002$ and $\lsim 0.0005$ respectively -- are much smaller than typical statistical errors in $\alpha$ and $\epsilon$ of order $\sim$1\% and $\sim$0.02 respectively inferred from existing galaxy catalogues \cite{Standardresults,wedgefit}. 

The upper bounds on $\abs*{\Delta_{\alpha}}$ and $\abs*{\Delta_{\epsilon}}$ -- valid for all models which obey the same constraints (\ref{eq:boundsLLtr}) and (\ref{eq:boundsRRtr}) as the tested models over the redshift range -- {are of order $\lsim 0.04$ and $\lsim 0.01$ respectively, and are in most cases comparable or smaller than typical statistical errors in $\alpha$ and $\epsilon$ when inferred from existing galaxy catalogues.}
Note that the bounds on $\abs*{\Delta_{\alpha}}$ and $\abs*{\Delta_{\epsilon}}$ quoted represent worst case scenarios, which are never realised in practice.

We conclude that in order to have a large difference between the two constant AP approximations, we must have models (\say{true} and fiducial) which differ more extremely in their distance measures (and derivatives of these) than is the case for the models presented in table \ref{table:alphaerror}. 
This can happen, for example, if the \say{true} underlying cosmological model has structure on a hierarchy of scales, with resulting small/intermediate scale wiggles in the distance-redshift relations $g_{\theta \theta}(z)$ and $g_{zz}(z)$.

\begin{table}[!htb]
\small
\centering
\scalebox{0.94}{
\begin{tabular}{| lllllllll |}
\hline
  & \multicolumn{2}{l}{$\Lambda\text{CDM}_{\Omega_M = 0.99}$}   & \multicolumn{2}{l}{\text{Milne}} & \multicolumn{2}{l}{$\text{FLRW}_{\Omega_M = 0.25}^{\Omega_{\Lambda} = 0.65}$} & \multicolumn{2}{l |}{$\text{Timescape}_{\Omega_M = 0.3}$}  \\ [0.5ex]
 & \textbf{LOWZ} & \textbf{CMASS}   &  \textbf{LOWZ} & \textbf{CMASS} &  \textbf{LOWZ} & \textbf{CMASS}&  \textbf{LOWZ} & \textbf{CMASS} \\ [0.5ex]
\hline
$\alpha(\bar{z})$ & 0.85& 0.78 & 0.93 & 0.92 &  0.9982   & 1.00052 & 0.95 & 0.94 \\
$\abs*{\Delta_{\alpha}}$ bound & 0.036& 0.0028 & 0.010 & 0.0016 &  0.0016    &  0.00034  & 0.0071 & 0.00094 \\
$\Delta_{\alpha}$ & 0.0022 & 0.00097 & 0.0015 &0.00066 & 0.00014 & 0.000060 &  0.0010 & 0.00043  \\
\hline
$\epsilon(\bar{z})$ & -0.037 & -0.052 & -0.020 & -0.027 &  -0.00078  &   -0.00084  & -0.023 & -0.029 \\
$\abs*{\Delta_{\epsilon}}$ bound & 0.010 & 0.00076  &0.0032  & 0.00049 &  0.00054  & 0.00011  & 0.0022 & 0.00029 \\
$\Delta_{\epsilon}$ & 0.00052 & 0.00022 & 0.00029 & 0.00012 &0.000022  & 0.0000071 & 0.00048 & 0.00019 \\
\hline
\end{tabular}
}
\caption{The AP scaling error terms $\Delta_{\alpha}$ and $\Delta_{\epsilon}$ computed from the artificial truncated Gaussian distributions $P(z)_{\rm CMASS}$ and
$P(z)_{\rm LOWZ}$. The corresponding upper bounds on $\abs*{\Delta_{\alpha}}$ and $\abs*{\Delta_{\epsilon}}$ obtained from (\ref{eq:alphaexpandbound}) and (\ref{eq:epsilonerrorbound}) respectively are also shown.}
\label{table:alphaerror}
\end{table}

We now consider a simple class of model cosmologies which can illustrate what might happen when gradients in the metric components become large.
We consider a simple three-parameter family of spatially flat unphysical models with metrics
\begin{align} \label{eq:metrictoy}
ds^2 = - c^2  dt^2 + \tilde{a}(t)^2 (d\tilde{D}^2 + \tilde{D}^2 d\Omega^2) \, , 
\end{align}
in coordinates adapted to a central observer.
The models are constructed by distorting the comoving distance--redshift relation $D(z)$ of a reference $\Lambda$CDM model with $\Omega_M = 0.3$ in the following way 
\begin{align} \label{eq:Dperturbed}
\tilde{D}(\tilde{z}) = D(\tilde{z}) \, \left(1 + A \cos(f\, \tilde{z} + \Phi) \right)  \, , 
\end{align}
where $A, \, f,$ and $\Phi$ are the amplitude, frequency and phase of the trigonometric distortion respectively. 
This form is chosen as a simple case of a bounded distance-redshift relation around the reference model relation $D(\tilde{z})$, but with the possibility of significant gradients of $\tilde{D}(\tilde{z})$ in redshift.
The Hubble distance function then reads
\begin{align} \label{eq:Hperturbed}
\hspace*{-0.4cm} \frac{c}{\tilde{H}} \equiv - \frac{c}{\tilde{a}} \frac{dt}{d\tilde{z}} = \frac{ d \tilde{D}(\tilde{z})}{d \tilde{z}} = \frac{ d D(\tilde{z})}{d \tilde{z}} \left(1 + A \cos(f\,\tilde{z} + \Phi) \right) -  D(\tilde{z}) \, f \, A \sin(f\,\tilde{z} + \Phi)  \,  , 
\end{align}
where $\tilde{z} \equiv 1/\tilde{a}$, and where the second equality follows from considering radially propagating null rays in the metric (\ref{eq:metrictoy}). 
We note that even though differences in the comoving distance scales $\tilde{D}(\tilde{z})$ and $D(\tilde{z})$ might be small, differences between the derivatives of the comoving distance scales in redshift can be large, if the frequency $f$ of the perturbation (\ref{eq:Dperturbed}) is large. 

The results for the tested unphysical models are shown in table \ref{table:alphaerrorpatho}. 
The error terms $\Delta_{\alpha}$, $\Delta_{\epsilon}$ and their bounds are in general significantly larger than for the model cosmologies in table \ref{table:alphaerror} -- especially for the large frequencies, $f$. 
The error terms $\Delta_{\alpha}$ are of order $1$\% for the two models with $f=30$ and $f=50$, and the error terms $\Delta_{\epsilon}$ are of order $0.01$ - which is similar in order to typical statistical errors in BAO analysis with current galaxy surveys. 
The upper bounds on $\abs*{\Delta_{\alpha}}$ and $\abs*{\Delta_{\epsilon}}$ are as high as $\sim$30\% and $\sim 0.1$ respectively. 
 
These results are intuitive; the more rapidly the \say{true} and fiducial models are varying with respect to each other, the more we expect evaluation at a single redshift and an average of $\alpha$ and $\epsilon$ to differ.
We expect the same tendencies to be present in models which are more complicated than the simple class of distorted models (\ref{eq:metrictoy})--(\ref{eq:Hperturbed}), but which possess the same features in terms of allowed gradients of the relevant metric components.

\begin{table}[!htb]
\small
\centering
\scalebox{0.94}{
\begin{tabular}{| lllllllll |}
\hline
  & \multicolumn{2}{l}{$\text{FLRW} \substack{A=0.01 \\ f=10 \\ \Phi=-0.3}$}   & \multicolumn{2}{l}{$\text{FLRW} \substack{A=0.005 \\ f=15 \\ \Phi=-0.3}$} & \multicolumn{2}{l}{$\text{FLRW} \substack{A=0.005 \\ f=30 \\ \Phi=-0.3}$}   &  \multicolumn{2}{l |}{$\text{FLRW} \substack{A=0.001 \\ f=50 \\ \Phi=-0.6}$}   \\ [0.5ex]
 & \textbf{LOWZ} & \textbf{CMASS}   &  \textbf{LOWZ} & \textbf{CMASS} &  \textbf{LOWZ} & \textbf{CMASS}&  \textbf{LOWZ} & \textbf{CMASS} \\ [0.5ex]
\hline
$\alpha(\bar{z})$ & 0.986  & 1.023 & 1.0046 & 0.983    &   0.986   & 1.0084 & 0.9947 & 0.990 \\ 
$\abs*{\Delta_{\alpha}}$ bound & 0.037  & 0.014  & 0.054  & 0.018  &  0.32   &  0.21  &  0.31  & 0.15   \\
$\Delta_{\alpha}$ & 0.0046 & -0.0047 & -0.0012   &  0.0064 &  0.016   &  -0.0085  &  0.0052  & 0.011 \\
\hline
$\epsilon(\bar{z})$ & -0.0047  & 0.019 & 0.0070 & -0.016  &  -0.011   & 0.013 & -0.0050 & -0.010 \\
$\abs*{\Delta_{\epsilon}}$ bound & 0.012 & 0.0047 &  0.018 &  0.0061  &   0.11   &  0.070 &  0.10  & 0.049 \\
$\Delta_{\epsilon}$ & 0.0026  & -0.0039 & -0.0024  &  0.0061 &   0.012  &  -0.013 &  0.0049 & 0.010 \\
\hline
\end{tabular}
}
\caption{The AP scaling error terms $\Delta_{\alpha}$ and $\Delta_{\epsilon}$ computed from the artificial truncated Gaussian distributions $P(z)_{\rm CMASS}$ and
$P(z)_{\rm LOWZ}$. The corresponding upper bounds on $\abs*{\Delta_{\alpha}}$ and $\abs*{\Delta_{\epsilon}}$ obtained from (\ref{eq:alphaexpandbound}) and (\ref{eq:epsilonerrorbound}) respectively are also shown.}
\label{table:alphaerrorpatho}
\end{table}

\section{The Alcock-Paczy\'nski scaling and the BAO feature}
\label{APBAO}
In order to quantify the impact of the redshift-dependent AP-scaling investigated in section \ref{errorAP} on the BAO feature as viewed in a fiducial cosmology, we must specify a model for the BAO feature. 

Let us investigate the example of the empirical model for the correlation function $\xi^{\rm tr}(D^{\rm tr},\mu^{\rm tr},z)$ as proposed in section \ref{empiricalBAOmain}, with a Gaussian function describing the BAO feature and polynomial terms describing the \say{background} featureless correlation function.
Using the identity derived in (\ref{eq:ximodel}) together with the form of the empirical model of the correlation function (\ref{eq:xifit}) we can write
\begin{align} \label{eq:ximodelempirical}
& \xi(D,\mu,z) =\xi^{\rm tr}\left(D^{\rm tr}(D,\mu,\alpha,\epsilon),\mu^{\rm tr}(\mu,\epsilon),z\right)    \nonumber
\\&\approx D^2 \alpha^2 \frac{  1 + \psi(\epsilon) \mu^2  }{(1 + \epsilon)^2}  A \e^{-\left( D \alpha  \sqrt{1 + \psi(\epsilon)  \mu^2  }/ (1 + \epsilon)  - r \right)^2\!/ \,(2 \sigma^2)} +\, C_0(\mu) + \frac{C_1(\mu) }{D} + \frac{C_2(\mu )}{D^2} \, , 
\end{align}
where the redshift dependence of $\alpha$, $\epsilon$, $A$, $\sigma$, and the polynomial coefficients $C_0(\mu)$, $C_1(\mu)$, and $C_2(\mu)$ is implicit, and where $\psi(\epsilon)$ is given by the second equation of (\ref{eq:DAP}).
The approximation in (\ref{eq:ximodelempirical}) follows from the approximations (\ref{eq:DAP}) and (\ref{eq:muAP}) for $D^{\rm tr}(D,\mu,\alpha,\epsilon)$ and $\mu^{\rm tr}(\mu,\epsilon)$ respectively. 
We note that (\ref{eq:ximodelempirical}) has the same form as (\ref{eq:xifit}) (Gaussian in $D$ plus a second-order polynomial function in $1/D$ for fixed $\mu$ and $z$), but the coefficients for each value of $\mu,z$ are redefined by the AP-scaling. 

We might further obtain $\xi(D,\mu)$ from (\ref{eq:ximodelempirical}) by applying the approximation (\ref{eq:xiapprox}), neglecting the second-order corrections from the non-multiplicatively separable parts $\delta(D,\mu,z)$ and $\delta_{{\text{Poisson}}}(D,\mu,z)$ of $f(D,\mu,z)$ and $f_{\text{Poisson}}(D,\mu,z)$ respectively. 
\begin{align} \label{eq:xiapproxexample}
 \xi(D,\mu)  &\approx  \int \, dz P(z) \xi(D,\mu,z)  \nonumber \\
& = \int \, dz P(z) \left( D^2 \alpha^2 \frac{  1 + \psi(\epsilon) \mu^2  }{(1 + \epsilon)^2}  A \e^{-\left( D \alpha  \sqrt{1 + \psi(\epsilon)  \mu^2  }/ (1 + \epsilon)  - r \right)^2\!/ \,(2 \sigma^2)} \right) \nonumber \\ 
&\hbox to 32mm{\hfil} +\, \overline{C}_0(\mu) + \frac{\overline{C}_1(\mu) }{D} + \frac{\overline{C}_2(\mu )}{D^2}  \, , 
\end{align}
where the overbar refers to the averaging operation in redshift $\overline{S} \equiv \int \, dz P(z) S(z)$. 
For $\alpha$ and $\epsilon$ for which deviations remain small ($\ll 1$) over the survey, we can use the approximation (\ref{eq:xiapproxexpand}), to simplify the Gaussian integral in (\ref{eq:xiapproxexample})
\begin{align} \label{eq:xiapproxexpandexample}
& \xi(D,\mu)  \approx  \int \, dz P(z) \xi(D,\mu,z) \nonumber \\
&\approx D^2 \bar{\alpha}^2 \frac{  1 + \psi(\bar{\epsilon}) \mu^2  }{(1 + \bar{\epsilon})^2}  A \e^{-\left( D \bar{\alpha}  \sqrt{1 + \psi(\bar{\epsilon}) \mu^2  }/ (1 +\bar{\epsilon})  - r \right)^2\!/ \,(2 \sigma^2)} + \, \overline{C}_0(\mu) + \frac{\overline{C}_1(\mu) }{D} + \frac{\overline{C}_2(\mu )}{D^2}  \, , 
\end{align}
where the Gaussian parameters $A(z)$, $\sigma(z)$ are now evaluated at the mean redshift of the survey $\bar{z}$. 
Note that the standard constant AP approximation $\alpha(z) \mapsto \alpha(\bar{z})$, $\epsilon(z) \mapsto \epsilon(\bar{z})$ yields the same form as the expression in (\ref{eq:xiapproxexpandexample}) but with $\bar{\alpha}, \bar{\epsilon}$ replaced by $\alpha(\bar{z}), \epsilon(\bar{z})$. 

From the definition of the wedge functions (\ref{eq:wedge2point}) we find that the wedges corresponding to the 2-point correlation function (\ref{eq:xiapproxexpandexample}) read 
\begin{align} \label{eq:xiwedgeexample}
& \xi_{ [\mu_{1}, \mu_{2}]} (D) = \frac{1}{\mu_{2}- \mu_{1}} \int_{\mu_{1}}^{\mu_{2}}\, d \mu\, \xi(D, \mu) \approx  \nonumber  \\
& \frac{\int_{\mu_{1}}^{\mu_{2}}\, d \mu\, D^2 \bar{\alpha}^2 \frac{  1 + \psi(\bar{\epsilon}) \mu^2  }{(1 + \bar{\epsilon})^2}  A \e^{-\left( D \bar{\alpha}  \sqrt{1 + \psi(\bar{\epsilon}) \mu^2  }/ (1 +\bar{\epsilon})  - r  \right)^2\!/ \,(2 \sigma^2)}      }{\mu_{2}- \mu_{1}} +   \overset{\mu_1, \mu_2}{C_0} + \frac{\overset{\mu_1, \mu_2}{C_1} }{D} + \frac{\overset{\mu_1, \mu_2}{C_2}}{D^2}  , 
\end{align}
with
\begin{align} \label{eq:Cwedgedef}
&\overset{\mu_1, \mu_2}{C_0} \equiv  \frac{\int_{\mu_{1}}^{\mu_{2}}\, d \mu\,  \overline{C}_0(\mu)  }{\mu_{2}- \mu_{1} } \, , \qquad \overset{\mu_1, \mu_2}{C_1} \equiv  \frac{\int_{\mu_{1}}^{\mu_{2}}\, d \mu\,  \overline{C}_1(\mu)  }{\mu_{2}- \mu_{1} }  \, , \qquad \overset{\mu_1, \mu_2}{C_2} \equiv  \frac{\int_{\mu_{1}}^{\mu_{2}}\, d \mu\,  \overline{C}_2(\mu)  }{\mu_{2}- \mu_{1} }  . 
\end{align}
For sufficiently small $\bar{\epsilon}$, the Gaussian part of (\ref{eq:xiwedgeexample}) can be expanded in $\psi \ll 1$ and $\psi(\bar{\epsilon})D/\sigma \ll 1$ to first order for relevant distance scales $D$ \footnote{See section 4.1 of \cite{HBLW} for details.}, such that (\ref{eq:xiwedgeexample}) reads 
\begin{align} \label{eq:xiwedgeexample2}
 \xi_{ [\mu_{1}, \mu_{2}]} (D) & \approx   D^2 \bar{\alpha}^2 \frac{  1 + \psi(\bar{\epsilon}) \kappa_{\mu_1}^{\mu_2} }{(1 + \bar{\epsilon})^2}  A \e^{-\left( D \bar{\alpha}  \sqrt{1 + \psi(\bar{\epsilon})  \kappa_{\mu_1}^{\mu_2}   }/ (1 +\bar{\epsilon})  - r \right)^2\!/ \,(2 \sigma^2)}      +   \overset{\mu_1, \mu_2}{C_0} + \frac{\overset{\mu_1, \mu_2}{C_1} }{D} + \frac{\overset{\mu_1, \mu_2}{C_2}}{D^2}  \nonumber \\ 
& \approx  D^2 \tilde{A} \e^{-\left( D - \tilde{r} \right)^2\!/ \,(2 \tilde{\sigma}^2)}      +   \overset{\mu_1, \mu_2}{C_0} + \frac{\overset{\mu_1, \mu_2}{C_1} }{D} + \frac{\overset{\mu_1, \mu_2}{C_2}}{D^2}     , 
\end{align}
where 
\begin{align} \label{eq:kappamu}
& \kappa_{\mu_1}^{\mu_2}  \equiv  \frac{1}{\mu_2 - \mu_1 } \int_{\mu_{1}}^{\mu_{2}}\, d \mu\, \mu^2 = \frac{1}{3} \frac{\mu_2^3 - \mu_1^3 }{\mu_2 - \mu_1} \, ,
\end{align}
and where the Gaussian parameters in the final line are kept to first order in $\psi(\bar{\epsilon})$ are given by\footnote{{The distorted BAO feature is equivalent to that derived in section 4.1 in \cite{HBLW}, but with the constant $\alpha(\bar{z}), \epsilon(\bar{z})$ parameters replaced by $\bar{\alpha}, \bar{\epsilon}$.}}
\begin{align} \label{eq:gausparams}
&\tilde{r} \equiv \frac{1 - \frac{1}{2} \kappa_{\mu_1}^{\mu_2} \psi(\bar{\epsilon})  }{\bar{\alpha} \,  / \,  (1+ \bar{\epsilon}) }  r  , \qquad \tilde{\sigma} \equiv \frac{1 - \frac{1}{2} \kappa_{\mu_1}^{\mu_2} \psi(\bar{\epsilon})  }{ \bar{\alpha} \,  / \,  (1+ \bar{\epsilon}) } \sigma , \qquad \widetilde{A} \equiv \frac{ \bar{\alpha}^2}{ (1+ \bar{\epsilon})^2}  (1 + \kappa_{\mu_1}^{\mu_2} \psi(\bar{\epsilon})  ) A .
\end{align}
This can be obtained from the first line of (\ref{eq:xiwedgeexample2}), by expanding to first order $\sqrt{1 + \psi(\bar{\epsilon})  \kappa_{\mu_1}^{\mu_2}   } \approx 1 + \psi(\bar{\epsilon})  \kappa_{\mu_1}^{\mu_2} / 2 $, and absorbing the coefficient multiplying $D$ in the exponential into a rescaling of $r$ and $\sigma$. 

For the isotropic, transverse, and radial wedges (\ref{eq:wedge2pointsc}) we can thus compute the scaled Gaussian parameters ($\tilde{r}$, $\tilde{\sigma}$, $\tilde{A}$) (\ref{eq:gausparams}) relative to the undistorted parameters ($r$, $\sigma$, $A$) as a function of $\bar{\alpha}$ and $\bar{\epsilon}$ by substituting the value of $\kappa_{\mu_1}^{\mu_2}$ corresponding to the given wedge 
\begin{align} \label{eq:kappamu}
& \kappa_{0}^{1}  = \frac{1}{3} \, , \qquad   \kappa_{0}^{0.5}  = \frac{1}{12} \, , \qquad   \kappa_{0.5}^{1}  = \frac{7}{12} \, .
\end{align}
We note that for the \say{isotropic wedge} ($\mu_1 = 0$, $\mu_2 = 1$), (\ref{eq:gausparams}) reduces to $\tilde{r} = r/\bar{\alpha} \,  , \, \sigma/\bar{\alpha} \, , \,  \widetilde{A} =  \bar{\alpha}^2  A$, since from the definition of $\psi$ (\ref{eq:DAP}) $\psi(\epsilon) \approx 1 + 6 \epsilon$ to first order in $\epsilon$ (or first order in $\psi$). Thus, to lowest order, the isotropic wedge contains information on the isotropic scaling $\alpha$ only. 

The exact result for the Gaussian parameter distortions (\ref{eq:gausparams}) is useful for gaining intuition about the appearance of the BAO feature in a cosmology which is distorted from the true one according to AP-scaling factors $\alpha(z)$ and $\epsilon(z)$. 
If the conditions for the expansions (\ref{eq:xiwedgeexample2}) and (\ref{eq:xiwedgeexample}) are not met, then the exact integrals in $\mu$ over (\ref{eq:xiapproxexample}) must be performed numerically. 

\section{Testing the predicted shift of the BAO feature with mock catalogues} 
\label{testsmocks}

In this section we test the predictions of section \ref{errorAP} and section \ref{APBAO} for the reparametrisation effects on the BAO feature using CMASS NGC mock catalogues.  The advantage of using mock catalogues is that by averaging many mock catalogues we can obtain arbitrarily small statistical variance in our correlation function estimators, meaning that we can detect small systematics which would otherwise be difficult to disentangle from noise.  A further advantage is that we know the true underlying cosmology of the mock catalogues.

By fitting the empirical Gaussian correlation function model described in section \ref{APBAO} to the mock data, we test the accuracy of the BAO scale recovered when using the standard constant AP scaling approximation $\alpha(z) \mapsto \alpha(\bar{z})$, $\epsilon(z) \mapsto \epsilon(\bar{z})$, the modified constant AP scaling approximation $\alpha(z) \mapsto \bar{\alpha}$, $\epsilon(z) \mapsto \bar{\epsilon}$, and the exact AP redshift-dependent scalings $\alpha(z)$ and $\epsilon(z)$ with no approximation respectively.  We calibrate the BAO scale against that measured in {the fiducial cosmology of the mock catalogues} in order to {account} for any systematic offsets that might occur between the peak of the empirical gaussian and the BAO scale.

\subsection{The mock catalogues}
\label{mocks}
In this analysis we use the Quick Particle Mesh (QPM) mock catalogues as described in
detail in \cite{BOSSsystematics}.
The QPM mock catalogues are generated from \LCDM\ $N$-body simulations, and are designed for
the BOSS clustering analysis. The number density in the mock
catalogues match the observed galaxy number density of
the BOSS catalogues, and follow the same radial and angular selection
functions. 
The QPM simulations are generated from the fiducial \LCDM\ cosmology 
\begin{align} \label{eq:qpmparams}
\Omega_M =0.29, \quad \Omega_{\Lambda} =0.71, \quad \Omega_b =0.048, \quad \sigma_8 =0.8, \quad h=0.7 ,
\end{align}
where $\Omega_M$, $\Omega_{\Lambda}$ and $\Omega_b$ are the present
epoch matter density parameter, dark energy density parameter, and baryonic
matter density parameter respectively, $\sigma_8$ is the
root mean square of the linear mass fluctuations at the present epoch
averaged at scales $8\hm$ given by the integral over the
\LCDM\ power spectrum, and $H_0 = 100\, h$ km/s/Mpc is the
Hubble parameter evaluated at the present epoch. The sound horizon at
the drag epoch within this model is $r_s = 103.05\hm$.

In this analysis we focus on the CMASS NGC catalogue. 
There are 1000 QPM mock catalogues available for the CMASS NGC catalogue. 
We use all 1000 QPM mock catalogues in calculating the correlation function of {the fiducial QPM model with $\Omega_{M} = 0.29$ which we use to calibrate the BAO peak position} (see section \ref{recoveringBAO}). For the remaining trial cosmologies we use 200 mock catalogues.
We use these, along with the associated random catalogues as described in \cite{randompoisson}, to construct the 2-point correlation in a number of different trial cosmologies.

\subsection{The likelihood function and the fitting procedure} 
\label{recoveringBAO}
We assume the likelihood function $\mathcal{L}$ of data given the empirical fitting model (\ref{eq:xifit}) $\xi_{\text{Fit}}$
\begin{align} \label{eq:likelihood}
\mathcal{L}\left(\left. \bar{\hat{\xi}} \, \right| \xi_{\text{Fit}} \right) \propto \exp(- \chi^2/2) ,
\end{align}
with
\begin{align} \label{eq:chi2}
\chi^2 = Z^\transpose \doubleunderline{C}_{\bar{\hat{\xi}}}^{-1} Z , \qquad Z = \bar{\hat{\xi}} - \xi_{\text{Fit}} , \qquad
\end{align}
where $\hat{\xi}$ is the binned estimate of the (isotropic or wedge)
2-point correlation function computed for each mock {using the Landy-Szalay estimator \cite{LSestimator}}, and where $\bar{\hat{\xi}}$ is the unweighted average
over the mock catalogues. In the anisotropic wedge analysis, the transverse and radial
estimates are combined into a single vector $\hat{\xi}$ in order to
perform a combined fit. 
$\xi_{\text{Fit}}$ is the fitting function, which in this analysis is taken to be the model described in section \ref{APBAO}. The covariance matrix of
$\bar{\hat{\xi}}$ is given by the covariance of the individual
measurements $\hat{\xi}$ scaled by the inverse of the number of mock catalogues used, $N_{\text{mocks}}$
\begin{align} \label{eq:covmat}
\doubleunderline{C}_{\bar{\hat{\xi}}} = \frac{1}{N_{\text{mocks}}} \doubleunderline{C}_{\hat{\xi}} , \qquad \doubleunderline{C}_{\hat{\xi}} = \overline{ (\hat{\xi} - \bar{\hat{\xi}}) (\hat{\xi} - \bar{\hat{\xi}})^\transpose } ,
\end{align}
where the overbar represents the averages over the mock catalogues.

As discussed in \cite{HBLW} and in section \ref{empiricalBAOmain} of this paper, the calibration of the BAO scale to the peak of the Gaussian model (\ref{eq:xifit}), $r$, is itself a source of error in empirical BAO investigations. 
In order to account for the calibration issue, we use the inferred peak position from the 2-point correlation function computed in the spatially-flat $\Lambda$CDM reference model with {$\Omega_M = 0.29$}.  We use the fitting function (\ref{eq:xiwedgeexample}) and the likelihood function (\ref{eq:likelihood}) for estimating the best fit of $\tilde{r}_{\text{isotropic}} = r/\bar{\alpha}$ and $\bar{\epsilon}$ for the spatially-flat $\Lambda$CDM reference model with $\Omega_M = 0.29$ using 1000 CMASS NGC QPM mock catalogues (see section \ref{mocks}). 
The fitting range is taken to be $[50,150]\,$Mpc/$h$. 

For the isotropic fit, $\mu_1 = 0, \mu_2 = 1$, we find a best fit value {$\hat{\tilde{r}}_{\text{isotropic}} = 102.47\,$}Mpc/$h$.
For the anisotropic analysis, consisting of a combined fit to the transverse wedge $\mu_1 = 0, \mu_2 = 0.5$ and the radial wedge $\mu_1 = 0.5, \mu_2 = 1$, we get best fit values $\hat{\tilde{r}}_{\text{isotropic}} = 102.46 \,$Mpc/$h$ and $\hat{\bar{\epsilon}} = -0.0004$. 
The discrepancy between the best fit estimates of the isotropic peak positions are $0.01\,$Mpc/$h$, while the errors in the individual estimates are of order $0.08\,$Mpc/$h$. Thus the isotropic and anisotropic peak positions are consistent within the level of uncertainty on the best fit. 
The error in the best fit epsilon is $0.0007$, and $\hat{\bar{\epsilon}}$ is thus consistent with zero at the level of precision of 1000 mock catalogues.  These findings are consistent with the reference model (\ref{eq:xifit}) with $r = 102.47\,$Mpc/$h$.
We thus use this empirical model as a reference, and consider reparametrisations (\ref{eq:ximodelempirical}) with respect to the reference $\Lambda$CDM model with $\Omega_M = 0.29$. 

We are now able to quantify the accuracy of the predictions of the constant AP scaling approximations $\alpha(z) \mapsto \alpha(\bar{z})$, $\epsilon(z) \mapsto \epsilon(\bar{z})$ and $\alpha(z) \mapsto \bar{\alpha}$, $\epsilon(z) \mapsto \bar{\epsilon}$ respectively, and the exact integral expression (\ref{eq:xiapproxexample}), under the assumptions of the empirical fitting function.  Let us first consider any constant AP-approximation $\alpha(z) \mapsto \mathcal{C}_{\alpha}$, $\epsilon(z) \mapsto \mathcal{C}_{\epsilon}$, where $\mathcal{C}_{\alpha}$ and $\mathcal{C}_{\epsilon}$ are constants. 
With this approximation (\ref{eq:xiapproxexample}) reads 
\begin{align} \label{eq:xiconstantAP}
& \xi(D,\mu)  \approx  \int \, dz P(z) \xi(D,\mu,z)  \nonumber \\
& =  D^2 \mathcal{C}_{\alpha}^2 \frac{  1 + \psi(\mathcal{C}_{\epsilon}) \mu^2  }{(1 + \mathcal{C}_{\epsilon})^2}  A \e^{-\left( D \mathcal{C}_{\alpha}  \sqrt{1 + \psi(\mathcal{C}_{\epsilon})  \mu^2  }/ (1 + \mathcal{C}_{\epsilon})  - r \right)^2\!/ \,(2 \sigma^2)}  +\, \overline{C}_0(\mu) + \frac{\overline{C}_1(\mu) }{D} + \frac{\overline{C}_2(\mu )}{D^2}  \, , 
\end{align}
which can be substituted into the definition of the wedges (\ref{eq:wedge2point}) to obtain 
\begin{align} \label{eq:xiconstantAPwedge}
\hspace*{-1cm}  \xi_{ [\mu_{1}, \mu_{2}]} (D) & \equiv  \frac{1}{\mu_2 - \mu_1} \int_{\mu_1}^{\mu_2}  d\mu \, \xi(D,\mu) \nonumber \\
& \approx  \frac{1}{\mu_2 - \mu_1} \int_{\mu_1}^{\mu_2}  d\mu \,  D^2 \mathcal{C}_{\alpha}^2 \frac{  1 + \psi(\mathcal{C}_{\epsilon}) \mu^2  }{(1 + \mathcal{C}_{\epsilon})^2}  A \e^{-\left( D \mathcal{C}_{\alpha}  \sqrt{1 + \psi(\mathcal{C}_{\epsilon})  \mu^2  }/ (1 + \mathcal{C}_{\epsilon})  - r \right)^2\!/ \,(2 \sigma^2)}\nonumber\\  &\hbox to 32mm{\hfil}
+  \overset{\mu_1, \mu_2}{C_0} + \frac{\overset{\mu_1, \mu_2}{C_1} }{D} + \frac{\overset{\mu_1, \mu_2}{C_2}}{D^2}    \, , 
\end{align}
where the polynomial coefficients $\overset{\mu_1, \mu_2}{C_0}$, $\overset{\mu_1, \mu_2}{C_1}$, and $\overset{\mu_1, \mu_2}{C_2}$ are given by (\ref{eq:Cwedgedef}). 
We perform the exact numerical integral (\ref{eq:xiconstantAPwedge}) for the three wedges $\xi_{ [0,1]}$, $\xi_{[0, 0.5]}$, $\xi_{ [0.5, 1]}$ and fit for the independent parameters $r/\mathcal{C}_{\alpha} , \, \mathcal{C}_{\alpha}^2 A, \, \sigma/\mathcal{C}_{\alpha}, \,  \mathcal{C}_{\epsilon},  \overset{\mu_1, \mu_2}{C_0},  \overset{\mu_1, \mu_2}{C_1}, \overset{\mu_1, \mu_2}{C_2}$ for a given model cosmology. 
Using the calibrated peak position $r = 102.47$ Mpc/$h$ for the reference spatially-flat $\Lambda$CDM model with $\Omega_M = 0.29$, we can compare the best fit estimates of $r/\mathcal{C}_{\alpha}$ and $\mathcal{C}_{\epsilon}$ with the standard constant AP scaling approximation $\mathcal{C}_{\alpha} = \alpha(\bar{z}), \, \mathcal{C}_{\epsilon} = \epsilon(\bar{z})$, and the modified constant AP scaling approximation $\mathcal{C}_{\alpha} = \bar{\alpha}, \, \mathcal{C}_{\epsilon} = \bar{\epsilon}$. 

We define the fractional error in any given constant AP scaling approximation as 
\begin{align} \label{eq:constantAPfiterrors}
& \text{APerror}_{r} = \frac{\widehat{r/\mathcal{C}_{\alpha}} - (r/\mathcal{C}_{\alpha})_{\text{th}}}{ (r/\mathcal{C}_{\alpha})_{\text{th}} } \, , \qquad  \text{APerror}_{\epsilon} = \hat{\mathcal{C}}_{\epsilon} - ( \mathcal{C}_{\epsilon} )_{\text{th}}  \qquad (\text{const. AP approx.}) \, , 
\end{align}
where $\widehat{r/\mathcal{C}_{\alpha}}$, $\hat{\mathcal{C}}_{\epsilon}$ are the best fit estimates of the parameters $r/\mathcal{C}_{\alpha}$, $\mathcal{C}_{\epsilon}$, and $(r/\mathcal{C}_{\alpha})_{\text{th}}$, $(\mathcal{C}_{\epsilon})_{\text{th}}$ are the corresponding theoretical predictions obtained from the calibration scale $r_{\text{th}} = 102.47$ Mpc/$h$, and the choice of constant AP approximation $\mathcal{C}_{\alpha}, \, \mathcal{C}_{\epsilon}$. 

For $\mathcal{C}_{\alpha} = \alpha(\bar{z}), \, \mathcal{C}_{\epsilon} = \epsilon(\bar{z})$, we can compute $(\mathcal{C}_{\alpha})_{\text{th}}, \, ( \mathcal{C}_{\epsilon})_{\text{th}}$ from the metric components $g_{\theta \theta}, \, g_{zz}$ of the tested model and the reference spatially-flat $\Lambda$CDM model with $\Omega_M = 0.29$ respectively evaluated at the mean redshift $z$ of the CMASS NGC catalogue.  For $\mathcal{C}_{\alpha} = \bar{\alpha}, \, \mathcal{C}_{\epsilon} = \bar{\epsilon}$, we can compute $(\mathcal{C}_{\alpha})_{\text{th}}, \, ( \mathcal{C}_{\epsilon})_{\text{th}}$ from the metric components $g_{\theta \theta}, \, g_{zz}$ of the tested model cosmology and the reference spatially-flat $\Lambda$CDM model with $\Omega_M = 0.29$ respectively and from the model redshift distribution $P(z)$ of the CMASS NGC catalogue. 

We also use this redshift distribution to evaluate the exact integral expression (\ref{eq:xiapproxexample}) where no constant AP approximation is made, using the knowledge of the exact AP scaling functions $\alpha(z), \, \epsilon(z)$ between the tested model cosmology and the reference spatially-flat $\Lambda$CDM model with $\Omega_M = 0.29$. 
Substituting the approximation (\ref{eq:xiapproxexample}) in the definition of the wedge functions (\ref{eq:wedge2point}) we find
\begin{align} \label{eq:xiapproxnoconstAPwedge}
 & \hspace*{-0.5cm} \xi_{ [\mu_{1}, \mu_{2}]} (D)  \equiv  \frac{1}{\mu_2 - \mu_1} \int_{\mu_1}^{\mu_2}  d\mu \, \xi(D,\mu) \nonumber \\
& \hspace*{-0.5cm}  \approx \frac{1}{\mu_2 - \mu_1} \int_{\mu_1}^{\mu_2}  d\mu \int dz\,P(z) \left( D^2 \alpha^2 \frac{  1 + \psi(\epsilon) \mu^2  }{(1 + \epsilon)^2}  A \e^{-\left( D \alpha  \sqrt{1 + \psi(\epsilon)  \mu^2  }/ (1 + \epsilon)  - r \right)^2\!/ \,(2 \sigma^2)} \right) \nonumber  \\
& \qquad  \qquad  \qquad \qquad \qquad  +  \overset{\mu_1, \mu_2}{C_0} + \frac{\overset{\mu_1, \mu_2}{C_1} }{D} + \frac{\overset{\mu_1, \mu_2}{C_2}}{D^2}   \, , 
\end{align}
where the free parameters are $r,\, A,\, \sigma, \overset{\mu_1, \mu_2}{C_0},  \overset{\mu_1, \mu_2}{C_1}, \overset{\mu_1, \mu_2}{C_2}$. 
Note that there is no \say{$\epsilon$} fitting parameter describing the anisotropic warping as in the corresponding fitting function (\ref{eq:xiconstantAPwedge}), since the exact AP scaling functions $\alpha(z), \, \epsilon(z)$ are integrated over in (\ref{eq:xiapproxnoconstAPwedge}). 
We can, however, artificially introduce a \say{warping} fitting parameter $\mathcal{K}_{\epsilon}$ by making the replacement $\epsilon(z) \rightarrow \epsilon(z) +  \mathcal{K}_{\epsilon} $ in (\ref{eq:xiapproxnoconstAPwedge}). $\mathcal{K}_{\epsilon} = 0$ gives back the exact expression (\ref{eq:xiapproxnoconstAPwedge}), and a non-zero $\mathcal{K}_{\epsilon}$ quantifies constant warping not accounted for in the expression (\ref{eq:xiapproxnoconstAPwedge}). 
We thus arrive at the 7 independent parameters $r,\, A,\, \sigma, \, \mathcal{K}_{\epsilon}, \overset{\mu_1, \mu_2}{C_0},  \overset{\mu_1, \mu_2}{C_1}, \overset{\mu_1, \mu_2}{C_2}$. 

We define fractional errors analogous to those of the constant AP approximation fitting function (\ref{eq:constantAPfiterrors}) for the \say{exact} fitting function (\ref{eq:xiapproxnoconstAPwedge}) as
\begin{align} \label{eq:constantAPfiterrorsexact}
& \text{APerror}_{r} = \frac{\hat{r} - r_{\text{th}}}{ r_{\text{th}} } \, , \qquad  \text{APerror}_{\epsilon} = \hat{\mathcal{K}}_{\epsilon}  \qquad (\text{exact $\alpha(z)$ and $\epsilon(z)$})
\end{align}
where $\hat{r}$ and $\hat{\mathcal{K}}_{\epsilon}$ are best fit estimates of the parameters $r$ and $\mathcal{K}_{\epsilon}$, and where $r_{\text{th}} = 102.47$ Mpc/$h$ is the calibration scale of the reference spatially-flat $\Lambda$CDM model with $\Omega_M = 0.29$.

\subsection{Large-scale model cosmologies}
\label{FLRWexamples}
First, we test the recovery of the BAO characteristic scale when using various large-scale cosmological models which differ substantially from the reference spatially-flat $\Lambda$CDM cosmological model with $\Omega_M = 0.29$. 
We are interested in testing models which are far from the reference $\Lambda$CDM model, rather than necessarily being candidates for accurately describing cosmological data. 

We consider the two-parameter family of spatially-flat FLRW models parameterised by the matter cosmological parameter $\Omega_M$ and the constant dark energy equation of state parameter $w$ for which we consider the values $\{-0.333, -1, -1.333\}$.
The dark energy cosmological parameter is given by $\Omega_{d.e.} = 1 - \Omega_M$, and all other cosmological parameters are zero. 
In addition, we consider three curved models: the Milne universe, the positively-curved $\Lambda$CDM universe with $\Omega_M = 1$ and $\Omega_K = -1$ and the timescape model with $\Omega_M = 0.3$.

For each test model we compute the mean of the (isotropic and wedge) 2-point correlation function of 200 mock catalogues $\bar{\hat{\xi}}$, and the associated covariance matrix (\ref{eq:covmat}).
For most models, the correlation function is calculated for the range of distances $[0,150]$ Mpc/$h$, however for models with $\bar{\alpha} < 1$ the correlation function is calculated out to values of $200$ Mpc/$h$. 
This is done in order to ensure that the full BAO feature lies within the calculated range, and to facilitate a broad enough physical fitting range. 

\begin{figure}[!htb]
\centering
\includegraphics[scale=0.55]{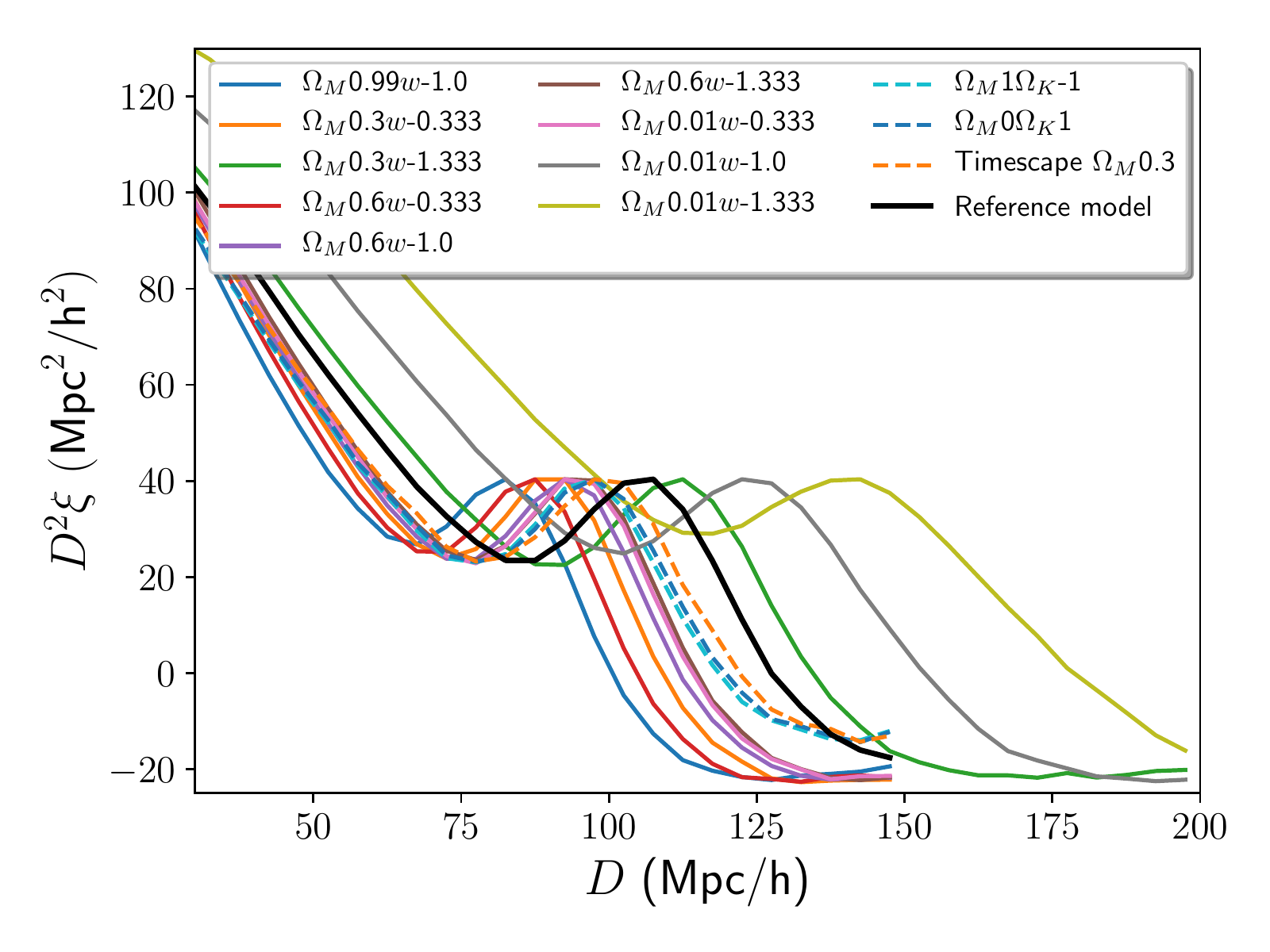}
\caption{The mean of the isotropic 2-point correlation function of 200 mock catalogues for each tested model. The reference spatially-flat $\Lambda$CDM cosmological model with $\Omega_M = 0.29$ is shown in black. In order to visualise the shift of the acoustic scale, the correlation function for each model has been normalised by a constant such that the local maxima are aligned with that of the reference $\Lambda$CDM model.}
\label{fig:meanmocks_Physical}
\end{figure}

The mean isotropic 2-point correlation function is shown in figure \ref{fig:meanmocks_Physical} for each model. Each correlation function has been normalised by a constant in order to align the local peaks of the correlation functions, to make the shift of the peak position more visible. 
As expected, the isotropic BAO feature shifts according to the magnitude of the isotropic scaling of the distance measures relative to the reference model $\alpha$. 
We shall investigate the shift of the peak in detail below. 

For 200 mock catalogues, the $1 \, \sigma$ error in the estimate of isotropic BAO scale $r$ is of order $\sim 0.2$ Mpc/$h$ corresponding to a $0.2$\% error, and the error in the estimate of the warping parameter is of order $\sim 0.002$.\footnote{The errors are reasonably robust between the test models and consistent within $\sim 0.1$\% for the peak position and $\sim 0.001$ for $\epsilon$.}
For perfect accuracy of any applied AP approximation and for perfect modelling assumptions in general, we expect recovery of the isotropic BAO scale and the warping parameter respectively at this level of accuracy.

In order to minimise systematic errors involved with the choices made in the fitting procedure, we fit to the range of distances $[50/\bar{\alpha},150/ \bar{\alpha}]$ Mpc/$h$, where $\alpha$ is the AP-scaling between the reference $\Lambda$CDM model and the model tested. 
In this way, we are approximately fitting to the same physical distance scale for all models involved, irrespective of the ruler with which we measure the distance between galaxies. 

For the isotropic fits, we fix the constant warping parameters $\mathcal{C}_{\epsilon}$ and $\mathcal{K}_{\epsilon}$ in (\ref{eq:xiconstantAPwedge}) and (\ref{eq:xiapproxnoconstAPwedge}) respectively, in order to avoid the degeneracies introduced by the quadratic contributions of $\epsilon$ to the isotropic wedge. 
We fix $\mathcal{C}_{\epsilon} = \bar{\epsilon}$ in the constant AP scaling approximation analysis \footnote{Changing the fixed value to $\mathcal{C}_{\epsilon} = \epsilon(\bar{z})$ does not significantly alter the results.} and $\mathcal{K}_{\epsilon} = 0$ in the exact analysis.

\begin{figure}[!htb]
\centering
\begin{subfigure}[b]{.6\textwidth}
\includegraphics[width=\textwidth]{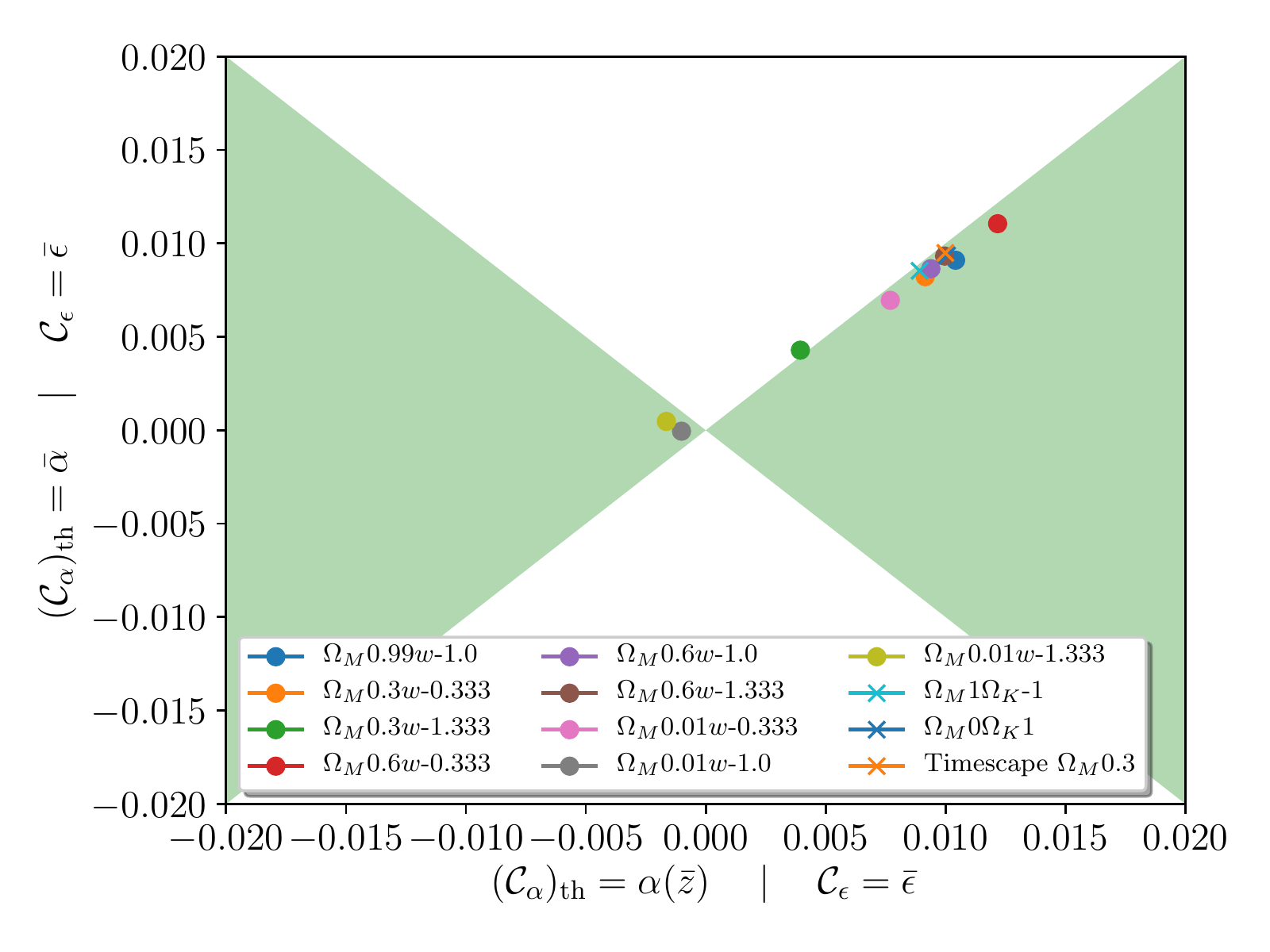}
\caption{The constant AP scaling approximation error $\text{APerror}_{r} = \left( \widehat{r/\mathcal{C}_{\alpha}} - (r/\mathcal{C}_{\alpha})_{\text{th}} \right) \, / \,(r/\mathcal{C}_{\alpha})_{\text{th}}$ with $(\mathcal{C}_{\alpha})_{\text{th}} = \alpha(\bar{z})$ (horizontal axis) and $(\mathcal{C}_{\alpha})_{\text{th}} = \bar{\alpha}$ (vertical axis) respectively.}
\label{fig:AP_vs_IntAP_Physical}
\end{subfigure}
\medskip
\begin{subfigure}[b]{.6\textwidth}
\includegraphics[width=\textwidth]{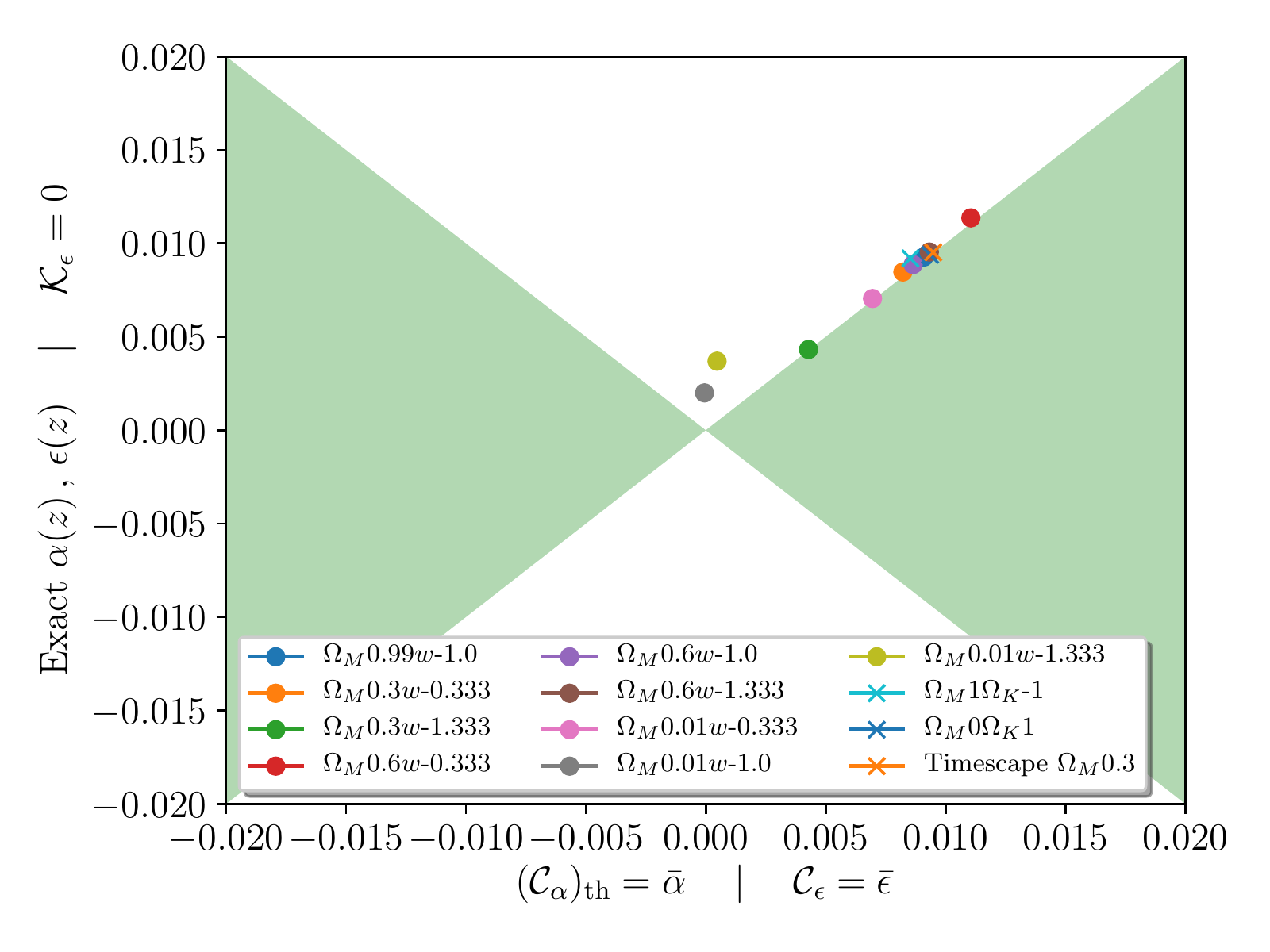}
\caption{The constant AP scaling error $\text{APerror}_{r} = \left( \widehat{r/\mathcal{C}_{\alpha}} - (r/\mathcal{C}_{\alpha})_{\text{th}} \right) \, / \,(r/\mathcal{C}_{\alpha})_{\text{th}}$ with $(\mathcal{C}_{\alpha})_{\text{th}} = \bar{\alpha}$ (horizontal axis) and the error in the \say{exact} AP scaling $\text{APerror}_{r} = \left( \hat{r} - (r)_{\text{th}} \right) \, / \,(r)_{\text{th}}$ (vertical axis).}
\label{fig:IntAP_vs_FullInt_Physical}
\end{subfigure}
\caption{The accuracy of the inferred isotropic peak position for the constant AP scaling approximations and for the exact AP scaling. For points within the green shaded region, the AP model on the vertical axis is more accurate, and for points in the unshaded region, the AP model on the horizontal axis is more accurate. The warping parameters are fixed such that $\mathcal{C}_{\epsilon} = \bar{\epsilon}$ in (\ref{eq:xiconstantAPwedge}) and $\mathcal{K}_{\epsilon} = 0$ in (\ref{eq:xiapproxnoconstAPwedge}).
Flat test models are represented by a dot, whereas curved models are represented by a cross. 
}
\label{fig:isotropicPhysical}
\end{figure}

The results of the isotropic analysis are shown in figure \ref{fig:isotropicPhysical} for nine models with different values of $\Omega_M$ and $w$, and for the three curved models described above. 
The figures show the fractional error in the recovery of the BAO scale for the various models. The labels on the $x$-axes and $y$-axes indicate the model used, and under which assumptions. 
The label \say{$(\mathcal{C}_{\alpha})_{\rm th} = \alpha(\bar{z})  \mid \mathcal{C}_{\epsilon} = \bar{\epsilon}$} indicates that the constant AP approximation (\ref{eq:xiconstantAPwedge}) has been used in the wedge functions under the assumption that $\mathcal{C}_{\alpha} = \alpha(\bar{z})$, and that $\mathcal{C}_{\epsilon}$ has been fixed to the theoretically computed value of $\bar{\epsilon}$.
\say{$(\mathcal{C}_{\alpha})_{\rm th} = \bar{\alpha} \mid \mathcal{C}_{\epsilon} = \bar{\epsilon}$} represents the same situation, only here it is assumed that $\mathcal{C}_{\alpha} = \bar{\alpha}$.
The label \say{Exact $\alpha(z) ,\epsilon(z)  \mid  \mathcal{K}_{\epsilon} = 0$} indicates that no constant AP approximation has been used, resulting in the wedge fitting function (\ref{eq:xiapproxnoconstAPwedge}). $\mathcal{K}_{\epsilon} = 0$ in the isotropic analysis,\footnote{See the motivation for introducing this parameter for the anisotropic analysis in the text below eq.~(\ref{eq:xiapproxnoconstAPwedge})} and $\mathcal{K}_{\epsilon}$ is only introduced as a free parameter in the anisotropic analysis.

In figure \ref{fig:AP_vs_IntAP_Physical}, $\text{APerror}_{r}$ (\ref{eq:constantAPfiterrors}) is shown for the standard constant AP scaling approximation analysis with $(\mathcal{C}_{\alpha})_{\text{th}} = \alpha(\bar{z})$ and the modified constant AP scaling approximation with $(\mathcal{C}_{\alpha})_{\text{th}} = \bar{\alpha}$. 
We see that the errors of both constant AP scaling approximations are of order $\lsim 1$\%.
The accuracy of the predictions from $(\mathcal{C}_{\alpha})_{\text{th}} = \alpha(\bar{z})$ and $(\mathcal{C}_{\alpha})_{\text{th}} = \bar{\alpha}$ are very close as expected from the results of section \ref{boundsexamples}. However, the modified constant AP scaling approximation $(\mathcal{C}_{\alpha})_{\text{th}} = \bar{\alpha}$ is marginally but systematically more accurate. 
For most models, the errors exceed the $0.2$\% level which is the magnitude of the $1 \sigma$ error bars of the individual best fit values of the peak positions. Thus, the instability of the best fits cannot account for the errors, and we conclude that systematic errors are dominating the error budget. 

In figure \ref{fig:IntAP_vs_FullInt_Physical} the accuracy of the modified constant AP scaling approximation $(\mathcal{C}_{\alpha})_{\text{th}} = \bar{\alpha}$ is compared to that of the exact AP scaling $\alpha(z), \, \epsilon(z)$. 
The plot shows no systematic improvements in accuracy when using the exact expression (\ref{eq:xiapproxnoconstAPwedge}) as compared to imposing the constant AP approximation $(\mathcal{C}_{\alpha})_{\text{th}} = \bar{\alpha}$. 
The errors thus remain of order $\lsim 1$\%, with most models exceeding $0.2$\%.
The inaccuracies in the recovery of the isotropic scale must thus be assigned to other systematic errors than those of any AP scaling approximation. 

\begin{figure}[!htb]
\centering
\begin{subfigure}[b]{.6\textwidth}
\includegraphics[width=\textwidth]{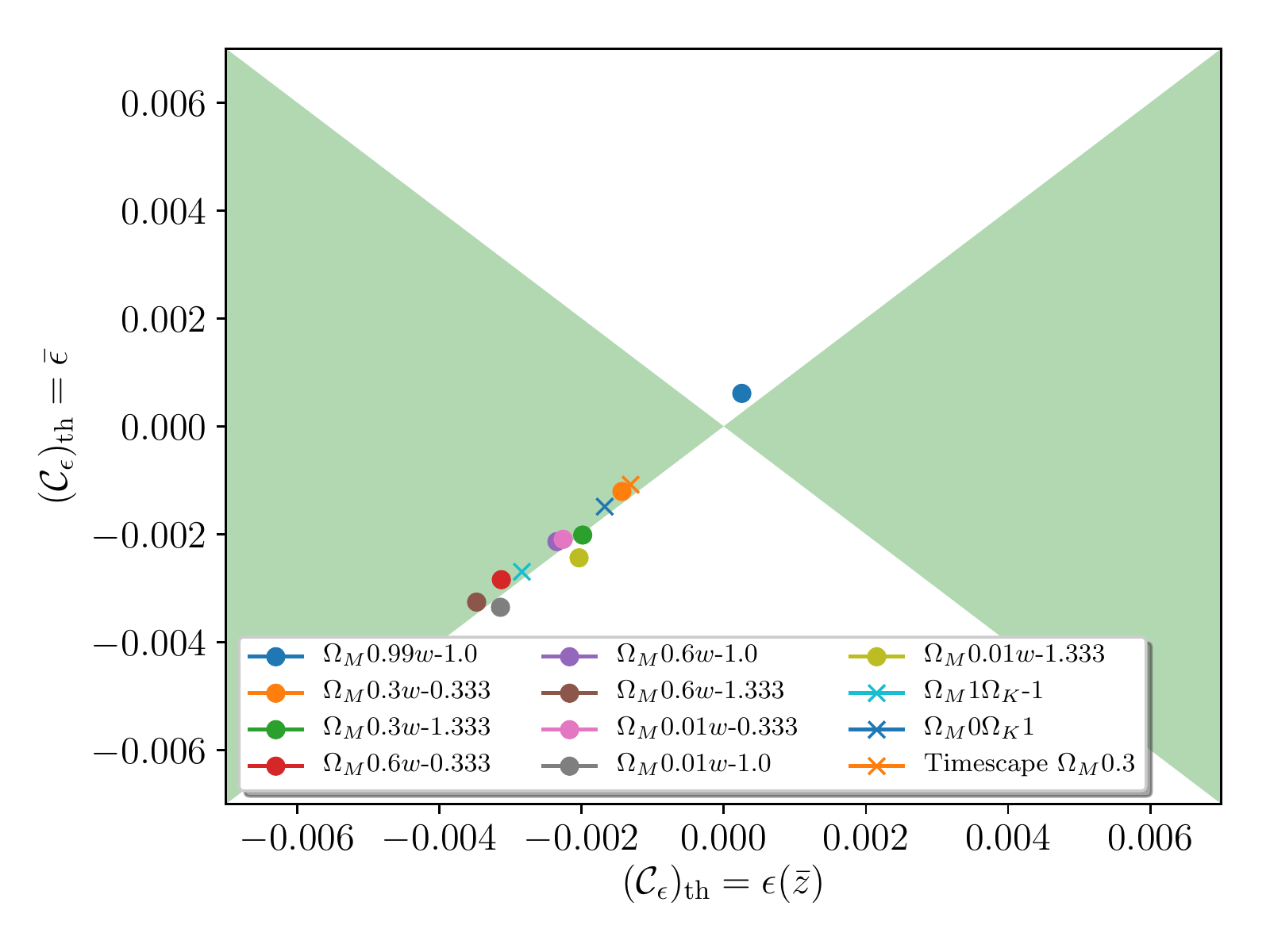}
\caption{The constant AP scaling approximation error $\text{APerror}_{\epsilon} = \hat{\mathcal{C}}_{\epsilon} - ( \mathcal{C}_{\epsilon} )_{\text{th}}$ for the constant AP scaling approximation analysis with $(\mathcal{C}_{\epsilon})_{\text{th}} = \epsilon(\bar{z})$ (horizontal axis) and with $(\mathcal{C}_{\epsilon})_{\text{th}} = \bar{\epsilon}$ (vertical axis) respectively.}
\label{fig:AP_vs_IntAP_Epsilon_Physical}
\end{subfigure}
\medskip
\begin{subfigure}[b]{.6\textwidth}
\includegraphics[width=\textwidth]{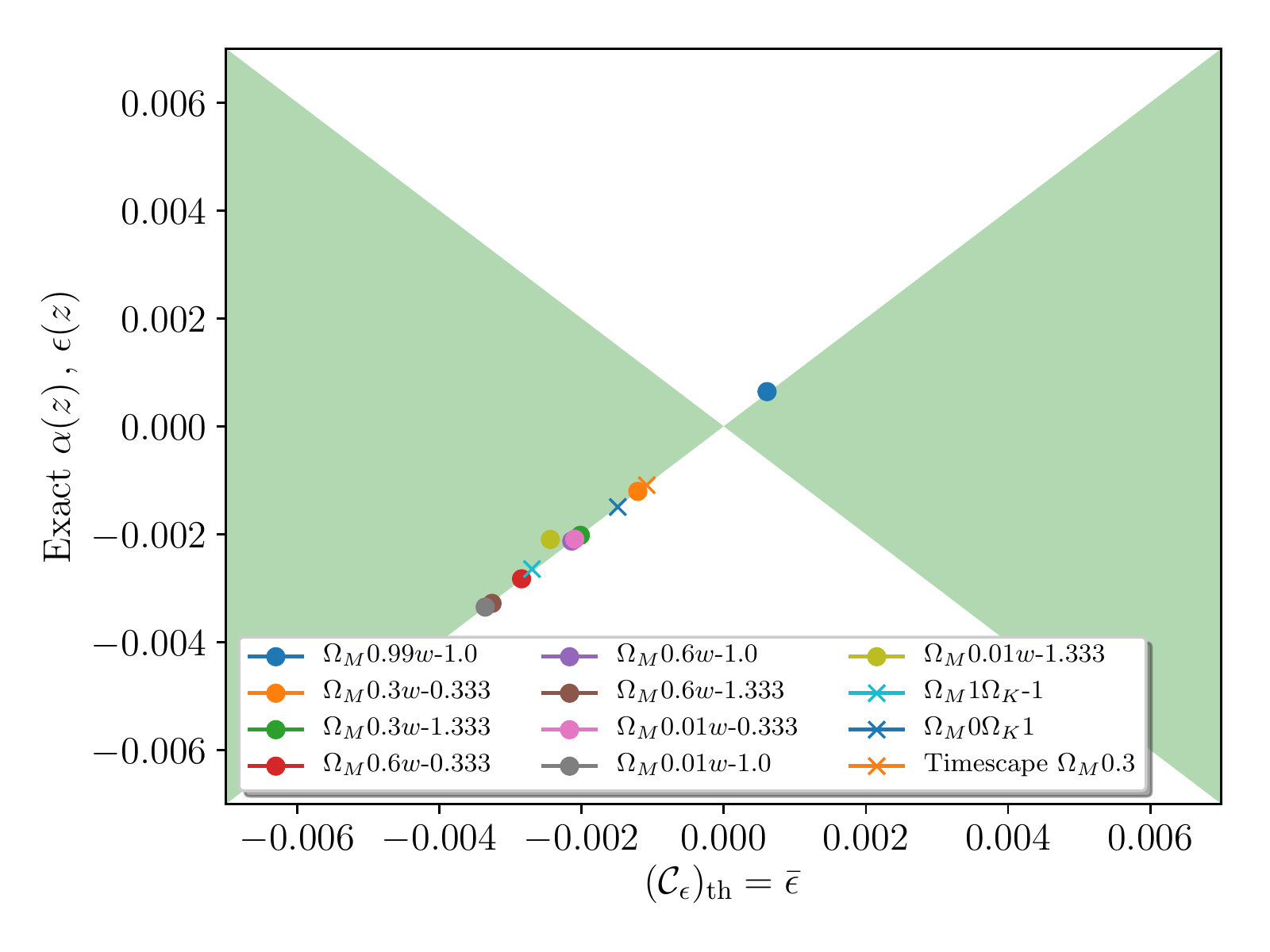}
\caption{The constant AP scaling approximation error $\text{APerror}_{\epsilon} = \hat{\mathcal{C}}_{\epsilon} - ( \mathcal{C}_{\epsilon} )_{\text{th}}$ with $(\mathcal{C}_{\epsilon})_{\text{th}} = \bar{\epsilon}$ (horizontal axis) and the error in the \say{exact} AP scaling $\text{APerror}_{\epsilon} = \hat{\mathcal{K}}_{\epsilon}$ (vertical axis).}
\label{fig:IntAP_vs_FullInt_Epsilon_Physical}
\end{subfigure}
\caption{The accuracy of the inferred warping parameters of the constant AP scaling approximations and for the exact AP scaling. For points within the green shaded region, the AP model on the vertical axis is more accurate, and for points in the unshaded region, the AP model on the horizontal axis is more accurate.
Flat test models are represented by a dot, whereas curved models are represented by a cross.}
\label{fig:anisotropicPhysical}
\end{figure}

One possible source of systematic error worth investigating is the decrease in the number of bins of the fit, since we are counting galaxy pairs in bins of a constant size of $5$ Mpc/$h$. 
Keeping the bin size constant in the reference cosmology, and scaling the bin-sizes accordingly by $1/\bar{\alpha}$ for the test models did, however, not improve the accuracy. 
Other possible sources of systematics might for instance include degeneracies of parameters in the fit or inaccuracies in the approximate integral relation (\ref{eq:xiapprox}) used in (\ref{eq:xiapproxexample}). 

In order to examine whether the systematic error in the determination of the isotropic scale is an artefact of our fitting procedure, we redid the analysis for the $\Lambda$CDM power-spectrum template fitting procedure for a few $\Lambda$CDM models with significantly differing $\Omega_M$ values. 
We used the conventionally employed 5-parameter $\Lambda$CDM template for the isotropic wedge-function, see e.g., \cite{wedgefit}. In each fit we kept the parameters $\Omega_M h^2$ and $\Omega_b h^2$ constant, in order to keep the template function fixed in each case and isolate the distortion due to the AP-scaling\footnote{{Note that the assumed value of $h$ is varying with $\Omega_M$ in this setting. However, the value of $h$ assumed does not affect the results of our analysis since distance scales are measured in units of Mpc/$h$.}}. 
We then measured the value of the constant AP-scaling parameter $\alpha(\bar{z})$ -- or interpreted via the modified constant AP scaling approximation\footnote{As noted several times in this paper the difference between $\alpha(\bar{z})$ and $\bar{\alpha}$ is negligible when the transformation is between smooth model cosmologies. We thus use $\alpha(\bar{z})$ and $\bar{\alpha}$ interchangeably for the $\Lambda$CDM models tested here.} $\bar{\alpha}$.

\begin{figure}[!htb]
\centering
\includegraphics[scale=0.6]{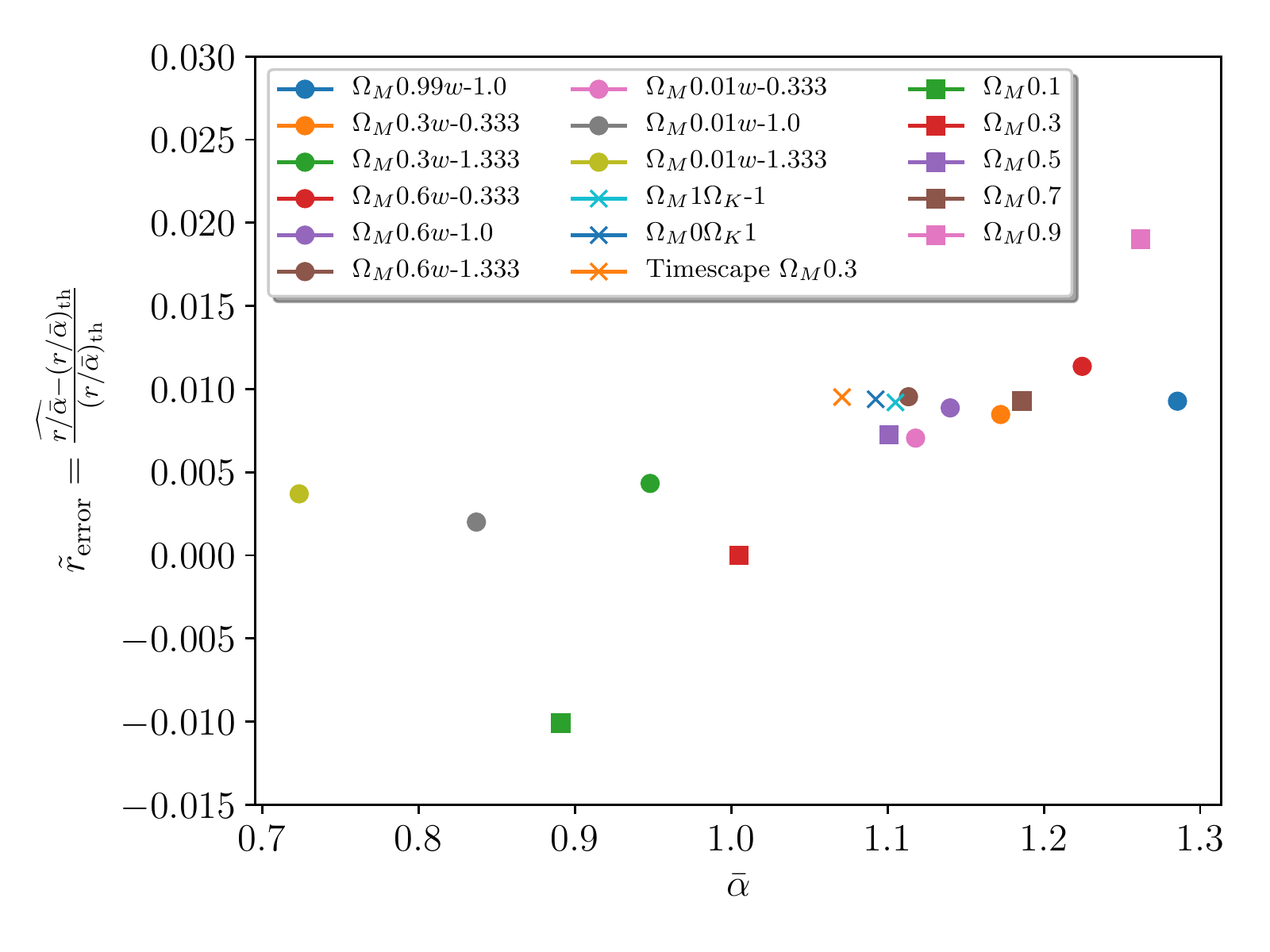}
\caption{The error of the inferred isotropic peak position as a function of the theoretical value of $\bar{\alpha}$. For the empirical fitting procedure, flat test models are represented by a dot, whereas curved models are represented by a cross. For the $\Lambda$CDM template fitting procedure, the models are represented by squares. Statistical errors of the individual best fit values are of order $0.2\%$.}
\label{fig:alpha_vs_FullInt_Physical}
\end{figure}

{In figure \ref{fig:alpha_vs_FullInt_Physical} the systematic error in the inferred best fit isotropic peak position $\tilde{r} = r/\bar{\alpha}$ -- eq.~(\ref{eq:gausparams}) with $\mu_1 = 0, \mu_2 = 1$ -- is shown for both the empirical fitting procedure and the $\Lambda$CDM template fitting procedure. 
Note that for the $\Lambda$CDM template fitting procedure the \say{true} BAO scale $r$ is considered known and fixed to a fiducial value in the  template model.}
The squares represent measurements done within the $\Lambda$CDM template fitting procedure, while the remaining measurements are the ones done within the empirical fitting procedure depicted also in figures (\ref{fig:isotropicPhysical}). 
The statistical errors of each measurement are of order $0.2\%$. 
The order of magnitude of the errors are similar between the fitting procedures, and are of order $\sim 1\%$ for models with $\abs{\bar{\alpha} - 1} \sim 0.1$. 
We note that trends in systematics as a function of $\bar{\alpha}$ are of the same sign between the two fitting procedures. 
Our results indicate that the level of systematics is robust to the exact choice of fitting procedure.
{We note that for test models with true value $\bar{\alpha} > 1$ the isotropic peak position $r/\bar{\alpha}$ is systematically overestimated equivalent to an underestimation of $\bar{\alpha}$ (when considering $r$ fixed). Conversely, for test models with $\bar{\alpha} < 1$ the isotropic peak position $r/\bar{\alpha}$ is systematically underestimated  equivalent to an overestimation of $\bar{\alpha}$. Thus the estimates are systematically shifted towards $\bar{\alpha} = 1$ and are artificially favouring the fiducial model used in the data reduction. Bearing in mind that these systematic effects might be small if the fiducial model is in fact reasonably close to the \say{true} cosmological model in terms of distance measures, they are important to take into account when considering models which are sufficiently far from the fiducial model.}

We now consider the anisotropic fits for the same models as for the isotropic analysis. 
The recovery of the isotropic peak positions in the anisotropic analysis closely resembles the results for the isotropic analysis shown in figure \ref{fig:isotropicPhysical}, and consequently we omit plots of these results. 
The recovery of the warping parameters is shown in figure \ref{fig:anisotropicPhysical}. 

The error in the constant AP scaling (\ref{eq:constantAPfiterrors}) $\text{APerror}_{\epsilon} = \hat{\mathcal{C}}_{\epsilon} - ( \mathcal{C}_{\epsilon} )_{\text{th}}$ is shown in figure \ref{fig:AP_vs_IntAP_Epsilon_Physical} for the standard constant AP scaling approximation analysis with $(\mathcal{C}_{\epsilon})_{\text{th}} = \epsilon(\bar{z})$ and the modified constant AP scaling approximation with $(\mathcal{C}_{\epsilon})_{\text{th}} = \bar{\epsilon}$. 
The errors for both constant AP scaling approximations are of order $\lsim 0.002$. The statistical $1\, \sigma$ error bars on the best fit values of the warping parameters are of order $0.002$. We conclude that the level of inaccuracy in the recovery of the warping parameters is consistent with the level of statistical noise expected for $\sim 200$ mock catalogues.
The accuracy of the two constant AP approximations is very close for each model cosmology as expected from the investigations in section \ref{boundsexamples}. 
There is no systematic improvement of the accuracy to be seen for the modified constant AP scaling approximation $(\mathcal{C}_{\epsilon})_{\text{th}} = \bar{\epsilon}$ as compared to the standard constant AP scaling approximation $(\mathcal{C}_{\epsilon})_{\text{th}} = \epsilon(\bar{z})$. 

In figure \ref{fig:IntAP_vs_FullInt_Epsilon_Physical} the accuracy of the modified constant AP scaling approximation $(\mathcal{C}_{\epsilon})_{\text{th}} = \bar{\epsilon}$ is compared to that of the exact AP scaling $\alpha(z), \, \epsilon(z)$. 
For each cosmological model, the recovery of the anisotropic warping parameter is almost identical for the constant AP scaling approximation and the exact case. 
In conclusion, the constant AP approximations tested work extremely well for the tested cosmological models for recovering the anisotropic warping parameter. 
The more accurate recovery of the warping parameter as compared to the isotropic peak position, suggests that the systematics governing the peak shift are similar between the wedges.

\subsection{Toy models with large metric gradients}
\label{Toyexamples}
Let us consider a class of unphysical model cosmologies, for which we can expect break-down of the standard AP scaling approximation with respect to the reference \LCDM\ model with $\Omega_M=0.29$.
 
As shown in general in section \ref{bounds} and detailed for a selection of model cosmologies in section \ref{boundsexamples}, the standard constant AP approximation $\alpha(z) \mapsto \alpha(\bar{z})$, $\epsilon(z) \mapsto \epsilon(\bar{z})$ is expected to be accurate between cosmologies which are of the same order of magnitude for gradient terms of the adapted metric components $\{g_{\theta \theta}, g_{zz}\}$ up to second order. 
This condition is fulfilled for essentially all cosmological metric theories designed for modelling the largest scales of our Universe. However, taking into account gradients in the geometry on smaller scales, we might arrive at physical models for which the usual constant AP scaling approximation breaks down when the fiducial cosmology used to formulate the 2-point correlation function is a large-scale metric. 
In this section we formulate some toy models which can illustrate how the usual constant AP scaling approximation might break down on account of gradients in the metric components $\{g_{\theta \theta}, g_{zz}\}$. 

In this analysis we consider the reference \LCDM\ model with $\Omega_M = 0.29$ as the \say{true} cosmological model, whereas the toy models with significant metric gradients are fiducial models used to formulate the 2-point correlation function by the observer who does not know about the true cosmological model. 
Our conclusions are expected to hold for the reversed scenario where the reference \LCDM\ model with $\Omega_M = 0.29$ plays the role of the fiducial model. 

\begin{figure}[!htb]
\centering
\begin{subfigure}[b]{.49\textwidth}
\includegraphics[width=\textwidth]{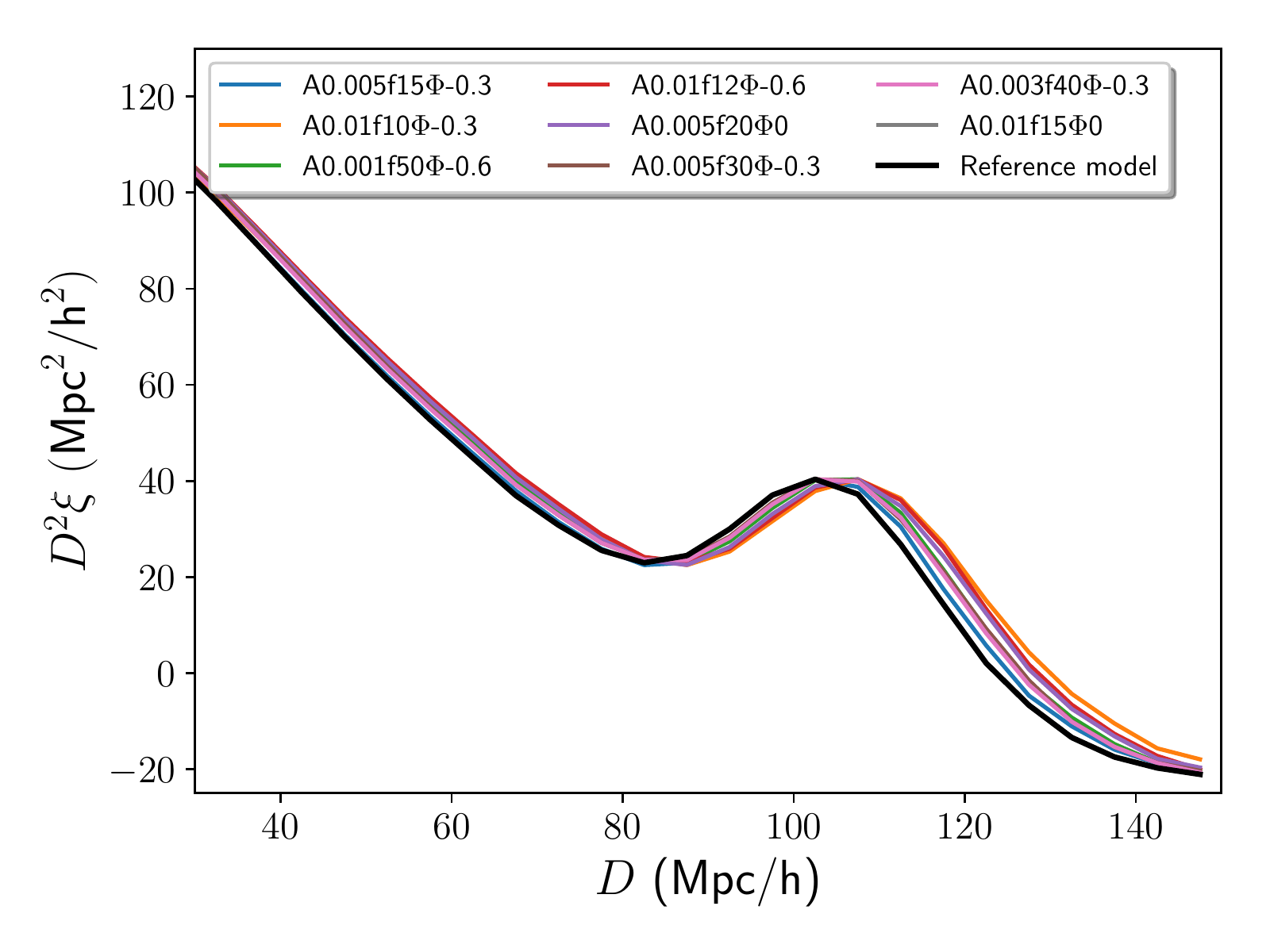}
\caption{The correlation functions as plotted over the full range.}
\label{fig:meanmocks_full}
\end{subfigure}
\medskip
\begin{subfigure}[b]{.49\textwidth}
\includegraphics[width=\textwidth]{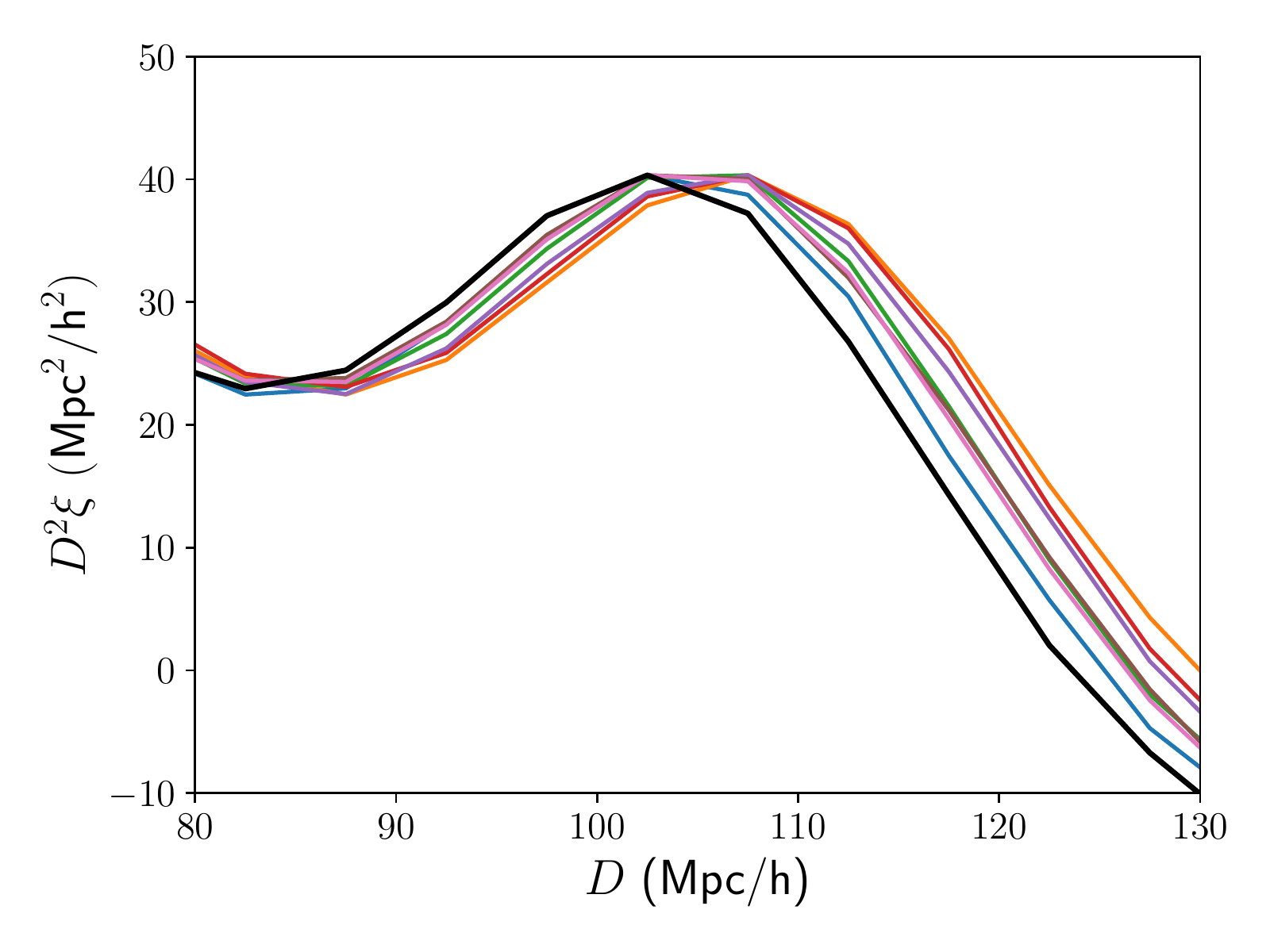}
\caption{Zoomed in version around the local maxima of the correlation functions.}
\label{fig:meanmocks_zoom}
\end{subfigure}
\caption{The mean of the isotropic 2-point correlation function of 200 mock catalogues for each tested toy model. The reference spatially-flat $\Lambda$CDM cosmological model with $\Omega_M = 0.29$ is shown in black. In order to visualise the shift of the acoustic scale, the correlation function for each model has been normalised by a constant such that the local maxima are aligned with that of the reference $\Lambda$CDM model.}
\label{fig:meanmocks}
\end{figure}

We consider the simple three parameter family of spatially-flat toy model metrics (\ref{eq:metrictoy})--(\ref{eq:Hperturbed}), which are perturbations around a spatially flat \LCDM\ model with $\Omega_M = 0.3$ with a trigonometric distortion parameterised by an amplitude $A$, frequency $f$, and phase $\Phi$.  We repeat the analysis of section \ref{FLRWexamples} for eight test models of varying $A$, $f$, and $\Phi$. 
The mean isotropic 2-point correlation function is shown in figure \ref{fig:meanmocks} for each model. 
The shifts of the BAO feature relative to the reference $\Lambda$CDM model are in general much smaller than for the models tested in section \ref{FLRWexamples}. 
This can be assigned to the fact that $\bar{\alpha}$ is close to the value 1 for all the models because of the cancellation of the relatively large fluctuation of $\alpha(z)$ by the averaging operation. Even though the mutual distance between many galaxy pairs are shifted significantly by changing from one model to the other, the overall count in each distance bin is largely robust, and the shape of the reference correlation function is largely preserved as seen in figure \ref{fig:meanmocks_full}.
The zoomed in version of the plot in figure \ref{fig:meanmocks_zoom} visualises the changes around the peak location.

\begin{figure}[!htb]
\centering
\begin{subfigure}[b]{.6\textwidth}
\includegraphics[width=\textwidth]{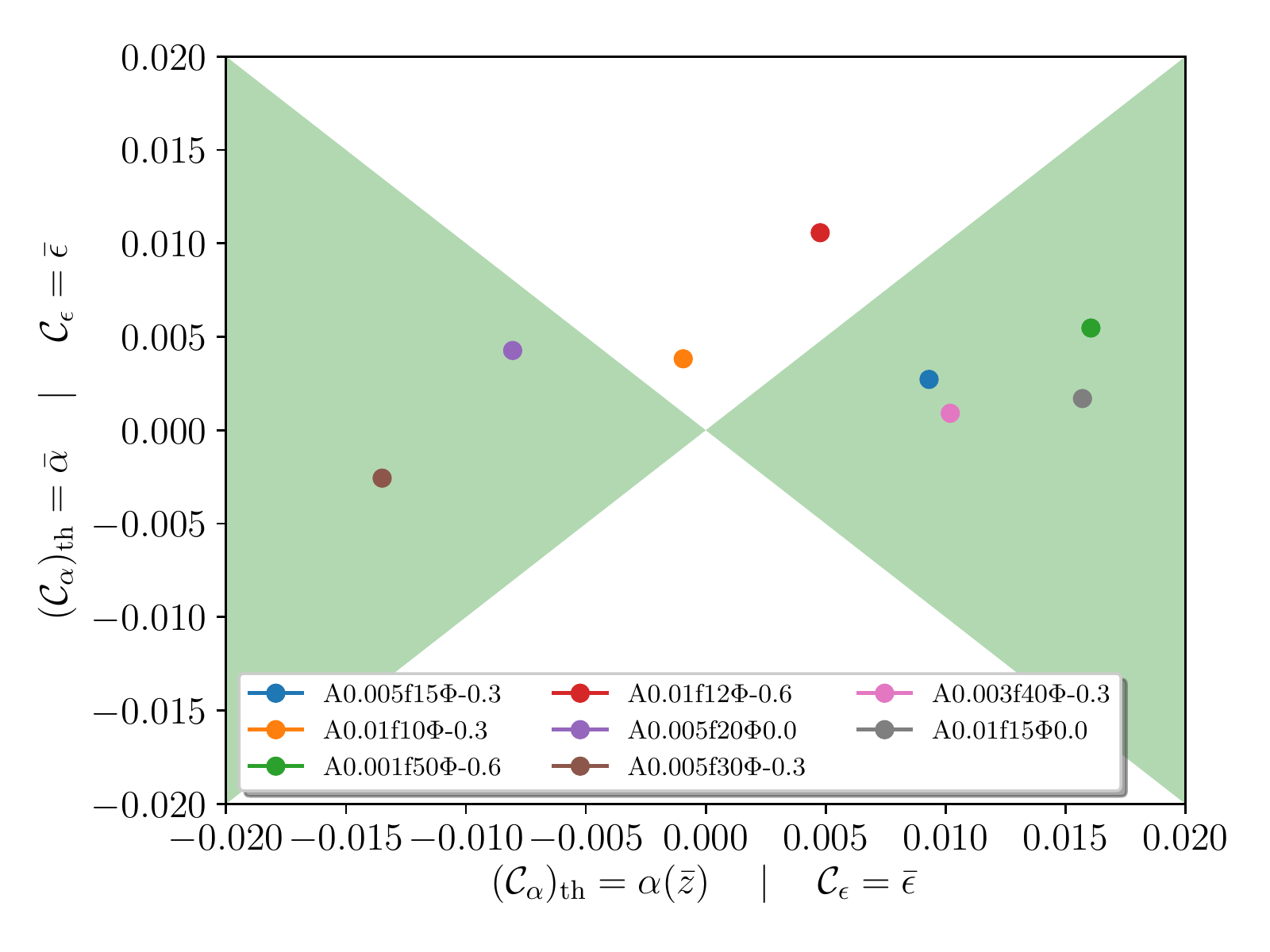}
\caption{The constant AP scaling approximation error $\text{APerror}_{r} = \left( \widehat{r/\mathcal{C}_{\alpha}} - (r/\mathcal{C}_{\alpha})_{\text{th}} \right) \, / \,(r/\mathcal{C}_{\alpha})_{\text{th}}$ with $(\mathcal{C}_{\alpha})_{\text{th}} = \alpha(\bar{z})$ (horizontal axis) and $(\mathcal{C}_{\alpha})_{\text{th}} = \bar{\alpha}$ (vertical axis) respectively.}
\label{fig:AP_vs_IntAP}
\end{subfigure}
\medskip
\begin{subfigure}[b]{.6\textwidth}
\includegraphics[width=\textwidth]{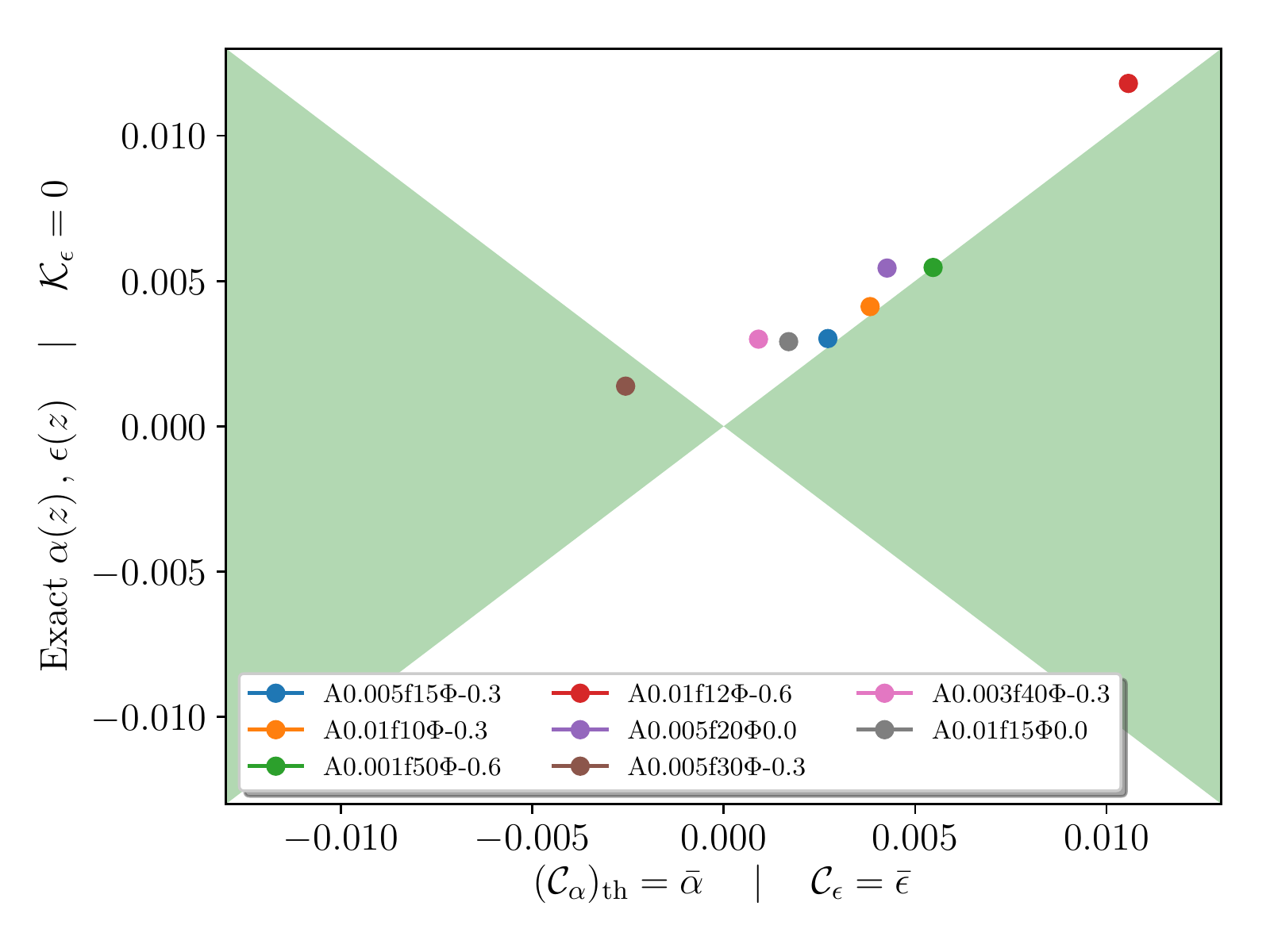}
\caption{The constant AP scaling error $\text{APerror}_{r} = \left( \widehat{r/\mathcal{C}_{\alpha}} - (r/\mathcal{C}_{\alpha})_{\text{th}} \right) \, / \,(r/\mathcal{C}_{\alpha})_{\text{th}}$ with $(\mathcal{C}_{\alpha})_{\text{th}} = \bar{\alpha}$ (horizontal axis) and the error in the \say{exact} AP scaling $\text{APerror}_{r} = \left( \hat{r} - (r)_{\text{th}} \right) \, / \,(r)_{\text{th}}$ (vertical axis).}
\label{fig:IntAP_vs_FullInt}
\end{subfigure}
\caption{The accuracy of the inferred isotropic peak position for the constant AP scaling approximations and for the exact AP scaling. For points within the green shaded region, the AP model on the vertical axis is more accurate, and for points in the unshaded region, the AP model on the horizontal axis is more accurate. The warping parameters are fixed such that $\mathcal{C}_{\epsilon} = \bar{\epsilon}$ in (\ref{eq:xiconstantAPwedge}) and $\mathcal{K}_{\epsilon} = 0$ in (\ref{eq:xiapproxnoconstAPwedge}).}
\label{fig:isotropic}
\end{figure}

The results of the isotropic analysis are shown in figure \ref{fig:isotropic}. 
In figure \ref{fig:AP_vs_IntAP} the error in the recovery of the isotropic peak position $\text{APerror}_{r}$ (\ref{eq:constantAPfiterrors}) is shown for the standard constant AP scaling approximation analysis with $(\mathcal{C}_{\alpha})_{\text{th}} = \alpha(\bar{z})$ and the modified constant AP scaling approximation with $(\mathcal{C}_{\alpha})_{\text{th}} = \bar{\alpha}$. 
The precision obtained with the modified constant AP scaling approximation is generally higher, with typical errors of order $\sim 0.5$ times those of the standard constant AP scaling approximation.
The errors associated with the modified constant AP scaling approximation are $\lsim 1$\% and comparable to those of the spatially-flat FLRW models investigated in section \ref{FLRWexamples}. 
As in the case of the spatially-flat FLRW models, the statistical errors are not sufficient to account for the errors, and we conclude that systematic uncertainties are dominating the error budget. 

In figure \ref{fig:IntAP_vs_FullInt_Physical} the accuracy of the modified constant AP scaling approximation $(\mathcal{C}_{\alpha})_{\text{th}} = \bar{\alpha}$ is compared to the accuracy of the exact AP scaling $\alpha(z), \, \epsilon(z)$. 
As for the spatially-flat FLRW models investigated in section \ref{FLRWexamples}, the plot shows no systematic improvements in accuracy when using the \say{exact} wedge function (\ref{eq:xiapproxnoconstAPwedge}) as compared to imposing the constant AP approximation $(\mathcal{C}_{\alpha})_{\text{th}} = \bar{\alpha}$. 

\begin{figure}[!htb]
\centering
\begin{subfigure}[b]{.6\textwidth}
\includegraphics[width=\textwidth]{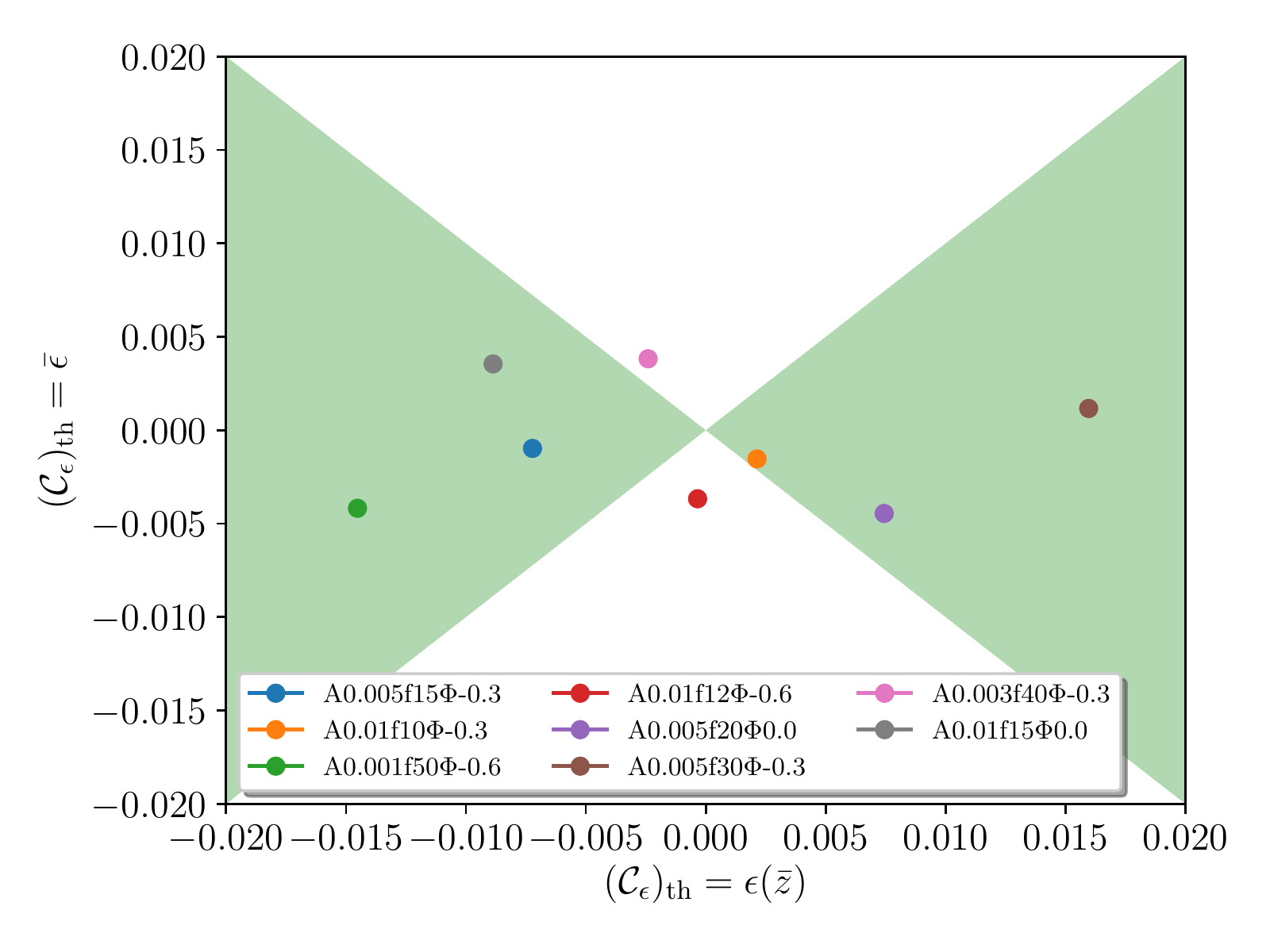}
\caption{The constant AP scaling approximation error $\text{APerror}_{\epsilon} = \hat{\mathcal{C}}_{\epsilon} - ( \mathcal{C}_{\epsilon} )_{\text{th}}$ for the constant AP scaling approximation analysis with $(\mathcal{C}_{\epsilon})_{\text{th}} = \epsilon(\bar{z})$ (horizontal axis) and with $(\mathcal{C}_{\epsilon})_{\text{th}} = \bar{\epsilon}$ (vertical axis) respectively.}
\label{fig:AP_vs_IntAP_Epsilon}
\end{subfigure}
\medskip
\begin{subfigure}[b]{.6\textwidth}
\includegraphics[width=\textwidth]{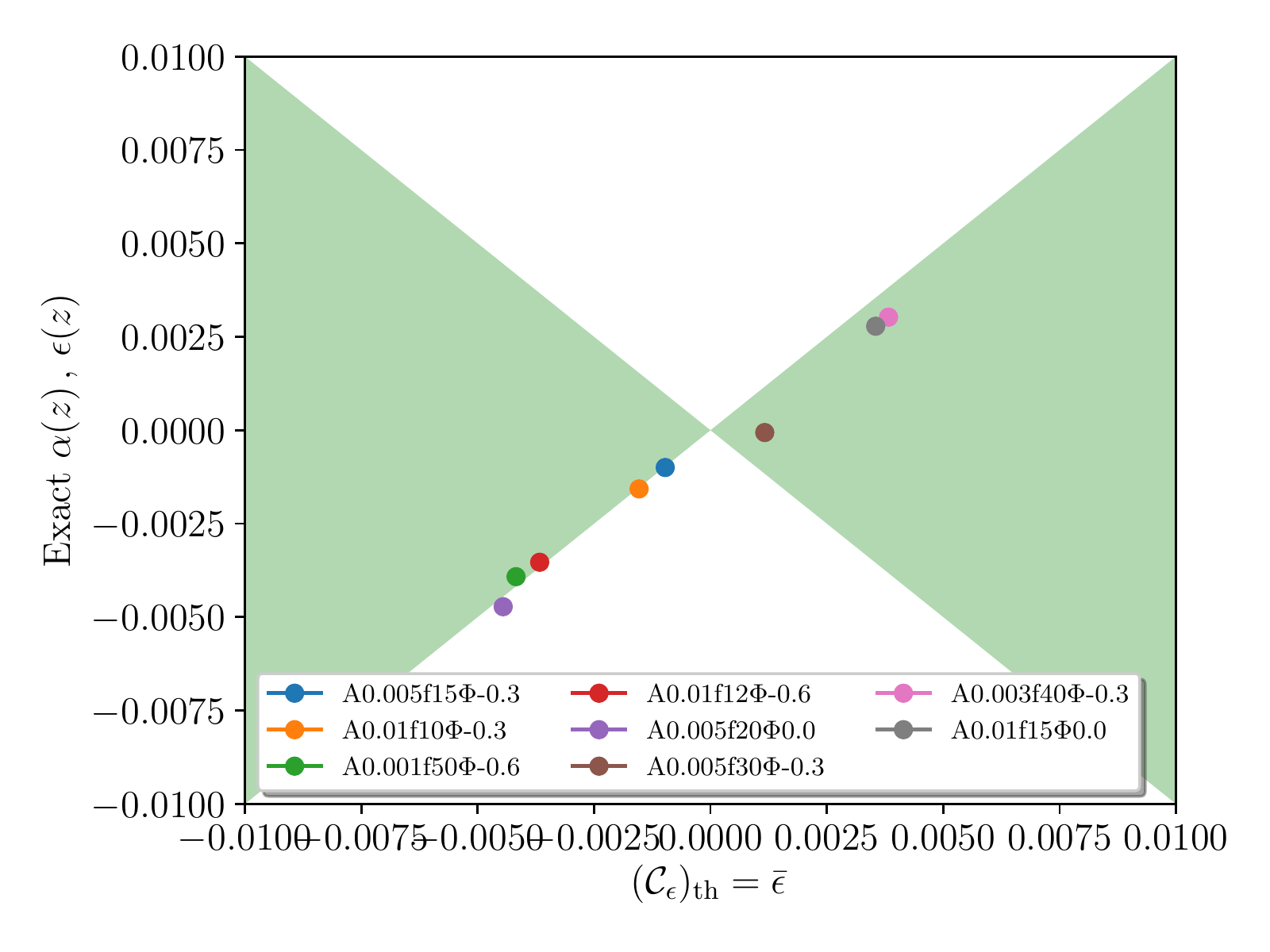}
\caption{The constant AP scaling approximation error $\text{APerror}_{\epsilon} = \hat{\mathcal{C}}_{\epsilon} - ( \mathcal{C}_{\epsilon} )_{\text{th}}$ with $(\mathcal{C}_{\epsilon})_{\text{th}} = \bar{\epsilon}$ (horizontal axis) and the error in the \say{exact} AP scaling $\text{APerror}_{\epsilon} = \hat{\mathcal{K}}_{\epsilon}$ (vertical axis).}
\label{fig:IntAP_vs_FullInt_Epsilon}
\end{subfigure}
\caption{The accuracy of the inferred warping parameters of the constant AP scaling approximations and for the exact AP scaling. For points within the green shaded region, the AP model on the vertical axis is more accurate, and for points in the unshaded region, the AP model on the horizontal axis is more accurate.}
\label{fig:anisotropic}
\end{figure}

The recovery of the warping parameters of the anisotropic wedge analysis is shown in figure \ref{fig:anisotropic}. 
The error term $\text{APerror}_{\epsilon} = \hat{\mathcal{C}}_{\epsilon} - ( \mathcal{C}_{\epsilon} )_{\text{th}}$ from (\ref{eq:constantAPfiterrors}) is shown in figure \ref{fig:AP_vs_IntAP_Epsilon} for the standard constant AP scaling approximation analysis $(\mathcal{C}_{\epsilon})_{\text{th}} = \epsilon(\bar{z})$ and the modified constant AP scaling approximation $(\mathcal{C}_{\epsilon})_{\text{th}} = \bar{\epsilon}$. 
The precision of the modified constant AP scaling approximation is in general higher than that of the standard constant AP scaling approximation. 
For the modified constant AP scaling approximation, errors in the inferred warping parameters are of order $\lsim 0.005$, and typical errors are roughly a factor of two higher than for the spatially-flat FLRW models investigated in section \ref{FLRWexamples}. Typical errors are slightly higher than $\sim 0.002$ for which we would expect most points to lie within, for the errors to be consistent with statistical noise. 

In figure \ref{fig:IntAP_vs_FullInt_Epsilon} the accuracy of the modified constant AP scaling approximation $(\mathcal{C}_{\epsilon})_{\text{th}} = \bar{\epsilon}$ is compared to that of the exact AP scaling $\alpha(z), \, \epsilon(z)$. 
The recovery of the anisotropic warping parameter is almost the same between the modified constant AP scaling approximation and the exact AP scaling.

In conclusion, the modified constant AP approximation works extremely well for the toy models considered here -- as well as for the models tested in section \ref{FLRWexamples}, where both constant AP scaling approximations were accurate -- and approximate the \say{exact} case, where no approximations are made for $\alpha(z), \epsilon(z)$, extremely well. 
However, additional systematic errors contribute to the error budget in the empirical fitting procedure, as discussed in context of the spatially-flat FLRW models in section \ref{FLRWexamples}.

\section{Discussion}
\label{discussion}
Since the mid 2000's when the first detections of the BAO scale were made \cite{EisensteinDetection,Cole}, galaxy surveys have increased in terms of sample size and volume coverage, which has led to an increased significance of the measured BAO peak in the concordance \LCDM\ cosmology \cite{wedgefit,Standardresults} and precisely mapped out the distance-redshift relation.
The increase in data has also facilitated (semi-)model-independent analysis such as \cite{sancheztransverse,HBLW}. Such analysis allows for determining characteristic scales in the 2-point correlation function without assuming a \LCDM\ fiducial model.   It is naturally of interest to what extent the measurements performed under the assumptions of conventional \LCDM\ BAO analysis can be expected to be accurate for a Universe which might be far from the \LCDM\ model in some respects. 

In this analysis we have investigated the accuracy of the standard constant AP scaling method, and a theoretically motivated modification of this, for applying BAO distance measurements in different cosmologies. 
We have quantified the difference between the two methods in section \ref{bounds}.
The two methods agree well when the \say{true} underlying cosmological model and the fiducial model have the same order of magnitude metric gradients. 
However, when large differences in metric gradients emerge -- which can happen in the scenario where a smooth fiducial model metric is used to extract information about the galaxy catalogue of a lumpy universe -- the methods can differ substantially.

In our mock-based tests in section \ref{testsmocks} we investigated the BAO peak shifts between different fiducial cosmologies. 
We avoid calibration issues in the extraction of the BAO feature by considering the shift of the BAO feature relative to {the reference model of the mock catalogues}.
We find that the standard constant AP method works well for recovering the BAO scale when the fiducial model and the reference (or \say{true}) cosmological model are close -- up to systematics which cannot be ascribed to the constant AP approximation.
As expected from the theoretical results of section \ref{bounds}, the modified constant AP method gives very similar results to those of the standard constant AP method when the fiducial and reference models are not differing substantially in terms of metric gradients.
When we introduce large differences in gradients between the fiducial model and the reference cosmological model, the standard constant AP scaling method becomes inaccurate, while our modified constant AP scaling method remains accurate. 
This is due to the fact the modified constant AP scaling takes into account the volume statistical aspect of the BAO feature, whereas the standard constant AP scaling method is based on evaluation at a single redshift. 

Our results can help understand the \say{effective distance scales} that we infer in BAO analysis. 
The conventionally \say{measured} AP parameters are better understood as averages $\overline{\alpha(z)}$ and $\overline{\epsilon(z)}$ over the survey. 
Thus, they do not represent the \emph{ratio of the mean} of the \say{true} model and the fiducial model distance scales evaluated a particular redshift -- but are more accurately thought of as the \emph{mean of the ratio} of the \say{true} model and the fiducial model distance scales, which vary over the galaxy survey. This difference in interpretation might be important, depending on the exact lumpy geometry that describes our universe. 

We find additional systematics in the recovery of the BAO scale of order $\sim 1\%$ for $\abs{\bar{\alpha} - 1} \sim 0.1$ in section \ref{testsmocks} which cannot be assigned to the constant AP approximation.
These systematics persists when we use a $\Lambda$CDM template fitting procedure instead of our empirical method, which indicate that the level of systematics is robust to the exact choice of fitting procedure. 
The systematics are in general larger than what is found in other examinations of systematic errors due to choice of fiducial cosmology, see e.g., \cite{BOSSsystematics,Carter}, which we hypothesise is due to the fact that such analysis are concerned with $\Lambda$CDM models which are close -- typically within a few percent in terms of cosmological parameters.
Our analysis reveal that larger systematic errors emerge when the fiducial and \say{true} cosmological models are not close in all respects.\footnote{We note that the systematics might be even larger if we omit the precaution of scaling the fitting range by a factor $1/\bar{\alpha}$ in order to approximately fit the tested models over the same physical distance range as the reference model. This is of course only possible to do since we know the \say{true} underlying model with respect to which we define $\alpha$, but is not possible when fitting to actual data where the \say{true} underlying model is unknown.}

Our analysis based on test cases indicate that the error budget in the standard literature is significantly underestimated when interpreting the herein measured distance scales as \say{model-independent} and using the results for constraining alternative models {which are not close to the fiducial model}.
{The additional systematics must either be included in the error budget, or alternatively it must be stressed that the results are not to be extrapolated to model cosmologies which differ more than a few percent from the fiducial model in terms of distance measures.
It is also worth noting that the fiducial model is typically chosen to be close to a concordance model, which is in practice constrained from the CMB and other cosmological probes -- and that caution must thus be taken about such implicit application of priors in BAO data reduction.}  

Our conclusions are twofold. On one hand, the standard constant AP scaling approximation works surprisingly well for a broad class of pairs of \say{true} and fiducial models. 
The fiducial model can be far from the \say{true} model in terms of the relevant distance measures -- as long as these and their gradients are bounded to be of similar order of magnitude to those of the \say{true} cosmological model -- while the constant AP approximation remains accurate for the purposes of BAO analysis. 
By reinterpreting the constant AP scaling parameters one can modify the standard constant AP scaling approximation to be accurate for an even larger class of pairs of models.
On the other hand, there are systematic uncertainties which are not directly related to the constant AP approximation. 
These systematic uncertainties of order $\sim 1\%$ for $\abs{\bar{\alpha} - 1} \sim 0.1$ -- which are independent of the fitting procedure chosen -- are comparable in size to the statistical errors often reported in BAO analysis. 

A limitation of our analysis is that it applies to spherically-symmetric template geometries only. Even though large-scale average template metrics are usually taken as spherically symmetric, the symmetry is broken at scales below that of statistical isotropy. 
{In fact, it is for models which apply general relativistic approximations on nested scales -- as, e.g., in the silent universe class of model space-times \cite{Bolejko} -- that we expect metric gradients to become significantly large to introduce significant errors to the standard constant AP scaling approximation as discussed in sections \ref{bounds} and \ref{boundsexamples}.}
Systematic effects of the anisotropy from smaller scales, which do not cancel on average in all respects and might feed into the large-scale estimators of the two-point correlation function, might be important for realistic lumpy space-times.  
One might attempt to generalise our methods to more generic geometries.
A challenge of this is that the AP scaling is designed for spherical symmetry. For generic spatial 3-metrics, one would need six generalised AP functions instead of two in order to account for the degrees of freedom involved.

\acknowledgments
We wish to thank Pierre Mourier for his insightful comments and contributions for improving this paper. 
{Thank you to Yong-Zhuang Li for his careful reading of this paper and for his useful suggestions for improvements.}
We wish to thank Thomas Buchert
for hospitality at the ENS, Lyon, France. This work was supported by
Catalyst grant CSG-UOC1603 administered by the Royal Society of New Zealand.
AH is grateful for the support given by the funds: `Knud H{\o}jgaards Fond', `Torben og Alice Frimodts Fond', and `Max N{\o}rgaard og Hustru Magda N{\o}rgaards Fond'.

\begin{appendices}

\section{The 2-point correlation function}
\label{2PCFestimatorsmain}
The 2-point correlation function in cosmology \cite{PeeblesTheory} describes the clustering of matter as a function of scale to lowest order. 
Here we give a review of the 2-point correlation function, and define a useful \say{reduced} form of the correlation function relevant for the present analysis.

The definition of the 2-point correlation function relies on considering ensemble averages of model universes as generated from a random process specified within the given cosmological model. 
Let us consider a fixed spatial domain $\domain$. We consider the
position of the galaxies within this domain random variables, and keep the total number of galaxies $N$ within the domain $\domain$ constant over the ensembles. We
use adapted coordinates $X^{i}$ on the spatial domain,
and denote the random position of the $a$'th particle $x_{a}^{i}$.
We define the ensemble averaged pair count density $f(X,Y)$ of galaxies as the ensemble average pair count per unit volume squared:     
\begin{align} \label{eq:excessprob}
&f(X,Y) dV_X \,dV_Y \equiv \braket{N(dV_{X}) N(dV_{Y})} ,
\end{align}
where $dV_X$ and $dV_Y$ are infinitesimal volume elements centred on coordinates $X$ and $Y$, and the indices on $X$ and $Y$ have been suppressed. (These volume elements need not be \say{physical} volume elements but might be conveniently defined as coordinate volumes, absorbing any volume measure into $f(X,Y)$.) The brackets $\braket{}$ denote the average over realisations of the ensemble, and
\begin{align} \label{eq:NEnsemble}
N(dV_{X}) \equiv \sum_{a}^{N} \mathbb{1}_{dV_{X}}(x^i_a) \, , \qquad   \mathbb{1}_{dV_{X}}(x^i_a)=\begin{cases}
1,&x^i_a\in dV_X,\\ 0,&x^i_a\notin dV_X, \end{cases}
\end{align}
is the number count in the volume element $dV_{X}$ in a given realisation, where $\mathbb{1}_{dV_{X}}$ is the indicator function of the volume $dV_{X}$. 
The ensemble averaged galaxy density function $f(X)$ can be expressed as an integral over (\ref{eq:excessprob})
\begin{align} \label{eq:probsinglegalaxy}
f(X)dV_X &\equiv \braket{N(dV_{X})} = \frac{1}{N} \left( \int f(X,Y) dV_Y \right) \, dV_X \, , 
\end{align}
where integration without limits indicate integration over the entire domain $\domain$, and where the normalisation $N = N(\domain)$ is the ensemble-fixed total number of galaxies in the domain $\domain$. 

The spatial 2-point correlation function is defined as
\begin{align} \label{eq:xidef1}
\xi(X,Y) \equiv \frac{ f(X,Y)}{f_{\mathrm{Poisson}}(X,Y)} - 1  = \frac{ f(X,Y)}{f(X)f(Y)} - 1 \, , 
\end{align}
and describes the excess ensemble number count over the ensemble number count in an artificial uncorrelated ensemble with factorising pair count density $f_{\mathrm{Poisson}}(X,Y) = f(X)f(Y)$. 

We can define a number of \say{reduced} 2-point correlation functions by integrating over $f(X,Y)dV_X \, dV_Y$ and $f_{\mathrm{Poisson}}(X,Y)dV_X \, dV_Y$ subject to a given constraint.\footnote{It is conventional to assume that the galaxy distribution is described by a homogeneous and isotropic point process, in which case the 2-point correlation function (\ref{eq:xidef1}) automatically reduces to a function of the geodesic distance $D$ between the galaxy pairs, where $D$ is defined within the \say{true} cosmological model. However, here we are relaxing the conventional assumptions of homogeneity and isotropy to potentially allow for asymmetric random processes describing the galaxy distribution, and to account for systematic observational effects, such as the redshift depth of the survey or survey-coverage. (The survey might be limited in parts of the sky as compared to others.) Even for the homogeneous and isotropic point process, this new formulation is relevant when the \say{wrong} fiducial cosmology is used for constructing the 2-point correlation function. }

Typically we might consider the geodesic distance to be fixed in the integration. For this purpose, it is useful to perform the change of variables 
$(X^{i},Y^{i}) \mapsto (X^{i}, \hat{n}_X^{i}, D)$, where $\hat{n}_X^{i}$ is a unit vector defined at $X^{i}$ representing a geodesic starting at $X^{i}$ and intersecting $Y^{i}$ and $D$ is the geodesic distance\footnote{The transformation $(X^{i},Y^{i}) \mapsto (X^{i}, \hat{n}_X^{i}, D)$ is bijective if there is a unique geodesic between the points $X^{i},Y^{i}$, or if some requirement is imposed to single out a unique geodesic. We assume bijectivity in the following.} from $X^{i}$ to $Y^{i}$.
Using the Jacobian of the transformation, we can then formulate the pair count density as a function of the new variables $f(X^{i}, \hat{n}_X^{i}, D)$. 

For the type of spherically-symmetric models specified in section \ref{models} we can decompose $\hat{n}_X$ into $\mu$, $\sgn(\delta z)$, and the normalised angular separation vector
$\hat{\Theta} = \frac{1}{ \left| \delta \Theta \right| }(\delta
\theta, \cos(\theta) \delta \phi)$. 
Furthermore for this class of models we can use the observer adapted functions $(z, \theta, \phi)$ as convenient coordinates on a spatial domain $\domain$ on a hypersurface defined by $t = \,$const., and take $X = (z, \theta, \phi)$.
We can then rewrite (\ref{eq:excessprob}) in terms of the new set of variables
\begin{align} \label{eq:excessprobtransmu}
&f(X,Y) dV_X \,dV_Y = f [z,\theta,\phi, \mu, \sgn(\delta z) , \hat{\Theta}, D] dz \, d\theta \, d\phi \,d\mu \,d \hat{\Theta} \, dD \, . 
\end{align}
Let us now define the \say{reduced} pair count density function in $(D,\mu,z)$ by integrating (\ref{eq:excessprobtransmu}) over the remaining variables $(\theta,\phi,\hat{\Theta}, \sgn(\delta z))$ 
\begin{align} \label{eq:freduced}
&f(D,\mu,z) \equiv \sum_{\sgn(\delta z)=\pm 1}  \int f(z,\theta,\phi, \mu, \sgn(\delta z) , \hat{\Theta}, D)  d\theta \, d\phi  \,d \hat{\Theta}  \, , 
\end{align}
and analogously define $f_{\mathrm{Poisson}}(D,\mu,z)$.
From the \say{reduced} pair count density functions, we can define the \say{reduced} 2-point correlation function 
\begin{align} \label{eq:ximuDz}
\xi(D,\mu,z) \equiv \frac{ f(D,\mu,z)}{f_{\mathrm{Poisson}}(D,\mu,z)} - 1  \, . 
\end{align}
We can further reduce the pair count density functions by defining 
\begin{align} \label{eq:freduceddmu}
&f(D,\mu) \equiv  \int f(D,\mu,z) dz \, ,  \qquad   f_{\mathrm{Poisson}}(D,\mu) \equiv  \int f_{\mathrm{Poisson}}(D,\mu,z) dz  \, , 
\end{align}
from which we can define the \say{reduced} 2-point correlation function in $D$ and $\mu$ as
\begin{align} \label{eq:ximudef1}
\xi(D,\mu)= \frac{f(D,\mu)}{f_{\text{Poisson}}(D,\mu) } - 1 \, . 
\end{align}
From (\ref{eq:ximudef1}) we can construct the following \say{wedge} 2-point correlation function
\begin{align} \label{eq:wedge2point}
\xi_{[\mu_{1}, \mu_{2}]} (D) = \frac{1}{\mu_{2}- \mu_{1}} \int_{\mu_{1}}^{\mu_{2}}\, d \mu\, \xi(D, \mu) ,
\end{align}
where we denote 
\begin{align} \label{eq:wedge2pointsc}
\xi (D) \equiv \xi_{[0, 1]} (D) , \qquad \xi_{\perp} (D) \equiv \xi_{[0, 0.5]} (D) , \qquad \xi_{\parallel} (D) \equiv \xi_{[0.5, 1]} (D) 
\end{align}
the isotropic wedge, the transverse wedge, and the radial wedge respectively.

It will be useful in the present analysis to approximate (\ref{eq:ximudef1}) as an integral over (\ref{eq:ximuDz}). 
We do this by defining the normalised density function in $z$ as 
\begin{align} \label{eq:numbercountz}
P(z) \equiv \frac{ \int f(z,\theta,\phi) d\theta \, d\phi}{\int f(z',\theta',\phi') dz' \, d\theta' \, d\phi'} = \frac{ \int f(D,\mu,z)  d \mu \, dD  }{  \int f(D',\mu',z')  dz' \, d \mu' \, dD' }  \, , 
\end{align}
where $f(z,\theta,\phi) dz \, d\theta \, d\phi = f(X)dV_X$, and where the equality follows from (\ref{eq:probsinglegalaxy}) and (\ref{eq:freduced}). Suppose that the pair count functions $f(D,\mu,z)$ and $f_{\mathrm{Poisson}}(D,\mu,z)$ are almost of a multiplicatively separable form, such that 
\begin{align} \label{eq:fDmuzapprox}
f(D,\mu,z) = f(D,\mu) P(z) (1 + \delta(D,\mu,z) ) \, &, \qquad  \delta(D,\mu,z) \ll 1 , \\
f_{\mathrm{Poisson}}(D,\mu,z) = f_{\mathrm{Poisson}}(D,\mu) P(z) (1 + \delta(D,\mu,z) ) \, &, \qquad  \delta_{\mathrm{Poisson}}(D,\mu,z) \ll 1 . 
\end{align}
Note that by the definitions (\ref{eq:freduceddmu}) we have the constraints
\begin{align} \label{eq:deltaint}
& \int P(z) \delta (D,\mu,z) dz = \int P(z) \delta_{\mathrm{Poisson}}(D,\mu,z) dz  = 0 \quad \forall \, D, \mu \, .
\end{align}
We can now use the decomposition (\ref{eq:fDmuzapprox}) to write the following integral over (\ref{eq:ximuDz}) in redshift as 
\begin{align} \label{eq:xidef1int}
& \int \xi(D,\mu,z) P(z) dz  = \int \frac{ f(D,\mu,z)}{f_{\mathrm{Poisson}}(D,\mu,z)} P(z) dz    - 1  \nonumber \\
&\approx  \frac{ f(D,\mu)}{f_{\mathrm{Poisson}}(D,\mu)}   \int\left[ 1 + \delta - \delta_{\mathrm{Poisson}} - \delta_{\mathrm{Poisson}} (\delta - \delta_{\mathrm{Poisson}} ) \right] P(z) dz    - 1    \nonumber \\
&=  \frac{ f(D,\mu)}{f_{\mathrm{Poisson}}(D,\mu)} \int  \left[ 1 - \delta_{\mathrm{Poisson}} (\delta - \delta_{\mathrm{Poisson}} ) \right] P(z) dz    - 1   \nonumber \\
&= \xi(D,\mu) - (1 + \xi(D,\mu)) \int \delta_{\mathrm{Poisson}} (\delta - \delta_{\mathrm{Poisson}} )  P(z) dz  \, , 
\end{align}
where the second line follows from substituting (\ref{eq:fDmuzapprox}) and expanding around $\delta = 0$ to second order, the third line follows from (\ref{eq:deltaint}), and the last equality follows from (\ref{eq:ximudef1}) and the condition that the integral of $P(z)$ is normalised to 1. 
Thus, under the assumptions (\ref{eq:fDmuzapprox}), $ \int \xi(D,\mu,z) P(z) dz =  \xi(D,\mu)$ to first order in $\delta$ and $\delta_{\mathrm{Poisson}}$.

\section{Conjecture of improved AP scaling}
\label{conjecture}
In this section we conjecture that the modified constant AP scaling $\alpha(z) \mapsto \bar{\alpha}$, $\epsilon(z) \mapsto \bar{\epsilon}$ is typically better for extracting characteristic features of a 2-point correlation function as compared to the standard constant AP scaling approximation $\alpha(z) \mapsto \alpha(\bar{z})$, $\epsilon(z) \mapsto \epsilon(\bar{z})$.

Let us consider the function $\Xi \, : \, X \subset {\rm I\!R}^{5} \rightarrow Y \subset {\rm I\!R}$, such that $\Xi$ assigns a unique real number 
\begin{align} \label{eq:reparapp1}
\Xi(D,\mu,z,\alpha,\epsilon) = \xi^{\rm tr}\left(D^{\rm tr}(D,\mu,\alpha,\epsilon),\mu^{\rm tr}(\mu,\epsilon),z\right) \, ,
\end{align}
to each point $\{D,\mu,z,\alpha,\epsilon\} \in X$. $\xi^{\rm tr}$ is the 2-point correlation function as given in the \say{true} underlying cosmology, and $D^{\rm tr}(D,\mu,\alpha,\epsilon)$ and $\mu^{\rm tr}(\mu,\epsilon)$ are given in (\ref{eq:DAP}) and (\ref{eq:muAP}). 
The parameters $\alpha$ and $\epsilon$ might take any values, but we shall often be interested in identifying $\alpha$ and $\epsilon$ with the AP scaling functions $\alpha(z)$ and $\epsilon(z)$, given by the \say{true} cosmological model and the choice of fiducial model respectively. 
When identifying $\alpha$ and $\epsilon$ with the AP scaling functions $\alpha(z)$ and $\epsilon(z)$, $\Xi$ reduces to the redshift dependent 2-point correlation function $\xi$ in (\ref{eq:ximodel}) 
\begin{align} \label{eq:reparapp}
\Xi(D,\mu,z,\alpha(z),\epsilon(z)) = \xi(D,\mu,z)  = \xi^{\rm tr}\left(D^{\rm tr}(D,\mu,\alpha(z),\epsilon(z)),\mu^{\rm tr}(\mu,\epsilon(z)),z\right) \, .
\end{align}
Consider the situation where the condition of almost multiplicative separability (\ref{eq:fDmuzapprox}) is satisfied. 
Then the result (\ref{eq:xidef1int}) holds, and we might write 
\begin{align} \label{eq:approxxi}
\xi(D,\mu)  = \overline{\xi(D,\mu,z)} = \overline{ \Xi(D,\mu,z,\alpha(z),\epsilon(z)) } \, , 
\end{align}
to first order in the \say{non-separability terms} $\delta$, $\delta_{\mathrm{Poisson}}$ defined in (\ref{eq:fDmuzapprox}). The overbar denotes the averaging operation $\overline{f(z)} \equiv \int \, dz \, P(z) f(z)$ for an arbitrary function $f(z)$, and the second equality follows from the re-parametrisation (\ref{eq:reparapp}).  
Let us for simplicity suppose that separability is a good approximation, such that the second order correction terms are so small that we can for all practical purposes consider the first order approximation in \ref{eq:approxxi} exact.

We consider what we denote the standard constant AP approximation of $\xi(D,\mu)$ by performing the mapping $\{z \mapsto \bar{z}, \alpha \mapsto \alpha(\bar{z}), \epsilon \mapsto \epsilon(\bar{z}) \}$ in (\ref{eq:reparapp1}) to obtain 
\begin{align} \label{eq:approxxiS}
\mathcal{I}_{\text{standard AP}}(D,\mu)  \equiv \Xi(D,\mu,\bar{z},\alpha(\bar{z}),\epsilon(\bar{z})) = \xi(D,\mu,\bar{z})  \,  .
\end{align}
In addition we consider the analogous modified constant AP approximation $\{z \mapsto \bar{z}, \alpha \mapsto \bar{\alpha}, \epsilon \mapsto \bar{\epsilon} \}$ of $\xi(D,\mu)$
\begin{align} \label{eq:approxxiM}
\mathcal{I}_{\text{modified AP}}(D,\mu)  \equiv  \Xi(D,\mu,\bar{z},\bar{\alpha},\bar{\epsilon}) \, ,  
\end{align}
where we use the short hand notation $\bar{f} = \overline{f(z)}$. 
We want to estimate which of the functions (\ref{eq:approxxiS}) and (\ref{eq:approxxiM}) provide the better approximation of $\xi(D,\mu)$. 

We assume that $\Xi(D,\mu,z,\alpha,\epsilon)$ is three times differentiable in $z,\alpha,\epsilon$ and that $\alpha,\epsilon$ are twice differentiable in $z$. We consider the first order Taylor expansions around $\{z=\bar{z}, \alpha = \alpha(\bar{z}), \epsilon = \epsilon(\bar{z})\}$ 
\begin{align} \label{eq:G}
G(D,\mu,z) &\equiv \xi(D,\mu,\bar{z}) + \left. \frac{d  \xi(D,\mu,z) }{dz}  \right|_{\bar{z}} \, (z- \bar{z})  \nonumber  \\
&=   \xi(D,\mu,\bar{z}) +  \left. \left( \frac{\partial  \Xi }{\partial z}  + \frac{d  \alpha }{dz}  \frac{\partial  \Xi }{\partial \alpha}  +\frac{d  \epsilon }{dz} \frac{\partial  \Xi }{\partial \epsilon}    \right) \right|_{\bar{z}} \, (z- \bar{z})   \,  , 
\end{align}
and 
\begin{align} \label{eq:H}
H(D,\mu,z,\alpha,\epsilon) \equiv \xi(D,\mu,\bar{z}) + \left.  \frac{\partial  \Xi }{\partial z} \right|_{\bar{z}} \, (z- \bar{z})  +  \left.  \frac{\partial  \Xi }{\partial \alpha} \right|_{\bar{z}} \, (\alpha - \alpha(\bar{z}))   +  \left.  \frac{\partial  \Xi }{\partial \epsilon}  \right|_{\bar{z}} \, (\epsilon - \epsilon(\bar{z}))    \,  . 
\end{align} 
Let us write the error term associated with (\ref{eq:H}) as an approximation of (\ref{eq:reparapp1}) as 
\begin{align} \label{eq:HErrorterm}
&\Xi(D,\mu,z,\alpha,\epsilon) - H(D,\mu,z,\alpha,\epsilon) = {}^{(2)}\Xi(D,\mu,z,\alpha,\epsilon) + {}^{(2)}\mathcal{R}(D,\mu,z,\alpha,\epsilon) \, ,
\end{align}
where 
\begin{align} \label{eq:HError2ndorder}
&{}^{(2)}\Xi(D,\mu,z,\alpha,\epsilon) \equiv
 \frac{1}{2} \left.  \frac{\partial^2  \Xi }{\partial z^2} \right|_{\bar{z}} \, (z- \bar{z})^2  + \frac{1}{2} \left.  \frac{\partial^2  \Xi }{\partial \alpha^2} \right|_{\bar{z}} \, (\alpha - \alpha(\bar{z}))^2 + \frac{1}{2} \left.  \frac{\partial^2  \Xi }{\partial \epsilon^2} \right|_{\bar{z}} \, (\epsilon - \epsilon(\bar{z}))^2 + \nonumber  \\
&  \left.  \frac{\partial^2  \Xi }{\partial \alpha \partial z} \right|_{\bar{z}} \, (z- \bar{z})(\alpha - \alpha(\bar{z})) +  \left.  \frac{\partial^2  \Xi }{\partial \epsilon \partial z} \right|_{\bar{z}} \, (z- \bar{z})(\epsilon - \epsilon(\bar{z})) +  \left.  \frac{\partial^2  \Xi }{\partial \alpha \partial \epsilon} \right|_{\bar{z}} \, (\epsilon - \epsilon(\bar{z}))(\alpha - \alpha(\bar{z})) \, , 
\end{align}
is the second order contribution and ${}^{(2)}\mathcal{R}$ is the remainder at second order.\footnote{ 
Following Taylor's theorem one might express ${}^{(2)}\mathcal{R}$ as an integral-expression where the integrand is a linear combination of third order derivatives of $\Xi$ evaluated at $\{z=\bar{z}, \alpha = \alpha(\bar{z}), \epsilon = \epsilon(\bar{z})\}$.} 
Combining (\ref{eq:G}) and (\ref{eq:H}) we have
\begin{align} \label{eq:GminusH}
& G(D,\mu,z) - H(D,\mu,z,\alpha(z), \epsilon(z)) \nonumber \\
&= \left( \left. \frac{d\alpha}{dz} (z-\bar{z}) \right|_{\bar{z}} - (\alpha(z) - \alpha(\bar{z}))  \right) \left. \frac{\partial  \Xi }{\partial \alpha}  \right|_{\bar{z}}  + \left( \left. \frac{d\epsilon}{dz} \right|_{\bar{z}} (z-\bar{z}) - (\epsilon(z) - \epsilon(\bar{z})) \right) \left. \frac{\partial  \Xi }{\partial \epsilon}  \right|_{\bar{z}}   \,  , 
\end{align}
and taking the average we obtain 
\begin{align} \label{eq:GminusHbar}
& \overline{G(D,\mu,z)} - \overline{H(D,\mu,z,\alpha(z), \epsilon(z))} = - (\bar{\alpha} - \alpha(\bar{z})) \left. \frac{\partial  \Xi }{\partial \alpha}  \right|_{\bar{z}}  - (\bar{\epsilon} - \epsilon(\bar{z}))  \left. \frac{\partial  \Xi }{\partial \epsilon}  \right|_{\bar{z}}   \,  . 
\end{align}
We might now conveniently rewrite (\ref{eq:approxxiS}) as
\begin{align} \label{eq:approxxiSG}
\mathcal{I}_{\text{standard AP}}(D,\mu)  &= \overline{G(D,\mu,z) } \nonumber \\
&= \overline{H(D,\mu,z,\alpha(z), \epsilon(z))} - (\bar{\alpha} - \alpha(\bar{z})) \left. \frac{\partial  \Xi }{\partial \alpha}  \right|_{\bar{z}}  - (\bar{\epsilon} - \epsilon(\bar{z}))  \left. \frac{\partial  \Xi }{\partial \epsilon}  \right|_{\bar{z}}     \,  , 
\end{align}
where the first equality follows from the definition (\ref{eq:G}) and the last equality follows from (\ref{eq:GminusHbar}). 
Similarly we might express (\ref{eq:approxxiM}) in terms of the average of (\ref{eq:H}) and its error terms as 
\begin{align} \label{eq:approxxiMH}
\mathcal{I}_{\text{modified AP}}(D,\mu) &= H(D,\mu,\bar{z},\bar{\alpha},\bar{\epsilon}) + {}^{(2)}\Xi(D,\mu,\bar{z},\bar{\alpha},\bar{\epsilon}) + {}^{(2)}\mathcal{R}(D,\mu,\bar{z},\bar{\alpha},\bar{\epsilon}) \nonumber  \\
&= \overline{H(D,\mu,z,\alpha(z),\epsilon(z))} + {}^{(2)}\Xi(D,\mu,\bar{z},\bar{\alpha},\bar{\epsilon}) + {}^{(2)}\mathcal{R}(D,\mu,\bar{z},\bar{\alpha},\bar{\epsilon})  \, . 
\end{align}
Let us now quantify the accuracy of $\mathcal{I}_{\text{standard AP}}(D,\mu)$ and $\mathcal{I}_{\text{modified AP}}(D,\mu)$ as estimates of $\xi(D,\mu)$. 
From (\ref{eq:approxxi}), (\ref{eq:HErrorterm}), and (\ref{eq:approxxiSG}) we have 
\begin{align} \label{eq:approxxiDeltaS}
& \abs*{\xi(D,\mu) - \mathcal{I}_{\text{standard AP}}(D,\mu)}  \nonumber \\
&= \abs*{ \overline{{}^{(2)}\Xi(D,\mu,z,\alpha(z),\epsilon(z))} +\overline{ {}^{(2)}\mathcal{R}(D,\mu,z,\alpha(z),\epsilon(z))} + (\bar{\alpha} - \alpha(\bar{z})) \left. \frac{\partial  \Xi }{\partial \alpha}  \right|_{\bar{z}}  +  (\bar{\epsilon} - \epsilon(\bar{z}))  \left. \frac{\partial  \Xi }{\partial \epsilon}  \right|_{\bar{z}}  }   \,  , 
\end{align}
and from (\ref{eq:approxxi}), (\ref{eq:HErrorterm}), and (\ref{eq:approxxiMH}) we have 
\begin{align} \label{eq:approxxiDeltaM}
& \abs*{\xi(D,\mu) - \mathcal{I}_{\text{modified AP}}(D,\mu)} \nonumber    \\
&= \abs*{ \overline{{}^{(2)}\Xi(D,\mu,z,\alpha(z),\epsilon(z))} +\overline{ {}^{(2)}\mathcal{R}(D,\mu,z,\alpha(z),\epsilon(z))}  -  {}^{(2)}\Xi(D,\mu,\bar{z},\bar{\alpha},\bar{\epsilon}) - {}^{(2)}\mathcal{R}(D,\mu,\bar{z},\bar{\alpha},\bar{\epsilon})  }   \,  .
\end{align}
We might note that (\ref{eq:approxxiDeltaS}) contains terms which are first order in $\alpha(z) - \alpha(\bar{z})$ and $\epsilon(z) - \epsilon(\bar{z})$ respectively, while (\ref{eq:approxxiDeltaM}) contain only second and higher order terms in these separations. 

So far we have made no assumption on $\Xi(D,\mu,z,\alpha,\epsilon)$, $\alpha(z)$, and $\epsilon(z)$ as functions, other than assuming regularity conditions to be fulfilled. 
Let us suppose that $\alpha(z)$ and $\epsilon(z)$ are sufficiently bounded in terms of size of the variations $\alpha(z) - \alpha(\bar{z})$ and $\epsilon(z) - \epsilon(\bar{z})$ respectively within this redshift interval.
Further assume that the third order derivatives of $\Xi(D,\mu,z,\alpha,\epsilon)$ in $z$, $\alpha$, and $\epsilon$ can be sufficiently bounded within the redshift interval $\mathcal{Z}$ of integration, in such a way that the error term (\ref{eq:HErrorterm}) is dominated by its second order contribution and that ${}^{(2)}\mathcal{R}$ can be neglected. 
In this case we have 
\begin{align} \label{eq:approxxiDeltaSapprox}
& \abs*{\xi(D,\mu) - \mathcal{I}_{\text{standard AP}}(D,\mu)}  = \abs*{ \overline{{}^{(2)}\Xi(D,\mu,z,\alpha(z),\epsilon(z))} + (\bar{\alpha} - \alpha(\bar{z})) \left. \frac{\partial  \Xi }{\partial \alpha}  \right|_{\bar{z}}  +  (\bar{\epsilon} - \epsilon(\bar{z}))  \left. \frac{\partial  \Xi }{\partial \epsilon}  \right|_{\bar{z}}  }   \,  , 
\end{align}
and from (\ref{eq:approxxi}), \ref{eq:HErrorterm}, and (\ref{eq:approxxiMH}) we have 
\begin{align} \label{eq:approxxiDeltaMapprox}
& \abs*{\xi(D,\mu) - \mathcal{I}_{\text{modified AP}}(D,\mu)}  = \abs*{ \overline{{}^{(2)}\Xi(D,\mu,z,\alpha(z),\epsilon(z))}  -  {}^{(2)}\Xi(D,\mu,\bar{z},\bar{\alpha},\bar{\epsilon}) }   \,  .
\end{align}
If $\alpha(z)$ and $\epsilon(z)$ are varying sufficiently slowly that the remainder at second order of their expansion can be ignored along with the remainder ${}^{(2)}\mathcal{R}$, then the first order term in $\alpha(z) - \alpha(\bar{z})$ and $\epsilon(z) - \epsilon(\bar{z})$ in (\ref{eq:approxxiDeltaSapprox}) reduces to the second order term $ \frac{1}{2} \left.  \frac{\partial^2  \alpha }{\partial z^2} \frac{\partial  \Xi }{\partial \alpha}  \right|_{\bar{z}}  \overline{(z-\bar{z})^2} +   \left.  \frac{\partial^2  \epsilon }{\partial z^2}  \frac{\partial  \Xi }{\partial \epsilon}  \right|_{\bar{z}} \overline{(z-\bar{z})^2}$.
In this case, the competing terms in (\ref{eq:approxxiDeltaSapprox}) and (\ref{eq:approxxiDeltaMapprox}) can be considered to be of the same order. 
As long as there are no chance cancellations we therefore expect the approximations to be accurate at the same order. 
We will now consider cases where gradients of $\alpha(z)$ and $\epsilon(z)$ are not necessarily small. 

Let us consider the case where $\abs*{\overline{{}^{(2)}\Xi(D,\mu,z,\alpha(z),\epsilon(z))} } \neq 0$. In this case we can write $\abs*{ \overline{{}^{(2)}\Xi(D,\mu,z,\alpha(z),\epsilon(z))}  -  {}^{(2)}\Xi(D,\mu,\bar{z},\bar{\alpha},\bar{\epsilon}) } \leq \mathcal{K}_{D,\mu} \, \abs*{\overline{{}^{(2)}\Xi(D,\mu,z,\alpha(z),\epsilon(z))} }$, where $\mathcal{K}_{D,\mu}$ is some positive number which might be chosen differently for different values of $D,\mu$ and for each test model.
Then the modified AP scaling approximation is guaranteed to be better if $\abs*{(\bar{\alpha} - \alpha(\bar{z})) \left. \frac{\partial  \Xi }{\partial \alpha}  \right|_{\bar{z}}  +  (\bar{\epsilon} - \epsilon(\bar{z}))  \left. \frac{\partial  \Xi }{\partial \epsilon}  \right|_{\bar{z}}  } >  (\mathcal{K}_{D,\mu}+1) \abs*{\overline{{}^{(2)}\Xi(D,\mu,z,\alpha(z),\epsilon(z))} }$. 
The left and right hand side of this inequality are just the averages of the first and second order term respectively in the expansion of $\Xi$ (where the latter is scaled by $\mathcal{K}_{D,\mu}+1 \geq 1$). 
In general we expect the first order term to dominate of a well behaved expansion. We expect $\mathcal{K}_{D,\mu} \lesssim 1$ for most values of $D, \mu$ for model 2-point correlation functions which do not have extreme variations with redshift -- i.e., where systematics such as galaxy evolution and the distortion due to the choice of trial cosmology are not disturbing the 2-point correlation function by more than order unity. 

For a given test model with some set of specified AP functions $\alpha(z)$ and $\epsilon(z)$, we expect that there will be values of $D,\mu$ in the physical range of interest for which the second order term of the expansion of $\Xi$ dominates over the first order term, and we might even expect (\ref{eq:approxxiDeltaSapprox}) to be zero for some region of the domain of negligible measure. 
However, for most of the domain of $D,\mu$ we expect first order terms to dominate over second order terms.
In the special case where $\overline{{}^{(2)}\Xi(D,\mu,z,\alpha(z),\epsilon(z))} = 0$ the condition for the modified constant AP scaling approximation to work better than the standard constant AP scaling approximation reduces to $\abs*{(\bar{\alpha} - \alpha(\bar{z})) \left. \frac{\partial  \Xi }{\partial \alpha}  \right|_{\bar{z}}  +  (\bar{\epsilon} - \epsilon(\bar{z}))  \left. \frac{\partial  \Xi }{\partial \epsilon}  \right|_{\bar{z}}  } > \abs*{  {}^{(2)}\Xi(D,\mu,\bar{z},\bar{\alpha},\bar{\epsilon})  }$. 
This is a direct constraint on the relative size of the first and second order term of the expansion of $\Xi$ as evaluated at $\bar{z}, \bar{\alpha}, \bar{\epsilon}$. Again we expect the first order term to dominate except for cases where chance cancellations occur.

We conclude without rigorous proof that it is reasonable to assume that the modified constant AP scaling $\alpha(z) \mapsto \bar{\alpha}$, $\epsilon(z) \mapsto \bar{\epsilon}$ approximation is in general better or -- in case of sufficiently slowly varying $\alpha(z)$ and $\epsilon(z)$ -- equally good, as compared to the standard constant AP scaling approximation $\alpha(z) \mapsto \alpha(\bar{z})$, $\epsilon(z) \mapsto \epsilon(\bar{z})$.

\section{Bounds on the AP error terms}
\label{boundsAP}
In this appendix, we discuss bounds on the magnitude of the error terms $\Delta_{\alpha}$ and $\Delta_{\epsilon}$ defined in (\ref{eq:alphaerror}) and (\ref{eq:epsilonerror}) respectively. 

\subsection{Bounds on the magnitude of $\Delta_{\alpha}$}
\label{deltaalpha}
We shall be interested in bounding $\Delta_{\alpha}$ from above for various situations. Obviously, from its definition in (\ref{eq:alphaerror}), $\Delta_{\alpha}$ can be bounded if a bound on $\mathcal{R}^{\alpha}_{1}(z)$ of the first order expansion (\ref{eq:taylor}) is obtained. 
From Taylor's theorem the remainder term can be written on the form
\begin{align} \label{eq:remainderalpha}  
\mathcal{R}^{\alpha}_{1}(z) = \frac{1}{2}  \frac{ \partial^2 \alpha}{\partial z^2} (b_{\bar{z}}(z)) \, (z-\bar{z})^2  \,    
\end{align}
for each value of $z$, where $b_{\bar{z}}(z)$ is a real number between $\bar{z}$ and $z$. 

Bounding the remainder term $\mathcal{R}^{\alpha}_{1}(z)$, amounts to bounding the second derivative of $\alpha$
\begin{align} \label{eq:alphadz2}
\hspace*{-0.3cm} \frac{ \partial^2 \alpha}{\partial z^2} =  \frac{L^{\rm tr}}{L} \left[  \frac{\frac{ \partial^2 L^{\rm tr}}{\partial z^2}  }{L^{\rm tr}}  -  \frac{\frac{ \partial^2 L}{\partial z^2}  }{L}   \, - \, 2 \, \frac{\frac{ \partial L}{\partial z}  }{L} \left(   \frac{\frac{ \partial L^{\rm tr}}{\partial z}  }{L^{\rm tr}}  -  \frac{\frac{ \partial L}{\partial z}  }{L}     \right)   \right]  \, ,  \quad  L\equiv  \left( g^2_{\theta \theta} g_{z z}\right)^{\frac{1}{6}}  , \, L^{\rm tr} \equiv  \left( (g^{\rm tr}_{\theta \theta})^2 g^{\rm tr}_{z z}  \right)^{\frac{1}{6}}  
\end{align}
where $\alpha = L^{\rm tr} / L$ which follows from the definition of $\alpha$ (\ref{eq:alphaepsilon}).

Constraints on $L$ and its derivatives can now be turned into constraints on (\ref{eq:alphadz2}). 
Let us for instance assume that we are considering a class of model cosmologies which are bounded with respect to the \say{true} cosmology over the redshift range of the survey in the following sense 
\begin{align} \label{eq:boundsLLtrapp}
\hspace*{-0.1cm} M^{\text{min}}_{L \, 0} \leq \frac{L^{\rm tr}}{L} \leq M^{\text{max}}_{L \,0} , \qquad  \abs*{ \frac{  \left(  \frac{\partial L^{\rm tr}/ \partial z  }{L^{\rm tr}} \right)   }{  \left( \frac{ \partial L / \partial z  }{L} \right) } \, - \, 1 }  \leq M_{L \,1} \, , \qquad \abs*{ \frac{  \left(  \frac{\partial^2 L^{\rm tr}/ \partial z^2  }{L^{\rm tr}} \right)   }{  \left( \frac{ \partial^2 L / \partial z^2  }{L} \right) } \, - \, 1 }  \leq M_{L \,2} \, , 
\end{align}
while $L$ is bounded in its first and second derivatives 
\begin{align} \label{eq:boundsLapp}
\abs*{ \frac{\frac{ \partial L}{\partial z}  }{L} } \leq  \beta_{L \,1}  \, , \qquad  \abs*{ \frac{\frac{ \partial^2 L}{\partial z^2}  }{L} } \leq  \beta_{L \,2} , 
\end{align}
where $M^{\text{min}}_{L \, 0}$, $M^{\text{max}}_{L \, 0}$, $M_{L \,1}$, $M_{L \,2}$, $\beta_{L \,1}$, and $\beta_{L \,2}$ are all positive dimensionless constants. 
We then obtain the following upper bound on (\ref{eq:alphadz2}) expressed in terms of these constants 
\begin{align} \label{eq:boundalpharemainder}
\abs*{\frac{ \partial^2 \alpha}{\partial z^2}}  \, & \leq \, \frac{L^{\rm tr}}{L} \left(   \abs*{  \frac{\frac{ \partial^2 L}{\partial z^2}  }{L} \left(    \frac{   \frac{\partial^2 L^{\rm tr}/ \partial z^2  }{L^{\rm tr}}    }{   \frac{ \partial^2 L / \partial z^2  }{L} }   -  1  \right)  } \,  +  \, 2 \,   \left( \frac{\frac{ \partial L}{\partial z}  }{L}\right)^2   \abs*{ \frac{  \frac{\partial L^{\rm tr}/ \partial z  }{L^{\rm tr}}   }{   \frac{ \partial L / \partial z  }{L}  } \, - \, 1 }   \right) \nonumber \\ 
& \leq  \,  M^{\text{max}}_{L \, 0} \left(  \beta_{L \,2} M_{L \,2} + 2 \beta^2_{L \,1} M_{L \,1}   \right)  \, .
\end{align}
The first inequality follows from the triangle inequality and rearranging of the terms of (\ref{eq:alphadz2}), and the second inequality follows from (\ref{eq:boundsLLtrapp}) and (\ref{eq:boundsLapp}). 
The inequality in (\ref{eq:boundalpharemainder}) implies the following bound on the remainder $\mathcal{R}^{\alpha}_{1}(z)$ in (\ref{eq:remainderalpha})
\begin{align} \label{eq:remainder1}
\abs*{\mathcal{R}^{\alpha}_{1}(z)} = \frac{1}{2} \abs*{ \frac{ \partial^2 \alpha}{\partial z^2} (b_{\bar{z}}(z)) } \, (z-\bar{z})^2  \, \leq  \frac{1}{2}  M^{\text{max}}_{L \, 0} \left(  \beta_{L \,2} M_{L \,2} + 2 \beta^2_{L \,1} M_{L \,1}   \right) \, (z-\bar{z})^2 \, .
\end{align}
Finally we can use the bound (\ref{eq:remainder1}) to obtain bounds on the error term in (\ref{eq:alphaerror})
\begin{align} \label{eq:alphaexpandboundappendix}
\abs*{\Delta_{\alpha} } &=  \frac{1}{\alpha(\bar{z})} \abs*{ \int \, dz P(z) \mathcal{R}^{\alpha}_{1}(z) } \, \leq  \, \frac{1}{\alpha(\bar{z})} \int \, dz P(z) \abs*{ \mathcal{R}^{\alpha}_{1}(z) }   \nonumber  \\
& \, \leq \,  \frac{1}{2} \frac{ M^{\text{max}}_{L \, 0}}{ M^{\text{min}}_{L \, 0} } \left(  \beta_{L \,2} M_{L \,2} + 2 \beta^2_{L \,1} M_{L \,1}   \right) \, \overline{(z-\bar{z})^2}  , 
\end{align}
where the lower bound on $\alpha = L^{\rm tr} / L$ in (\ref{eq:boundsLLtrapp}) has been used in the final inequality. 

\subsection{Bounds on the magnitude of $\Delta_{\epsilon}$}
\label{deltaepsilon}
We shall now bound the magnitude of the error term $\Delta_{\epsilon}$ (\ref{eq:epsilonerror}) in a similar fashion as done for $\Delta_{\alpha}$ in appendix \ref{deltaalpha}. 
We can write the remainder term $\mathcal{R}^{\epsilon}_{1}(z)$ (\ref{eq:taylor}) of the first order expansion
\begin{align} \label{eq:remainderepsilon}  
\mathcal{R}^{\epsilon}_{1}(z)  =\frac{1}{2}  \frac{ \partial^2 \epsilon}{\partial z^2} (c_{\bar{z}}(z)) \, (z-\bar{z})^2 \, 
\end{align}
for each value of $z$, where $c_{\bar{z}}(z)$ is a real number between $\bar{z}$ and $z$.

In a similar way to (\ref{eq:alphadz2}) we write the second derivative of $\epsilon$ in terms of first and second derivatives of metric combinations of the models 
\begin{align} \label{eq:epsilondz2}
\frac{ \partial^2 \epsilon}{\partial z^2} = \frac{1}{3} \left( \frac{R^{\rm tr}}{R} \right)^{\frac{1}{3}} \left(  \frac{\frac{ \partial^2 R^{\rm tr}}{\partial z^2}  }{R^{\rm tr}}  -  \frac{\frac{ \partial^2 R}{\partial z^2}  }{R}   \, - \, 2 \, \frac{\frac{ \partial R}{\partial z}  }{R} \left(   \frac{\frac{ \partial R^{\rm tr}}{\partial z}  }{R^{\rm tr}}  -  \frac{\frac{ \partial R}{\partial z}  }{R}     \right) - \frac{2}{3} \left(   \frac{\frac{ \partial R^{\rm tr}}{\partial z}  }{R^{\rm tr}}  -  \frac{\frac{ \partial R}{\partial z}  }{R}     \right)^2   \right)  \, , 
\end{align}
where $R \equiv ( g_{z z} / g_{\theta \theta} )^{1/2}$ and $R^{\rm tr} \equiv ( g^{\rm tr}_{z z} / g^{\rm tr}_{\theta \theta} )^{1/2}$ are relative distance scales of the models, and where $(1+\epsilon)^3 = \frac{R^{\rm tr}}{R}$ from the definition of epsilon in (\ref{eq:alphaepsilon}).

Similarly to the case of $\alpha$, the second derivative of $\epsilon$ can be bounded as a function of bounds on the metric combination $R$ and its first and second derivatives. 
Let us consider a class of models which are bounded with respect to the \say{true} cosmological model within the redshift interval of the survey in the following way:
\begin{align} \label{eq:boundsRRtrapp}
\hspace*{-0.3cm} \left( \frac{R^{\rm tr}}{R} \right)^{\frac{1}{3}}\leq M^{\text{max}}_{R \, 0}  , \qquad  \abs*{ \frac{  \left(  \frac{\partial R^{\rm tr}/ \partial z  }{R^{\rm tr}} \right)   }{  \left( \frac{ \partial R / \partial z  }{R} \right) } \, - \, 1 }  \leq M_{R \,1} \, , \qquad \abs*{ \frac{  \left(  \frac{\partial^2 R^{\rm tr}/ \partial z^2  }{R^{\rm tr}} \right)   }{  \left( \frac{ \partial^2 R / \partial z^2  }{R} \right) } \, - \, 1 }  \leq M_{R \,2} \, , 
\end{align}
and where $R$ is bounded in its first and second derivative as 
\begin{align} \label{eq:boundsRapp}
\abs*{ \frac{\frac{ \partial R}{\partial z}  }{R} } \leq  \beta_{R \,1}  \, , \qquad  \abs*{ \frac{\frac{ \partial^2 R}{\partial z^2}  }{R} } \leq  \beta_{R \,2} \, . 
\end{align}
We can now bound the second derivative of $\epsilon$ (\ref{eq:epsilondz2}) based on the above bounds as follows 
\begin{align} \label{eq:epsilondz2bound}
\abs*{\frac{ \partial^2 \epsilon}{\partial z^2}}  & \leq \frac{1}{3} \left( \frac{R^{\rm tr}}{R} \right)^{\frac{1}{3}}   \abs*{  \frac{\frac{ \partial^2 R}{\partial z^2}  }{R} \left(    \frac{  \frac{\partial^2 R^{\rm tr}/ \partial z^2  }{R^{\rm tr}}  }{ \frac{ \partial^2 R / \partial z^2  }{R} }   -  1  \right)  } \nonumber\\  &\hbox to 70pt{\hfil}+  \, \frac{2}{3} \,\left( \frac{R^{\rm tr}}{R} \right)^{\frac{1}{3}}   \left( \frac{\frac{ \partial R}{\partial z}  }{R}\right)^2 \left(  \abs*{ \frac{  \frac{\partial R^{\rm tr}/ \partial z  }{R^{\rm tr}}    }{ \frac{ \partial R / \partial z  }{R}  } \, - \, 1 }  + \frac{1}{3} \abs*{ \frac{ \frac{\partial R^{\rm tr}/ \partial z  }{R^{\rm tr}}  }{ \frac{ \partial R / \partial z  }{R} } \, - \, 1 }^2   \right) \nonumber \\
&\leq \frac{1}{3} M^{\text{max}}_{R \, 0} \left( \beta_{R \,2} M_{R \,2}  + 2 \beta^2_{R \,1} \left(  M_{R \,1} + \frac{1}{3} M_{R \,1}^2  \right)     \right) \, ,
\end{align}
where the first inequality follows from the triangle inequality, and the second inequality follows from (\ref{eq:boundsRRtrapp}) and (\ref{eq:boundsRapp}). 
We can now bound the remainder $\mathcal{R}^{\epsilon}_{1}(z)$ (\ref{eq:remainderepsilon}) in a similar manner to (\ref{eq:remainder1}), and use the result for bounding the error term $\Delta_{\epsilon}$.
The result reads 
\begin{align} \label{eq:epsilonerrorboundapp}
\hspace*{-0.3cm} \abs*{\Delta_{\epsilon}} \, \leq \, \int \, dz P(z) \abs*{\mathcal{R}^{\epsilon}_{1}(z)} \, \leq \,  \frac{1}{6} M^{\text{max}}_{R \, 0} \left( \beta_{R \,2} M_{R \,2}  + 2 \beta^2_{R \,1} \left(  M_{R \,1} + \frac{1}{3} M_{R \,1}^2  \right)   \right)  \, \overline{(z-\bar{z})^2}        \, , 
\end{align}
which follows from the triangle inequality, the bound (\ref{eq:epsilondz2bound}), and the definition of the remainder term (\ref{eq:remainderepsilon}).

\end{appendices}

\end{document}